\documentclass{aa}  

\usepackage{graphicx}
\usepackage{txfonts}
\usepackage{comment}
\usepackage{color}
\usepackage[usenames,dvipsnames,svgnames,table]{xcolor}
\usepackage{ulem}

\usepackage{orcidlink}
\begin{document}

   \title{Monitoring the large-scale magnetic field of AD~Leo with SPIRou, ESPaDOnS, and Narval}
   \subtitle{Towards a magnetic polarity reversal?}

   \titlerunning{Near-infrared Zeeman-Doppler imaging of AD Leo with SPIRou}

   \author{S. Bellotti\inst{1,2}\orcidlink{0000-0002-2558-6920}
           \and
           J. Morin \inst{3}\orcidlink{0000-0002-4996-6901}
           \and
           L. T. Lehmann\inst{1}\orcidlink{0000-0001-5674-2116}
           \and
           C. P. Folsom \inst{4}\orcidlink{0000-0002-9023-7890}
           \and 
           G. A. J. Hussain \inst{2}\orcidlink{0000-0003-3547-3783}
           \and 
           P. Petit \inst{1}\orcidlink{0000-0001-7624-9222}
           \and 
           J-F. Donati \inst{1}\orcidlink{0000-0001-5541-2887}
           \and
           A. Lavail\inst{1,5}\orcidlink{0000-0001-8477-5265}
           \and
           A. Carmona \inst{6}\orcidlink{0000-0003-2471-1299}
           \and
           E. Martioli \inst{7,8}\orcidlink{0000-0002-5084-168X}
           \and
           B. Romano Zaire\inst{9}\orcidlink{0000-0002-9328-9530}
           \and
           E. Alecian\inst{6}\orcidlink{0000-0001-5260-7179}
           \and
           C. Moutou \inst{1}\orcidlink{0000-0002-2842-3924}
           \and
           P. Fouqu\'e\inst{1}\orcidlink{0000-0002-1436-7351}
           \and
           S. Alencar\inst{9}\
           \and
           E. Artigau\inst{10}\orcidlink{0000-0003-3506-5667}
           \and
           I. Boisse\inst{11}\orcidlink{0000-0002-1024-9841}
           \and
           F. Bouchy\inst{12}\orcidlink{0000-0002-7613-393X}
           \and
           C. Cadieux\inst{10}\orcidlink{0000-0001-9291-5555}
           \and
           R. Cloutier\inst{15}\orcidlink{0000-0001-5383-9393}
           \and
           N.~J. Cook\inst{10}\orcidlink{0000-0003-4166-4121}
           \and
           X. Delfosse \inst{6}\orcidlink{0000-0001-5099-7978}
           \and
           R. Doyon\inst{10}\orcidlink{0000-0001-5485-4675}
           \and
           G. H\'ebrard\inst{8,13}\orcidlink{0000-0001-5450-7067}
           \and
           O. Kochukhov \inst{5}\orcidlink{0000-0003-3061-4591}
           \and 
           G. A. Wade \inst{14}
          }

   \institute{
            Institut de Recherche en Astrophysique et Plan\'etologie,
            Universit\'e de Toulouse, CNRS, IRAP/UMR 5277,
            14 avenue Edouard Belin, F-31400, Toulouse, France\\
            \email{stefano.bellotti@irap.omp.eu}
        \and
             Science Division, Directorate of Science, 
             European Space Research and Technology Centre (ESA/ESTEC),
             Keplerlaan 1, 2201 AZ, Noordwijk, The Netherlands
        \and
             Laboratoire Univers et Particules de Montpellier,
             Universit\'e de Montpellier, CNRS,
             F-34095, Montpellier, France
        \and 
            Tartu Observatory, 
            University of Tartu, 
            Observatooriumi 1, Tõravere, 61602 Tartumaa, Estonia
        \and
            Department of Physics and Astronomy,
            Uppsala University, 
            Box 516, SE-75120 Uppsala, Sweden
        \and
            Univ. Grenoble Alpes, CNRS, IPAG, 38000 Grenoble, France
        \and     
           Laborat\'{o}rio Nacional de Astrof\'{i}sica, Rua Estados Unidos 154, 37504-364, Itajub\'{a} - MG, Brazil
        \and 
           Institut d'Astrophysique de Paris, CNRS, UMR 7095, Sorbonne Universit\'{e}, 98 bis bd Arago, 75014 Paris, France
        \and
            Universidade Federal de Minas Gerais, Belo Horizonte, MG, 31270-901, Brazil
        \and
            Universit\'e de Montr\'eal, D\'epartement de Physique, IREX,
            Montr\'eal, QC H3C 3J7, Canada
        \and
            Aix Marseille Univ, CNRS, CNES, LAM, Marseille, France
        \and
            Observatoire de Gen\`eve, Universit\'e de Gen\`eve, Chemin Pegasi, 51, 1290 Sauverny, Switzerland
        \and
            Observatoire de Haute Provence, St Michel l'Observatoire, France
        \and
            Department of Physics \& Space Science,
            Royal Military College of Canada,
            PO Box 17000 Station Forces, Kingston, ON, Canada K7K 0C6
        \and 
            Department of Physics \& Astronomy, 
            McMaster University, 
            1280 Main St West, Hamilton, ON, L8S 4L8, Canada
}

   \date{Received ; accepted }

 
  \abstract
   {One clear manifestation of dynamo action on the Sun is the 22-yr magnetic cycle, exhibiting a polarity reversal and a periodic conversion between poloidal and toroidal fields. For M~dwarfs, several authors claim evidence of activity cycles from photometry and analyses of spectroscopic indices, but no clear polarity reversal has been identified from spectropolarimetric observations. These stars are excellent laboratories to investigate dynamo-powered magnetic fields under different stellar interior conditions, that is partly or fully convective.}
   {Our aim is to monitor the evolution of the large-scale field of AD~Leo, which has shown hints of a secular evolution from past dedicated spectropolarimetric campaigns. This is of central interest to inform distinct dynamo theories, contextualise the evolution of the solar magnetic field, and explain the variety of magnetic field geometries observed in the past.}
   {We analysed near-infrared spectropolarimetric observations of the active M~dwarf AD~Leo taken with SPIRou between 2019 and 2020 and archival optical data collected with ESPaDOnS and Narval between 2006 and 2019. We searched for long-term variability in the longitudinal field, the width of unpolarised Stokes profiles, the unsigned magnetic flux derived from Zeeman broadening, and the geometry of the large-scale magnetic field using both Zeeman-Doppler imaging and principal component analysis.}
   {We found evidence of a long-term evolution of the magnetic field, featuring a decrease in axisymmetry (from 99\% to 60\%). This is accompanied by a weakening of the longitudinal field ($-$300 to $-$50\,G) and a correlated increase in the unsigned magnetic flux (2.8 to 3.6\,kG). Likewise, the width of the mean profile computed with selected near-infrared lines manifests a long-term evolution corresponding to field strength changes over the full time series, but does not exhibit modulation with the stellar rotation of AD~Leo in individual epochs.}
   {The large-scale magnetic field of AD Leo manifested first hints of a polarity reversal in late 2020 in the form of a substantially increased dipole obliquity, while the topology remained predominantly poloidal and dipolar for 14\,yr. This suggests that low-mass M~dwarfs with a dipole-dominated magnetic field can undergo magnetic cycles.}


   \maketitle
%

\section{Introduction}\label{sec:introduction}

Studying stellar surface magnetic fields yields relevant insights into the internal structure of stars, as well as their essential role in stellar formation, evolution, and activity \citep[][]{Donati2009}. For cool stars, monitoring secular changes of the field's configuration provides useful feedback on the dynamo processes operating in the stellar interior and constraints on stellar wind models. The latter is fundamental to understanding atmospheric hydrodynamic escape of embedded planets since magnetic cycles modulate the star's activity level and thus its radiation output \citep{Vidotto2020,Hazra2020}.

The Sun is an important benchmark in this context: its long-term monitoring revealed a periodic variation in sunspot number, size, and latitude \citep{Schwabe1844, Maunder1904, Hathaway2010}, and a polarity reversal of the large-scale magnetic field over a timescale of 11\,yr \citep{Hale1919}. The proposed mechanism to reproduce these phenomena theoretically is the $\alpha\Omega$ dynamo \citep{Parker1955, Charbonneau2010}, namely the combination of differential rotation and cyclonic turbulence at the interface between the radiative and convective zones, known as tachocline. A different model is the Babcock-Leighton mechanism, which describes the conversion from a toroidal to poloidal field via a poleward migration of bipolar magnetic regions \citep{Babcock1961, Leighton1969}. However, there is still no model that can account for all the solar magnetic processes \citep{Petrovay2020}.

For other stars, magnetic field measurements can be performed with two complementary approaches \citep{Morin2012, Reiners2012}. One is to model the Zeeman splitting in individual unpolarised spectral lines and estimate the total unsigned magnetic field, which is insensitive to polarity cancellation. The other is to apply tomographic techniques that use the polarisation properties of the Zeeman-split components to recover the orientation of the local field. In addition to these well-established methods, \citet{Lehmann2022} show that fundamental properties of the large-scale field topology can be derived directly from the circularly polarised Stokes~$V$ time series using principal component analysis (PCA), without prior assumptions. This method allows us to qualitatively infer the predominant component of the field topology, as well as its complexity, axisymmetry, and evolution. Altogether, these observational constraints guide dynamo theories to a comprehensive description of the magnetic field generation and dynamic nature in the form of magnetic cycles \citep{Reiners2010,Gregory2012,See2016}.

Over the last three decades, Zeeman-Doppler imaging (ZDI, \citealt{Semel1989,DonatiBrown1997}) has been applied to reconstruct the poloidal and toroidal components of stellar magnetic fields, providing evidence of a wide variety of the large-scale magnetic topologies \citep[e.g.,][]{Morin2016}. Among rapidly rotating cool stars, the partly convective ones with masses above 0.5\,M$_\odot$ tend to have moderate, predominantly toroidal large-scale fields generally featuring a non-axisymmetric poloidal component \citep{Petit2008, Donati2008, See2015}. Those with masses between 0.2\,M$_\odot$ and 0.5\,M$_\odot$ -- close to the fully convective boundary at 0.35\,M$_\odot$ \citep{Chabrier1997} -- generate stronger large-scale magnetic fields, dominated by a poloidal and axisymmetric component. For fully convective stars with M$<$0.2\,M$_\odot$, spectropolarimetric analyses have revealed a dichotomy of field geometries: either strong, mostly axisymmetric dipole-dominated or weak, non-axisymmetric multipole-dominated large-scale fields are observed \citep{Morin2010}. The latter findings could be understood either as a manifestation of dynamo bistability \citep{Morin2011, Gastine2013}, that is two dynamo branches that coexist over a range of stellar rotation periods and masses, or of long magnetic cycles, implying that different topologies correspond to different phases of the cycle \citep{Kitchatinov2014}. Yet, no firm conclusion has been reached. In parallel, studies relying on the analysis of unpolarised spectra have shown that the average (unsigned) surface magnetic field of cool stars follows a classical rotation-activity relation including a non-saturated and a saturated (or quasi-saturated) regime, without a simple relation with the large-scale magnetic geometry \citep{ReinersBasri2009,Shulyak2019,Kochukhov2021,Reiners2022}. Similarly, recent dynamo simulations conducted by, for instance, \citet{Zaire2022} confirm that the influence of rotation on convective motions alone could not explain the observed variety of magnetic geometry. Only in the case of fully convective very fast rotators, \cite{Shulyak2017} found that the strongest average fields were measured for stars with large-scale dipole-dominated fields. \citet{Kochukhov2021} show that the fraction of magnetic energy contained in the large-scale field component is also the highest for these stars.

Cyclic trends for Sun-like stars were found via photometric and chromospheric activity (i.e. Ca \textsc{II} H\&K lines) monitoring, and timescales shorter (e.g., 120\,d for $\tau$~Boo, \citealt{Mittag2017}) or longer ($\simeq$ 20\,yr for HD\,1835; \citealt{BoroSaikia2018}) than the solar magnetic cycle were reported \citep{Wilson1968, Baliunas1995, BoroSaikia2018}. Moreover, polarity flips of the large-scale field were detected for a handful of stars based on optical spectropolarimetric observations \citep{Donati2008b,Petit2009, Fares2009, Morgenthaler2011, BoroSaikia2016, Rosen2016, Jeffers2018, BoroSaikia2022, Jeffers2022}. For M~dwarfs, numerous studies relying on photometry and spectroscopic indices claimed evidence of activity cycles \citep[e.g.,][]{GomesDaSilva2012,Robertson2013,Mignon2023}, and radio observations suggest the occurrence of polarity reversal at the end of the main sequence \citep[][]{Route2016},  but no polarity reversal has been directly observed with spectropolarimetry so far. This motivates long-term spectropolarimetric surveys, to reveal secular changes in the field topology and shed more light on the dynamo processes in action.

A well-known active M~dwarf is AD~Leo (GJ~388), whose mass (0.42\,M$_\odot$) falls at the boundary between the domains where toroidal- and dipole-dominated magnetic topologies have previously been identified, and thus represents an interesting laboratory to study stellar dynamos. \citet{Morin2008} analysed the large-scale magnetic field from spectropolarimetric data sets collected with Narval at T\'elescope Bernard-Lyot in 2007 and 2008 and reported a stable, axisymmetric, dipole-dominated geometry. Later, \citet{Lavail2018} examined data collected with ESPaDOnS at Canada-France-Hawaii Telescope (CFHT) from 2012 and 2016, and showed an evolution of the field in the form of a global weakening (about 20\%) and small-scale enhancement. The latter was quantitatively expressed by a decrease in the magnetic filling factor (from 13\% to 7\%), meaning that the field was more intense on local scales. No polarity reversal was reported on AD~Leo \citep{Lavail2018}. The large-scale magnetic topology has remained stable since spectropolarimetric observations of AD~Leo have been initiated (2007--2016): dominated by a strong axial dipole, the visible pole corresponding to negative radial field (magnetic field vector directed towards the star).

Here, we extend the magnetic analysis of AD~Leo using  both new optical ESPaDOnS observations collected in 2019 and near-infrared spectropolarimetric time series collected with SPIRou at CFHT in 2019 and 2020 under the SPIRou Legacy Survey (SLS), which adds to the previous optical data sets collected with ESPaDOnS and Narval between 2006 and 2016. The aim is to apply distinct techniques to search for long-term variations that may or may not resemble the solar behaviour. 

The paper is structured as follows: in Sec.~\ref{sec:observations} we describe the observations performed in the near-infrared and optical domains, in Sec.~\ref{sec:mag_analysis} we outline the temporal analysis of the longitudinal magnetic field, the Full-Width at Half Maximum (FWHM) of the Stokes $I$ profile, and the total magnetic flux inferred from Zeeman broadening modelling. Then, we describe the magnetic geometry reconstructions by means of ZDI and PCA. In Sec.~\ref{sec:discussion} we discuss the wavelength dependence of magnetic field measurements and in Sec.~\ref{sec:conclusion} we present our conclusions.

\section{Observations}\label{sec:observations}

AD~Leo is an M3.5~dwarf with a $V$ and $H$ band magnitude of 9.52 and 4.84, respectively \citep{Zacharias2013}, at a distance of 4.9651$\pm$0.0007~pc \citep{GaiaEDR3}. Its age was estimated to be within 25 and 300~Myr by \citet{Shkolnik2009}. AD~Leo has a rotation period of 2.23\,days \citep{Morin2008,Carmona2023} and an inclination $i = 20^\circ$, implying an almost pole-on view \citep{Morin2008}. Its high activity level is seen in frequent flares \citep{Muheki2020,Namekata2020} and quantified by an X-ray-to-bolometric luminosity ratio ($\log(\mathrm{L}_\mathrm{X}/\mathrm{L}_\mathrm{bol})$) of -3.62 \citep{Wright2011} and a mean Ca\textsc{II} H\&K index ($\mathrm{logR'}_\mathrm{HK}$) of -4.00 \citep{BoroSaikia2018}.

AD~Leo's mass is 0.42\,$M_\odot$ \citep{Mann2015,Cristofari2023}, which places it above the theoretical fully convective boundary at 0.35\,$M_\odot$ \citep{Chabrier1997}. The latter value is in agreement with observations, as it has been invoked to explain the dearth of stars with $M_G\sim$10.2, known as {\it Gaia} magnitude gap \citep{Feiden2021}. However, it is not an absolute limit: age \citep{Maeder2000} and metallicity affect the depth of the convective envelope \citep{VanSaders2012,Tanner2013}, and the presence of strong magnetic fields quenches convection and could push the theoretical boundary towards later spectral type \citep{Mullan2001}.

\begin{figure}[t]
    \centering
    \includegraphics[width=\columnwidth]{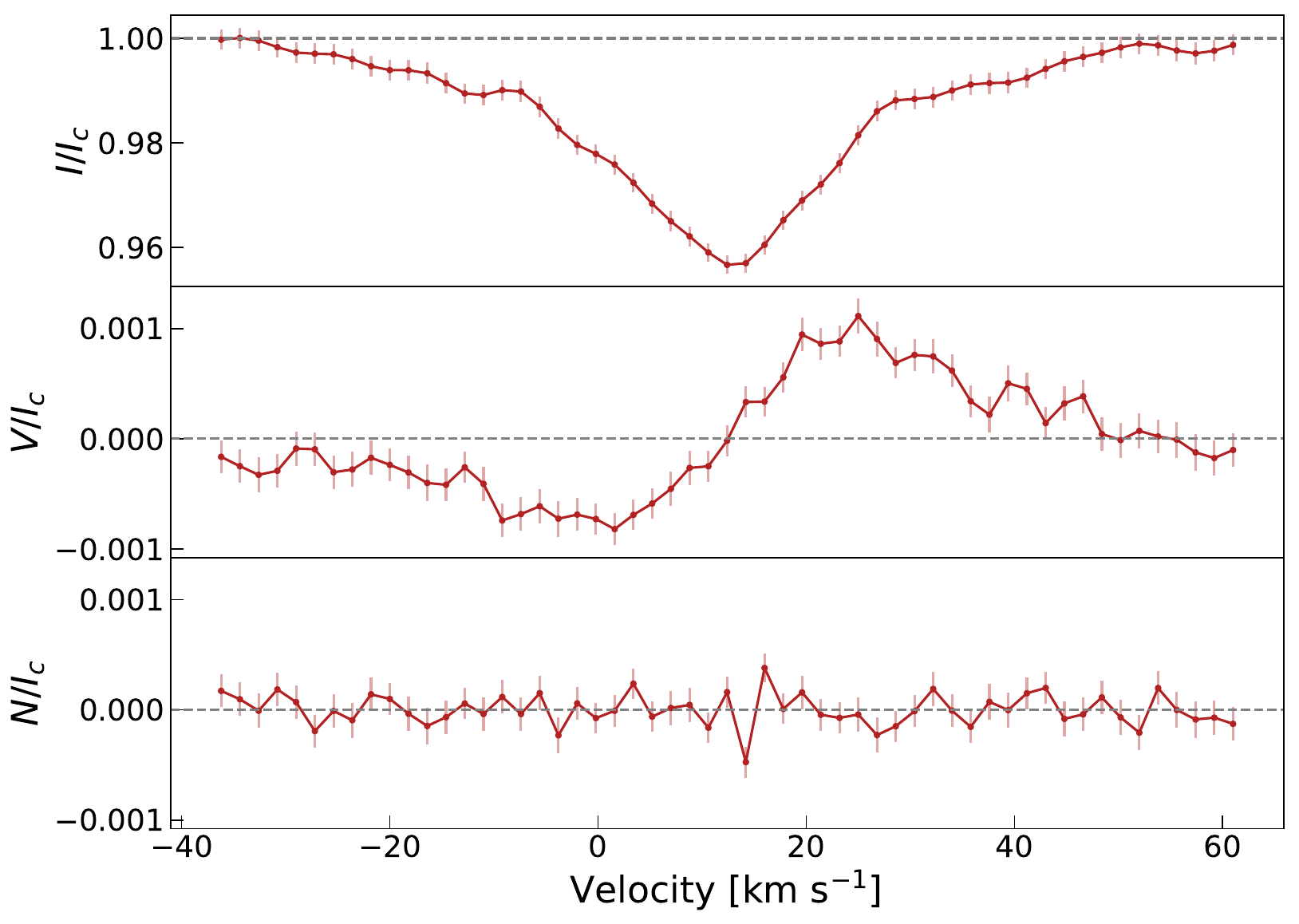}
    \caption{Example Stokes profiles for AD~Leo for the 9th~December~2019 observation collected with SPIRou. From the top: Stokes~$I$ profile (unpolarised); Stokes~$V$ profile (circularly polarised) with a clear Zeeman signature of S/N = 6\,500, and Null $N$~profile, used for quality check of the Stokes profiles \citep{Donati1997,Bagnulo2009}. The LSD profiles were obtained using a mask containing 838 lines. In all panels, the units are relative to the unpolarised continuum.}
    \label{fig:Stokes}%
\end{figure}

\subsection{Near-infrared}
\label{sec:observationsNIR}

A total of 77 spectropolarimetric observations in the near-infrared were collected with the SpectroPolarim\`etre InfraRouge (SPIRou) within the SLS. SPIRou is a stabilised high-resolution near-infrared spectropolarimeter \citep{Donati2020} mounted on the 3.6~m CFHT atop Maunakea, Hawaii. It provides a full coverage of the near-infrared spectrum from 0.96 to $2\mu$m at a spectral resolving power of $R \sim 70,000 $. Optimal extraction of SPIRou spectra was carried out with {\it A PipelinE to Reduce Observations} (\texttt{APERO} v0.6.132), a fully automatic reduction package installed at CFHT \citep{Cook2022}. The same data set was used in \citet{Carmona2023} to perform a velocimetric study and reject the hypothesis of a planetary companion by \citet{Tuomi2018} in favour of activity-induced variations, in agreement with \citet{Carleo2020}.

Observations were performed in circular polarisation mode between February 2019 and June 2020, spanning 482 days in total; the journal of observations is available in Table~\ref{tab:log}. The mean airmass is 1.32 and the signal-to-noise ratio (S/N) at $1,650$~nm per spectral element ranges from 68 to 218, with an average of 168. We applied least-squares deconvolution (LSD) to atomic spectral lines to derive averaged-line Stokes~$I$ (unpolarised) and $V$ (circularly polarised) profiles \citep{Donati1997,Kochukhov2010}. This numerical technique assumes the spectrum to be the convolution between a mean line profile and a line mask, that is to say a series of Dirac delta functions centred at each absorption line in the stellar spectrum, with corresponding depths and Land\'e factors (i.e. sensitivities to the Zeeman effect at a given wavelength). The output mean line profile gathers the information of thousands of spectral lines and, because of the consequent high S/N, enables the extraction of polarimetric information from the spectrum. The adopted line mask was generated using the Vienna Atomic Line Database\footnote{\url{http://vald.astro.uu.se/}} \citep[VALD,][]{Ryabchikova2015} and a MARCS atmosphere model \citep{Gustafsson2008} with $T_{\mathrm{eff}}=3,500$\,K, $\log g=$ 5.0\,[cm s$^{-2}$] and $v_{\mathrm{micro}}=$ 1~km s$^{-1}$. It contains $1,400$ atomic lines between 950--$2,600$~nm and with known Land\'e factor (ranging from 0 to 3) and with depth larger than 3\,\% of the continuum level.

We discarded six observations in February 2019 since one optical component of the instrument was not working nominally, one observation in November 2019 because likely affected by a flare (the corresponding radial velocity is $>$8 sigma lower than the bulk of the measurements) and two observations in 2020 as they led to noisier (by a factor of 10) LSD profiles. Therefore, the data set analysed in this work comprises 68 polarimetric sequences, whose characteristics are reported in Table~\ref{tab:log}.

The near-infrared observations were performed monthly between 2019 and 2020, except for two gaps of approximately two and three months. There is also a gap of 1.5 month between the end of 2019 and beginning of 2020. We thus split the time series in four epochs to maintain coherency of magnetic activity over short time scales and for clearer visualisation: 2019a (15th April 2019 to 21st June 2019, i.e. 2019.29 to 2019.47), 2019b (16th October 2019 to 12th December 2019, i.e. 2019.79 to 2019.95) 2020a (26th January 2020 to 12th March 2020, i.e. 2020.07 to 2020.19), and 2020b (8th May 2020 to 10th June 2020, i.e. 2020.35 to 2020.44).

The near-infrared domain covered by SPIRou is polluted by strong and wide telluric bands due to Earth's atmospheric absorption. Their contribution to the stellar spectra is corrected using a telluric transmission model (which is built from observations of standard stars since the start of SPIRou operations, and using Transmissions of the AtmosPhere for AStromomical data (TAPAS) atmospheric model \citealt{Bertaux2014}) and a PCA method implemented in the \texttt{APERO} pipeline \citep{Artigau2014}. To account for potential residuals in the telluric correction, we ignored the following intervals of the spectrum when computing the LSD profiles: [950, 979], [1116, 1163], [1331, 1490], [1790, 1985], [1995,2029], [2250, 2500]~nm. These intervals correspond to H$_2$O absorption regions, with transmission typically smaller than 40\%. We assessed whether removing these telluric intervals optimises the quality of the Stokes~$V$ profiles. In a first test, we searched for stellar absorption lines deeper than 75\,\% of the continuum level and within $\pm100$\,km\ s$^{-1}$ from telluric lines included in the transmission model of \texttt{APERO}. This approach allowed us to identify stellar lines that are contaminated by the telluric lines throughout the year. When the telluric-affected spectral lines were removed, no significant improvement was reported in the final LSD profiles, indicating a robust telluric correction as already reported in \citep{Carmona2023}. In a second test, we extended the intervals by 25 and 50~nm or reduced them by 10~nm and noticed an increment of the noise level in LSD profiles up to 20\,\%, so we proceeded with the previous intervals.

Accounting for the ignored telluric intervals, the number of spectral lines used in LSD is 838. We show the LSD Stokes profiles for one example observation in Fig.~\ref{fig:Stokes}. The average noise level in Stokes $V$ for the entire time series is $1.6\cdot10^{-4}$ relative to the unpolarised continuum, similar to the optical domain \citep{Morin2008}. We also note that the profiles are broader than in the optical by more than $10$ km s$^{-1}$, owing to a stronger Zeeman effect in the near-infrared domain \citep{Zeeman1897}.

\subsection{Optical}

For most of the analyses presented here, we considered all archival observations collected with ESPaDOnS and Narval, and studied previously in \citet{Morin2008} and \citet{Lavail2018}. We also included six new observations taken in November 2019 (from 2019.87 to 2019.89) with ESPaDOnS for CFHT programme 19BC06, PI A. Lavail  (reported in Table~\ref{tab:log_opt}). They are contemporaneous to the SPIRou ones for the same period, hence enabling us to study the dependence of the measured magnetic field strength on the wavelength domain employed (see Sec.~\ref{sec:discussion}).

ESPaDOnS is the optical spectropolarimeter on the 3.6~m CFHT located atop Mauna Kea in Hawaii, and Narval is the twin instrument on the 2~m TBL at the Pic du Midi Observatory in France \citep{Donati2003}. The data reduction was performed with the \texttt{LIBRE-ESPRIT} pipeline \citep{Donati1997}, and the reduced spectra were retrieved from the PolarBase archive \citep{Petit2014}.

The LSD profiles were computed similar to the near-infrared, but using an optical VALD mask containing 3330 lines in range 350-1080 nm and with depths larger than 40\% the continuum level, similar to \citet{Morin2008,Bellotti2021}. The number of lines used is 3240 and accounts for the removal of the following wavelength intervals, which are affected by telluric lines or in the vicinity of H$\alpha$: [627,632], [655.5,657], [686,697], [716,734], [759,770], [813,835], and [895,986] nm. For the 2019 observations, the average noise in Stokes $V$ is $3\cdot10^{-4}$ relative to the unpolarised continuum.

In the next sections, the near-infrared and optical observations will be phased with the following ephemeris:
\begin{equation}
    \mathrm{HJD} = 2458588.7573 + \mathrm{P}_\mathrm{rot}\cdot n_\mathrm{cyc} \text{ , }
    \label{eq:ephemeris}
\end{equation}
where we used the first SPIRou observation taken in April 2019 as reference, P$_\mathrm{rot}=$2.23\,days is the stellar rotation period \citep{Morin2008}, and $n_\mathrm{cyc}$ corresponds to the rotation cycle (see Table~\ref{tab:log}).

\section{Magnetic analysis}\label{sec:mag_analysis}

\subsection{Longitudinal magnetic field}\label{sec:B_lon}

\begin{figure}[!t]
    \centering
    \includegraphics[width=\columnwidth]{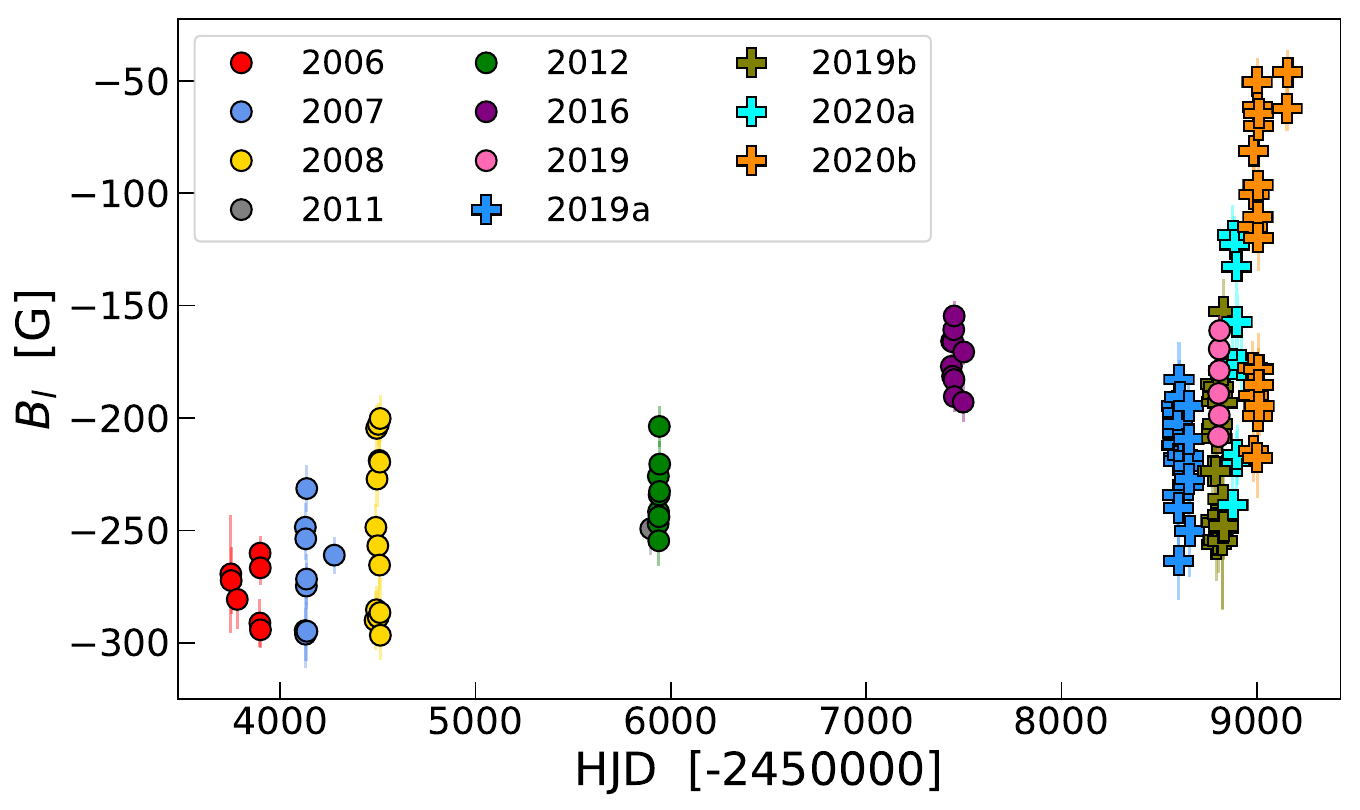}
    \includegraphics[width=\columnwidth]{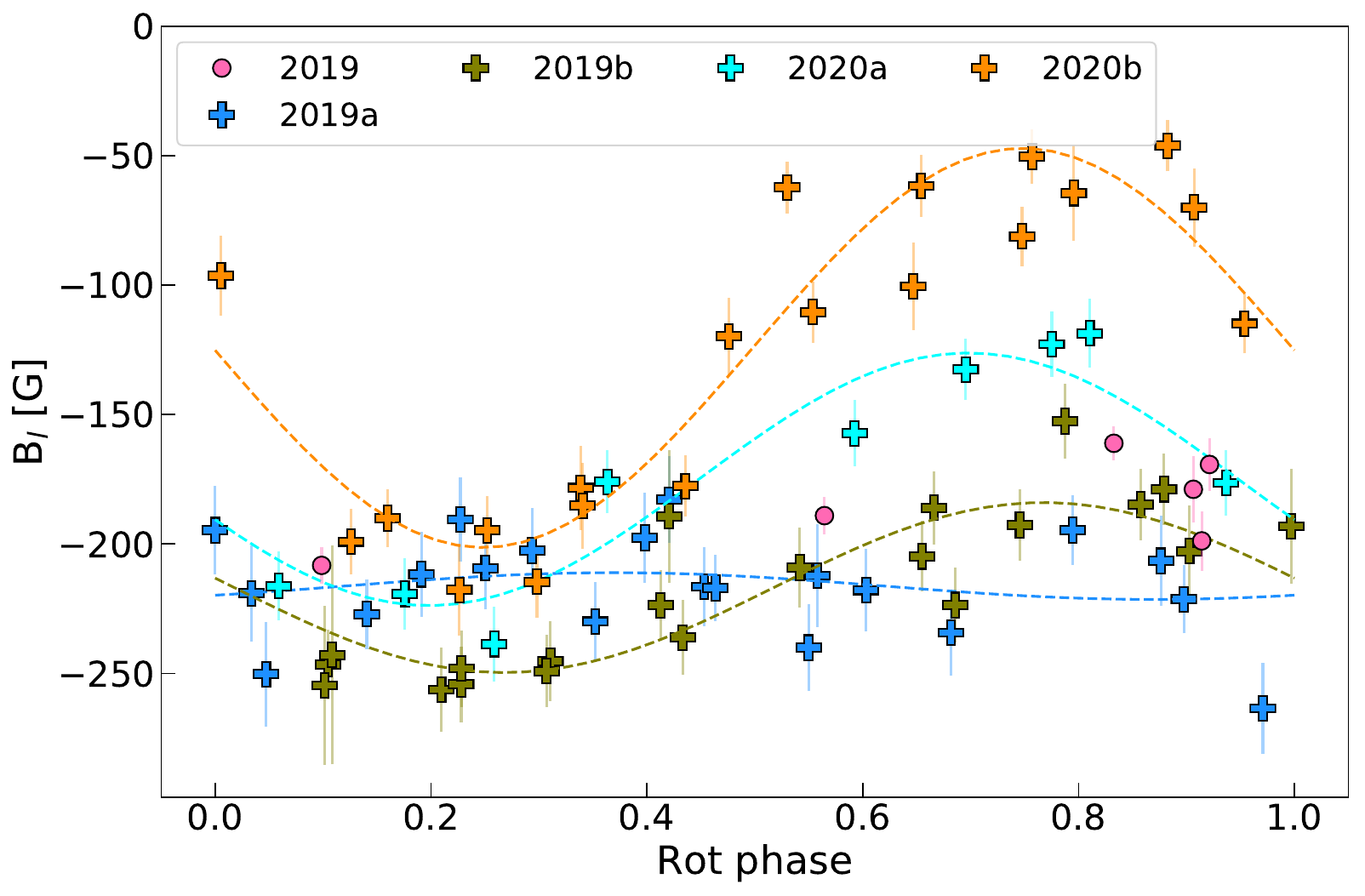}
    \caption{Temporal evolution of the longitudinal magnetic field (B$_l$). Top: full time series of measurements between 2006 and 2020 with ESPaDOnS, Narval and SPIRou. The shape of the data points corresponds to the wavelength domain, optical (circles) or near-infrared (pluses), and the colour to the epoch in which the observations were performed. Bottom: SPIRou time series split in four epochs phase-folded according to Eq.~\ref{eq:ephemeris}. The least-square sine fit corresponding to Eq.~\ref{eq:stibbs} \citep{Stibbs1950}, to assess the change in magnetic obliquity, is shown as dashed lines. The six ESPaDOnS observations taken in 2019 are plotted as pink circles.
    }
    \label{fig:Bl_evol}%
\end{figure}

\begin{table*}[!t]
\caption{Best fit values of the polar field and obliquity of Eq.~\ref{eq:stibbs} obtained for optical and near-infrared epochs.} 
\label{tab:stibbs}     
\centering                       
\begin{tabular}{l | c c c c c | c c c c}      
\hline     
& 2006 & 2007 & 2008 & 2012 & 2016 & 2019a & 2019b & 2020a & 2020b\\
\hline
$B_p$ [G] & $-868\pm20$ & $-842\pm21$ & $-873\pm15$ & $-736\pm19$ &  $-555\pm10$ & $-814\pm17$ & $-882\pm22$ & $-828\pm55$ & $-923\pm70$\\
$\beta$  [$^{\circ}$] & $13\pm6$ & $12\pm4$ & $26\pm1$ & $7\pm4$ & $12\pm1$ & $3\pm4$ & $23\pm2$ & $37\pm4$ & $59\pm2$\\
\hline                                 
\end{tabular}
\end{table*}

\begin{table*}[ht]
\caption{Comparison of a constant line vs a sine fit for the FWHM phase variations.} 
\label{tab:fwhm}     
\centering                       
\begin{tabular}{l l r r r | r r | r r}      
\hline     
Epoch & Mask & Mean & Mean Error & STD & RMS$_\mathrm{const}$ & $\chi^2_{r,\mathrm{const}}$ & RMS$_\mathrm{sine}$ & $\chi^2_{r,\mathrm{sine}}$\\
& & [km s$^{-1}$] & [km s$^{-1}$] & [km s$^{-1}$] & [km s$^{-1}$] & & [km s$^{-1}$] &\\
\hline
2019a      & default          & 19 & 0.29 & 0.40  & 0.40  & 1.8  & 0.39  & 2.0 \\
           & g$_\mathrm{eff}>1.2$ & 23 & 0.46 & 1.16 & 1.16 & 6.9 & 1.13 & 8.0\\
2019b      & default          & 21 & 0.41 & 0.63 & 0.63 & 2.3  & 0.62 & 2.5 \\
           & g$_\mathrm{eff}>1.2$ & 27 & 0.84 & 1.77  & 1.77  & 6.7 & 1.71  & 6.9\\
2020a      & default          & 21 & 0.41 & 0.68 & 0.68 & 3.2 & 0.61 & 3.4\\
           & g$_\mathrm{eff}>1.2$ & 27 & 0.82 & 2.47 & 2.47 & 12.2 & 2.29 & 13.3\\
2020b      & default          & 19 & 0.30 & 0.47  & 0.47  & 3.0  & 0.47  & 3.4\\
           & g$_\mathrm{eff}>1.2$ & 22 & 0.46 & 0.83  & 0.83  & 5.3 & 0.79  & 7.3\\
\hline                                 
\end{tabular}
\tablefoot{The columns are: 1) subset of the time series, 2) line list used in LSD computation, 3) mean value of FWHM, 4) mean error bar on FWHM, 5) dispersion of the data set, 6) RMS (Root Mean Square) residual of a constant line fit equal to the average of the data set, 7) reduced $\chi^2$ of the constant fit, 8) RMS residual of a sine fit at the stellar rotation period, 9) reduced $\chi^2$ of the sine fit.}
\end{table*}

We measured the line-of-sight component of the magnetic field integrated over the stellar disk (B$_l$) for all the available observations, in optical (2006--2019) and near-infrared (2019--2020). Since B$_l$ traces magnetic features present on the visible hemisphere, its temporal variations are modulated at the stellar rotation period and can be therefore used as a robust magnetic activity proxy \citep{Folsom2016,Hebrard2016,Fouque2023}. Formally, it is computed as the first-order moment of Stokes~V \citep{Donati1997}:
\begin{equation}
\mathrm{B}_l\;[G] = \frac{-2.14\cdot10^{11}}{\lambda_0 \mathrm{g}_{\mathrm{eff}}c}\frac{\int vV(v)dv}{\int(I_c-I)dv} \,,
\label{eq:Bl}
\end{equation}
where $\lambda_0$ and $\mathrm{g}_\mathrm{eff}$ are the normalisation wavelength and Land\'e factor of the LSD profiles, $I_c$ is the continuum level, $v$ is the radial velocity associated to a point in the spectral line profile in the star's rest frame and $c$ the speed of light in vacuum. For the near-infrared and optical Stokes profiles, the normalisation wavelength and Land\'e factor are 1700~nm and 1.2144, and 700~nm and 1.1420, respectively. In accordance with the fact that near-infrared lines are broader than optical ones, the integration was carried out within $\pm$ 50\,km\ s$^{-1}$ from line centre in the former case and $\pm$ 30\,km\ s$^{-1}$ in the latter case, to include the absorption ranges of both Stokes $I$ and $V$ profiles. 

The list of measurements is reported in Table~\ref{tab:Bl_log}. The values are of constant sign (negative), which is expected when observing one polarity of a dipole almost aligned with the stellar rotation axis over the entire stellar rotation, especially for a star observed nearly pole-on as AD~Leo \citep{Morin2008, Lavail2018}. The near-infrared measurements range between $-263$ and $-46$\,G, with an average of $-179$\,G and a median error bar of 15\,G. The optical measurements range between $-297$ and $-155$\,G, with an average of $-233$\,G and a median error bar of 10\,G. The lower error bar is likely due to the narrower velocity range over which the optical measurements are performed, since less noise is introduced in Eq.~\ref{eq:Bl}. A discussion about chromatic differences in the longitudinal field measurements is presented in Sec.~\ref{sec:discussion}.

We plot the temporal evolution of B$_l$ in Fig.~\ref{fig:Bl_evol}. In general, we note a secular weakening of the field strength over 14~yr, with an oscillation between 2016 and 2019 followed by a rapid decrease in strength (in absolute value). We also note that the intra-epoch dispersion increases for the last two epochs.

By phase-folding the near-infrared data at P$_\mathrm{rot}$, we observe a systematic increase in the rotational modulation towards 2020b, meaning that the axisymmetry level of the field has likely decreased (see Fig.~\ref{fig:Bl_evol}). For a first quantitative evaluation, we followed \citet{Stibbs1950} and \citet{Preston1967} to model the phase variations of the longitudinal field for a predominantly-dipolar magnetic configuration. Formally,
\begin{equation}
\mathrm{B}_l\;[G] = \frac{1}{20} \frac{15+\varepsilon}{3-\varepsilon}\mathrm{B}_p\left(\cos\beta\cos i+\sin\beta\sin i\cos(2\pi p)\right)\,,
\label{eq:stibbs}
\end{equation}
with $\varepsilon$ the limb darkening coefficient (set to 0.3; \citealt{Claret2011}), $p$ the rotational phase, B$_p$ the longitudinal field of the dipole, $i$ the stellar inclination and $\beta$ the obliquity between magnetic and rotation axes. The results are listed in Table~\ref{tab:stibbs}, for both near-infrared and optical time series for completeness. The six optical observations in November 2019 have poor coverage (three of them are clustered around phase 0.9) and lead to a less reliable sine fit. Nevertheless, they are compatible with the 2019b fit curve.
These clues clearly indicate that the magnetic field of AD~Leo is evolving, in agreement with \citet{Lavail2018}, and demonstrate the interest of long-term spectropolarimetric monitoring of active M~dwarf stars.

\subsection{The mean line width}\label{sec:fwhm}

\begin{figure*}[t]
    \centering
    \includegraphics[width=\textwidth]{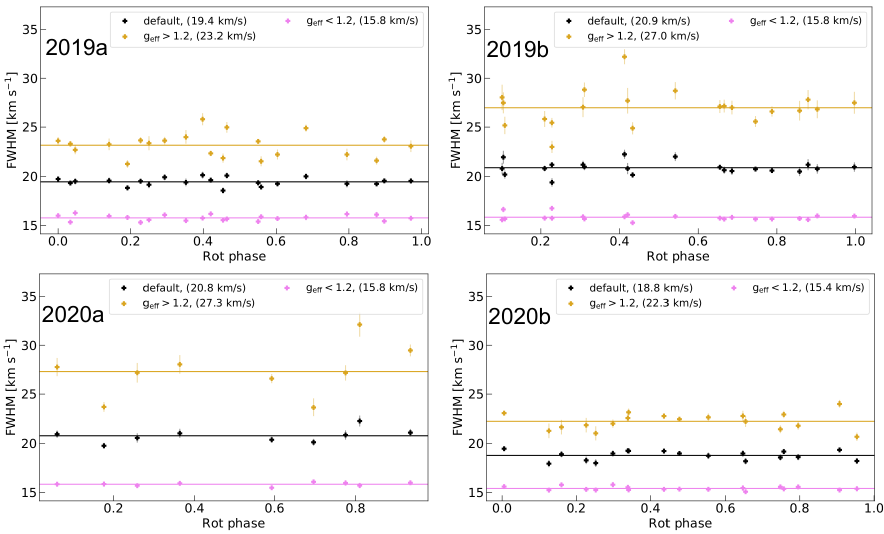}
    \caption{Phase variations of the FWHM, folded at a stellar rotation period of 2.23\,days. The data sets are computed with the default (black), high-g$_\mathrm{eff}$ (yellow), and low-g$_\mathrm{eff}$ (purple) mask for three epochs: 2019a (top left), 2019b (top right), 2020a (bottom left), 2020b (bottom right). Although no evident rotational modulation is observed, we notice a long-term oscillation of the mean value in a similar way to the longitudinal field (absolute) values.}
    \label{fig:fwhm}%
\end{figure*}

The width of near-infrared spectral lines of stars with intense fields and low equatorial velocity such as AD~Leo ($v_e\sin(i)=$ 3\,km\,s$^{-1}$; \citealt{Morin2008}) are sensitive to the Zeeman effect, given its proportionality to wavelength, field strength, and Land\'e factor. The rotationally-modulated line broadening correlates with the azimuthal distribution of the unsigned small-scale magnetic flux, a useful diagnostic for stellar activity radial velocity contamination, as shown for the Sun by \citet{Haywood2022}. In this context, \citet{Klein2021} adopted a selection of magnetically sensitive lines for the young star AU~Mic and saw a correspondence in the variations of RV and FWHM of the Stokes $I$ profiles at the stellar rotation period. This confirmed the sensitivity of the FWHM to the distortions induced by magnetic regions on the stellar surface.

Here, we proceeded analogously in an attempt to connect modulations of the FWHM with variations of the large-scale field. We applied LSD on the near-infrared data using a mask of 417 lines characterised by g$_\mathrm{eff}>1.2$, following \citet{Bellotti2021}. The near-infrared time series was divided in four epochs as in Sec.~\ref{sec:B_lon} for consistency.

In Table~\ref{tab:fwhm}, we compare the phase variations of the FWHM when adopting the default and high-g$_\mathrm{eff}$ masks, and we inspect whether they are more compatible with a sine fit or a constant fit equal to the mean of the data set. In all cases, there is no clear rotational modulation of the data points, as the sine fit does not provide a better description (i.e. lower $\chi^2_r$) of the variations than the constant fit. This is confirmed by a quick inspection of the periodogram applied to the FWHM data for each individual epoch. 
The $\chi^2_r$ increase when using a sine model rather than a constant is not statistically significant. The observed variations are attributable to dispersion, as illustrated in Fig.~\ref{fig:fwhm}. We observed that the FWHM is systematically larger in all epochs for the high-g$_\mathrm{eff}$ mask, as expected given the linear dependence of Zeeman effect to g$_\mathrm{eff}$, and the dispersion is between 1.8 and 3.0 times larger. The lack of rotational modulation prevents us from searching for correlations with other quantities such as RV and B$_l$ as done in \citet{Klein2021}.

From Fig.~\ref{fig:fwhm}, we also noticed an evident long-term evolution of the mean FWHM. Such evolution has a moderate correlation (Pearson $R$ coefficient of 0.5) with the variations of the mean B$_l$ for the same epochs, meaning that the FWHM is a reasonable proxy to trace long-term evolution of the field. This is consistent with the recent \citet{Donati2023} analysis analysis of AU\,Mic. When using the default mask, the mean FWHM oscillated from 19\,km\,s$^{-1}$ in 2019a to 21\,km\,s$^{-1}$ in 2019b and 2020a, and back to 19\,km\,s$^{-1}$ in 2020b. As expected, such oscillation is enhanced when considering the magnetically sensitive lines and goes from 23\,km\,s$^{-1}$ in 2019a to 27\,km\,s$^{-1}$ in 2019b and 2020a, and back to 22\,km\,s$^{-1}$ in 2020b. We performed the same analysis with low-Land\'e factor lines (i.e. g$_\mathrm{eff}<1.2$ and 406 lines) and noticed no appreciable variation of the mean FWHM, since it remained stable at $\sim$15\,km\,s$^{-1}$. A view of the Stokes~$I$ profiles computed with the three different line lists can be found in Appendix~\ref{sec:FWHM_StokesI}.

\begin{figure}[t]
    \centering
    \includegraphics[width=\columnwidth]{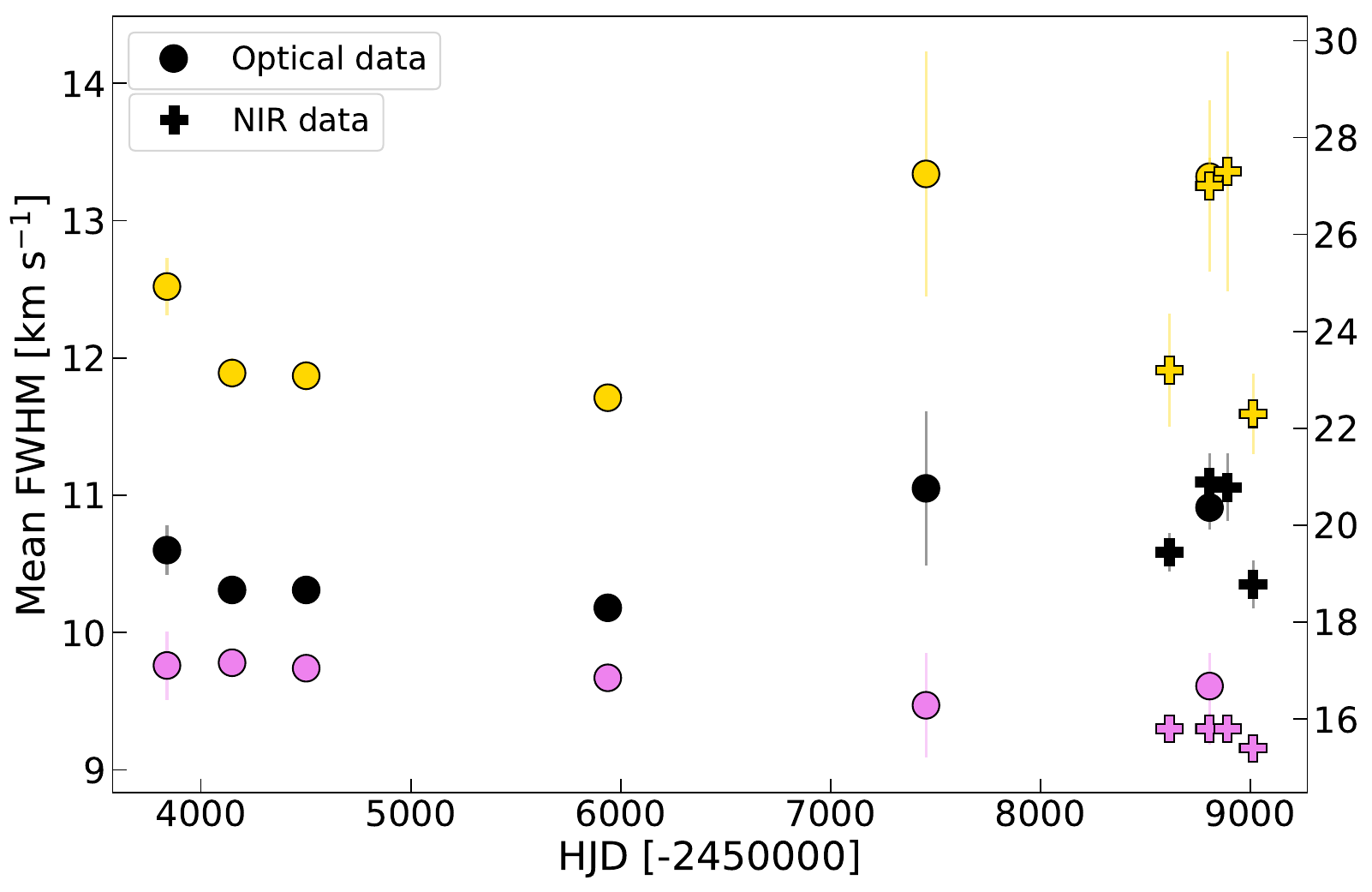}
    \caption{Secular evolution of the epoch-averaged FWHM of Stokes $I$. The scale on the left refers to optical data, while on the right to the near-infrared, and the colours indicate the use of different masks similar to Fig.~\ref{fig:fwhm}. Both optical (circles) and near-infrared (pluses) observations feature a long-term variation, which is enhanced (quenched) when using magnetically sensitive (insensitive) lines. We note that the optical data point of 2019 falls behind the near-infrared ones when using magnetically sensitive lines.}
    \label{fig:fwhm_full}%
\end{figure}

The FWHM analysis was also carried out on the ESPaDOnS and Narval data between 2006 and 2019. When using low-g$_\mathrm{eff}$ lines, the mean width of Stokes~$I$ is reasonably stable around 9.7\,km\,s$^{-1}$, stressing their potential for precise radial velocity measurements. The full (high-g$_\mathrm{eff}$) mask yields a mean value at 10\,km\,s$^{-1}$ (12\,km\,s$^{-1}$) between 2006 and 2012, which then increases to 11\,km\,s$^{-1}$ (13\,km\,s$^{-1}$) in 2016 and 2019. Such long-term evolution is only moderate compared to the one seen in the near-infrared time series. The entire evolution is illustrated in Fig.~\ref{fig:fwhm_full}.

The difference between the mean FWHM of low-g$_\mathrm{eff}$ lines in optical ($\sim$9.5\,km\,s$^{-1}$) and near-infrared ($\sim$16\,km\,s$^{-1}$) can be attributed to lines that have non-zero Land\'e factor. Indeed, the quadratic differential broadening between the two domains is $11.4$\,km\,s$^{-1}$, corresponding to a total magnetic field of 2.5\,kG for a line at 1700\,nm with g$_\mathrm{eff}$=0.96 (the normalisation values of the low-g$_\mathrm{eff}$ mask). Although we assumed that the Zeeman effect for low-g$_\mathrm{eff}$ lines is negligible in the optical with this exercise, the inferred value of total magnetic field is reasonably consistent with what is reported in the literature \citep{Saar1994,Shulyak2017,Shulyak2019}, indicating that the magnetic field accounts mostly for the difference in width between optical and near-infrared low-g$_\mathrm{eff}$ lines.

Our analysis confirms that the FWHM is capable of tracing secular changes in the total, unsigned magnetic field, which could be used to better understand stellar activity jitter. Activity-mitigating techniques would benefit from this information even for low-inclination stars such as AD~Leo, for which the phase modulation of the radial velocity jitter is more difficult to constrain. At the same time, the analysis highlights the presence of short-term variability producing scatter and that is not rotationally modulated.

\begin{figure*}[!t]
    \centering
    \includegraphics[width=\textwidth]{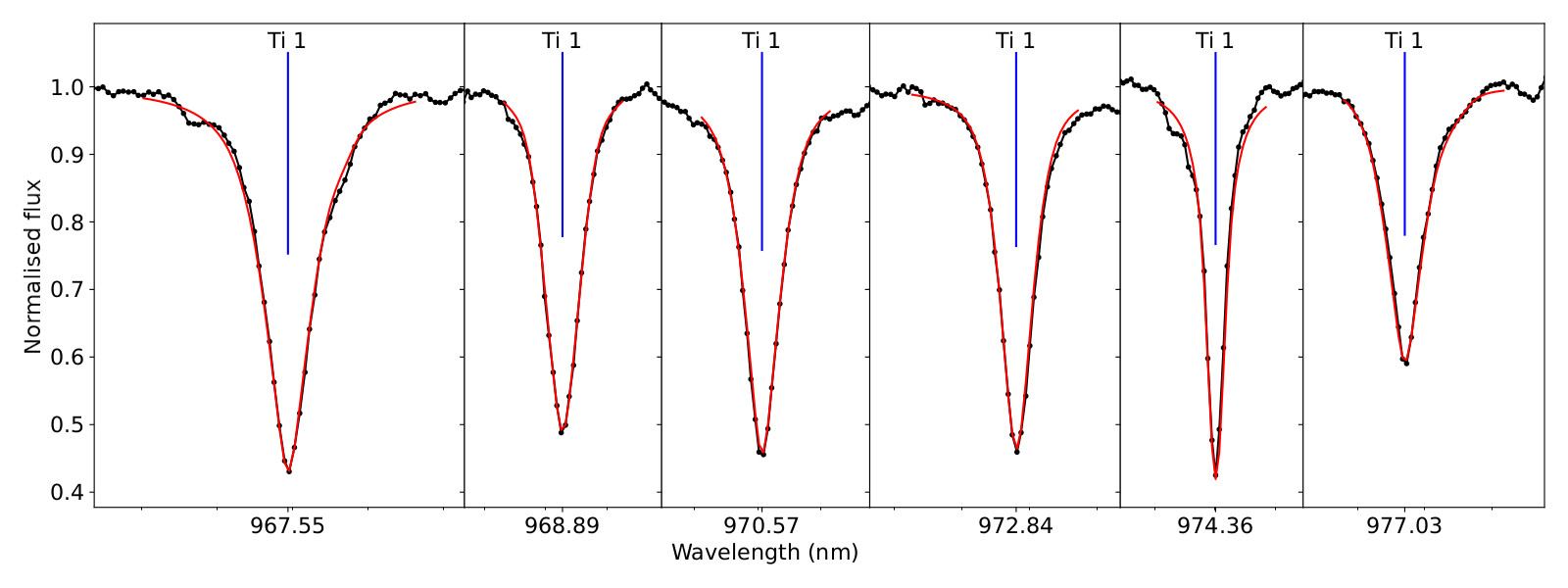}
    \includegraphics[width=\textwidth]{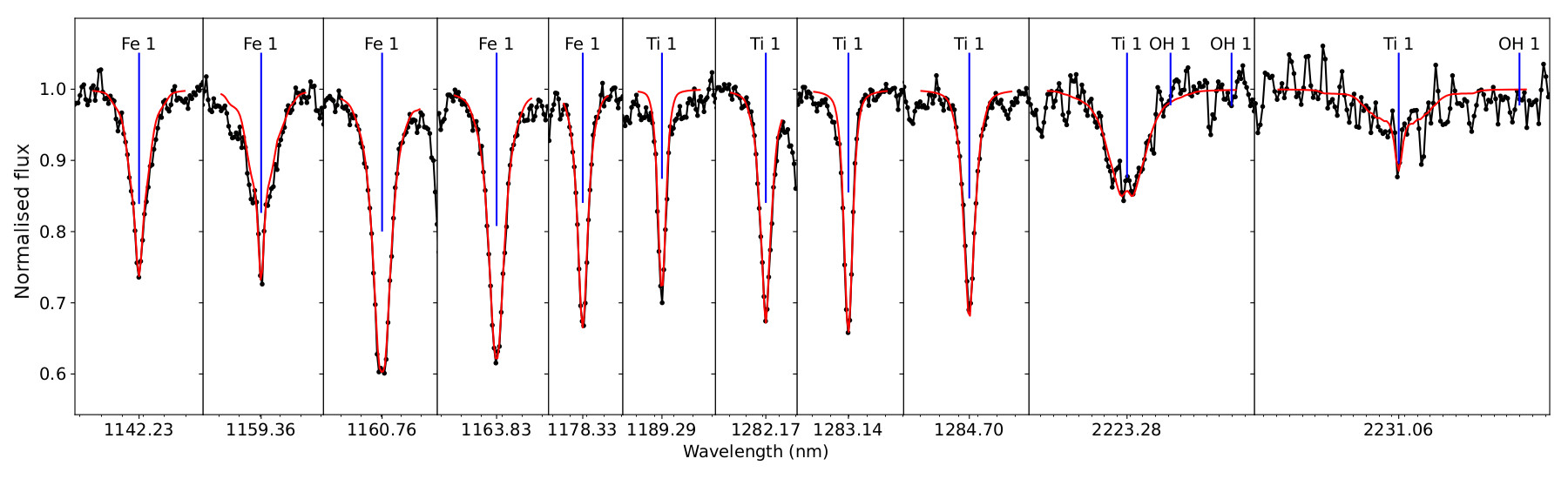}
    \caption{Example fits including Zeeman broadening resulting from the MCMC-based approach of radiative transfer modelling. Top: ESPaDOnS observation from 24 Febuary 2016. Bottom: SPIRou observation from 3 November 2020. The panels have the same wavelength scale with minor ticks at 0.1~nm.}
    \label{fig:zbroad-fitIR}
\end{figure*}

\subsection{Modelling Zeeman broadening}\label{sec:zeemanbroadening}

To further investigate the small-scale magnetic field of AD Leo we conducted a Zeeman broadening analysis.  For this analysis we used the full set of new and archival data, both in the near-infrared from SPIRou and in the optical from ESPaDOnS and Narval.  All the data sets require a telluric correction, since telluric lines are present in much of the SPIRou wavelength range, and the red end of the ESPaDOnS and Narval range. For the SPIRou data we relied on the telluric correction from the \texttt{APERO} pipeline (Sec~\ref{sec:observationsNIR}, \citealt{Cook2022} for more detail). 
For the ESPaDOnS and Narval data, we made a telluric correction using the {\tt molecfit}\footnote{\url{https://www.eso.org/sci/software/pipelines/}} pipeline, originally designed for handling spectra from ESO instruments \citep{molecfit2015A&A...576A..77S, molecfit2015A&A...576A..78K}. {\tt molecfit} retrieves weather conditions and other relevant information at the time of observation and models the atmosphere in the line of sight. It performs radiative transfer and iteratively models the telluric component in the input spectrum while also fitting the continuum and the wavelength scale of the spectrum. It finally corrects telluric lines and provides a telluric-corrected output spectrum. 
After telluric correction the spectra were re-normalised in the regions of interest using a low order polynomial fit through carefully selected continuum regions. A few ESPaDOnS and Narval spectra were affected by fringing effects, hence we adopted a higher-order polynomial fit to normalise to a flatter continuum. Finally, we discarded any observations where the telluric correction left a noticeable residual feature that was blended with the stellar lines of interest.

To characterise the magnetic field, we fitted synthetic spectra to the observed Stokes~$I$ spectra, incorporating both the Zeeman broadening and intensification effects. Synthetic spectra were calculated with {\sc Zeeman} \citep{Landstreet1988,Wade2001,Folsom2016}, using model atmospheres from {\sc marcs} \citep{Gustafsson2008}. {\sc Zeeman} performs polarised radiative transfer including the Zeeman effect. However, a major limitation for M-dwarfs is that the programme does not currently include molecular lines, which are not typically used in Zeeman broadening analyses. Weak molecular lines are blended with many atomic lines in the spectra of M-dwarfs. With careful attention we identified a set of atomic lines suitable for AD~Leo, with no evident distortion in the line shape by molecular blends. Thus the systematic error from this limitation is expected to be negligible, but the inclusion of molecular lines in the future would substantially simplify the selection of lines for Zeeman broadening analyses, as shown in the recent work of \citet{Cristofari2023} and previously applied to AD~Leo by \citet{Shulyak2014,Shulyak2017}. To check the validity of the analysis presented here, a second analysis of the ESPaDOnS and Narval spectra was carried out with the {\tt SYNMAST} code \citep{Kochukhov2010}. The analyses used nearly the same set of Ti~{\sc i} lines, and the results we obtained were consistent within uncertainty.

For the ESPaDOnS and Narval observations, we used the Ti~{\sc i} lines at 9675.54 \AA\ (g$_\mathrm{eff}=1.35$), 9688.87 \AA\ (g$_\mathrm{eff}=1.50$), 9705.66 \AA\ (g$_\mathrm{eff}=1.26$), 9728.40 \AA\ (g$_\mathrm{eff}=1.00$), 9743.61 \AA\ (g$_\mathrm{eff}=0.00$), and 9770.30 \AA\ (g$_\mathrm{eff}=1.55$). These lines have been used extensively for Zeeman broadening analysis \citep[e.g.,][and references therein]{Kochukhov-Lavail2017, Hill2019, Kochukhov2021} and have reliable oscillator strengths and Land\'e factors in VALD. These lines have relatively weak telluric blending, very little molecular blending, and a wide range of effective Land\'e factors.

For the SPIRou observations, we selected a set of lines using similar criteria, but also avoided lines with large pressure broadened wings, since small errors in the pressure broadening could cause larger errors in the Zeeman broadening estimation. In order to maximise the range of available effective Land\'e factors we used the Fe~{\sc i} lines at 11422.32 \AA\  (g$_\mathrm{eff}=1.98$), 11593.59 \AA\  (g$_\mathrm{eff}=2.50$), 11607.57 \AA\  (g$_\mathrm{eff}=1.66$), 11638.26 \AA\  (g$_\mathrm{eff}=1.58$), and 11783.26 \AA\  (g$_\mathrm{eff}=1.14$) and the Ti~{\sc i} lines at 11892.88 \AA\  (g$_\mathrm{eff}=0.75$), 12821.67 \AA\  (g$_\mathrm{eff}=1.26$), 12831.44 \AA\  (g$_\mathrm{eff}=0.67$), 12847.03 \AA\  (g$_\mathrm{eff}=1.08$), 22232.84 \AA\  (g$_\mathrm{eff}=1.66$), and 22310.61 \AA\  (g$_\mathrm{eff}=2.50$).  This provides multiple lines with both high and low effective Land\'e factors, but uses lines from two different ions, which we compensated for by using the Ti and Fe abundances as independent free parameters in our analysis. 
There are a few other Ti~{\sc i} lines near 22000 \AA\ with large effective Land\'e factors, but there is a relatively severe blending by many weak molecular lines in this region, hence we  did not include these lines.
Line data were extracted from VALD. In these line lists, experimental oscillator strengths for Ti~{\sc i} lines were from \citet{Lawler2013-TiI-lines} \citep[consistent with][]{Blackwell-Whitehead2006-TiI-lines}, except for 22232.84 \AA\ from \citet{Blackwell-Whitehead2006-TiI-lines}, and a theoretical value for 22310.61 \AA\ from the compilation of R. L. Kurucz\footnote{\url{http://kurucz.harvard.edu}}. Oscillator strengths for the Fe~{\sc i} lines were taken from \citet{OBrian1991-FeI-lines}.

The total magnetic field was modelled with a grid of field strengths and filling factors for the fraction of the surface area with the corresponding field strength \citep[e.g.][]{Johns-Krull1999}. A uniform radial orientation was assumed for the magnetic field, since Stokes $I$ spectra have little sensitivity to magnetic field orientation. This is also a reasonable assumption given the magnetic field maps reconstructed in Sec.~\ref{sec:magnetic_imaging}. For the optical spectra, we adopted magnetic fields of 0, 2, 4, 6, 8, and 10\,kG, and derived their filling factors. For the SPIRou spectra, we used a finer grid of 1\,kG from 0 to 10\,kG, since the sensitivity to Zeeman effect is larger at longer wavelengths, and a finer grid is needed to produce smooth line profiles.

To derive the magnetic filling factors we applied an MCMC-based approach, using the {\tt emcee} package \citep{ForemanMackey2013} integrated with {\tt Zeeman}. The filling factors for $B > 0$ were treated as free parameters, with the filling factor for $B=0$ ($f_{B=0}$) calculated from $1 - \sum_{B>0} f_{B}$. Proposed steps in the chain where $\sum_{B>0} f_{B} > 1$ were rejected to ensure that the filling factors sum to unity. The projected rotational velocity $v_e\sin(i)$ and the abundance of Ti (and Fe for SPIRou) were included as free parameters in the MCMC process. 
The modelling used $T_{\rm eff} = 3500$ K, $\log g = 5.0$, a microturbulence of 1\,km\,s$^{-1}$.
The chemical abundances may be unreliable since they do not account for elements bound in molecules, making them effectively nuisance parameters in this study. However, this provides the code with flexibility for fitting line strength and width in the absence of a magnetic field, reducing the sensitivity of the results to small errors in non-magnetic parameters. 

Example fits resulting from the MCMC-based approach are show in Figs.~\ref{fig:zbroad-fitIR} for ESPaDOnS and SPIRou. The shapes of the posterior distributions are generally similar for all observations using the same sets of lines, and are illustrated in Appendix \ref{sec:zbroad_appendix}. There are important anti-correlations between filling factors with adjacent magnetic field strengths, and weak correlations between filling factors spaced by two bins in field strength. Therefore, some caution should be taken in interpreting the uncertainties from this and similar analyses. The filling factor for $B=0$ and the quantity $\sum_i B_i f_i$ summed over magnetic field bins (abbreviated to $\sum Bf$), were calculated from samples in the MCMC chain. The resulting distribution was used to provide the median value from the 50th percentile, with uncertainties from the 16th and 84th percentile.

\begin{figure}[t]
    \centering
    \includegraphics[width=\columnwidth]{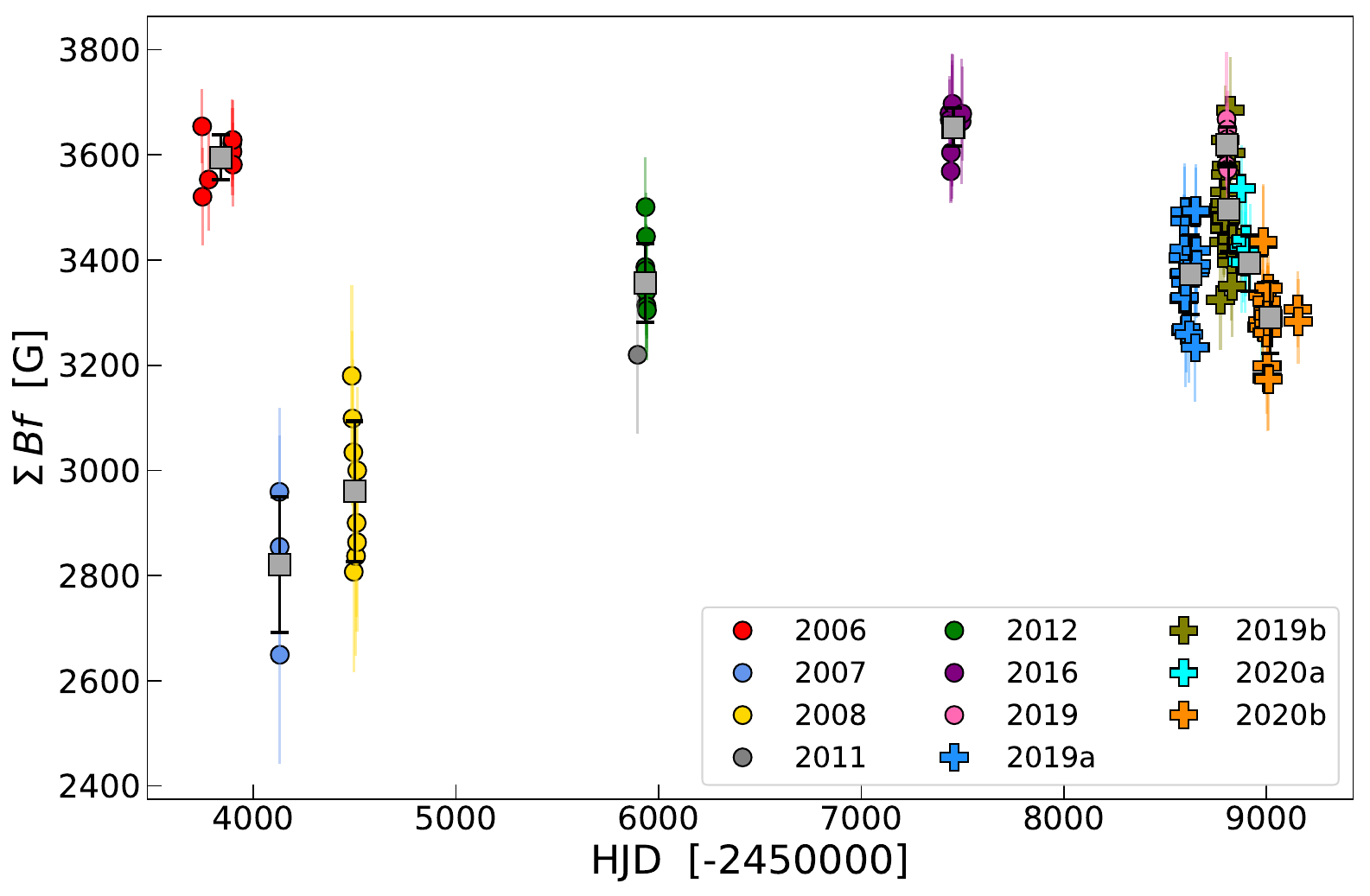}
    \caption{Long-term evolution of the magnetic flux density measured from Zeeman broadening modelling. The data point format is the same as Fig.~\ref{fig:Bl_evol}, and for each epoch we overplotted the mean value and standard deviation as error bar. There is an overall trend that reflects the variations of B$_l$ across both optical and near-infrared data. In addition, for the near-infrared SPIRou time series, we also observe the same oscillating behaviour found for the FWHM of Stokes~$I$.}
    \label{fig:zbroadening}%
\end{figure}

The results for all observations, and averages for each epoch, are presented in Fig.~\ref{fig:zbroadening}, and values for each epoch are provided in Tables \ref{tab:zbroad-vis} and \ref{tab:zbroad-ir}. The quantity $\sum Bf$ (sometimes called the magnetic flux, and analogous to a magnetic flux density) ranges between 2.6\,kG and 3.7\,kG, which is consistent with previous measurements \citep[e.g.,][]{Saar1994,Kochukhov2009,Shulyak2014,Shulyak2017,Shulyak2019,Cristofari2023}. We observe a long-term increase of the average $\Sigma\,Bf$ from 2.8\,kG in 2007 to 3.6\,kG in 2016, followed by a weakening towards 3.4\,kG with the latest SPIRou observations. Such behaviour correlates with the long-term decrease (in absolute value) of the longitudinal field (Pearson coefficient $R$=0.6, excluding the 2006 data point). Likewise, the average $\Sigma\,Bf$ time series correlates with the average FWHM of Stokes~$I$, demonstrating its capability at tracing the evolution of the total, unsigned magnetic field \citep{Donati2023}.

The $\Sigma\,Bf$ values for the ESPaDOnS optical data acquired in 2006 fall out from this trend. This could stem from residuals of the telluric correction blending with the lines used in the modelling and/or instrumental effects such as fringing, for which the results are sensitive to the choice of continuum normalisation.  Attempts were made to correct for these potential systematic errors: rejecting observations where the telluric correction left residual features in the used portion of the spectrum, and careful continuum normalisation to remove any weak fringing. However, it is possible these attempts were not fully successful, and thus the departure from the general trend of the 2006 result should be treated with caution. 

\subsection{Magnetic imaging}\label{sec:magnetic_imaging}

\begin{figure*}[!ht]
    \centering
    \includegraphics[width=0.96\textwidth]{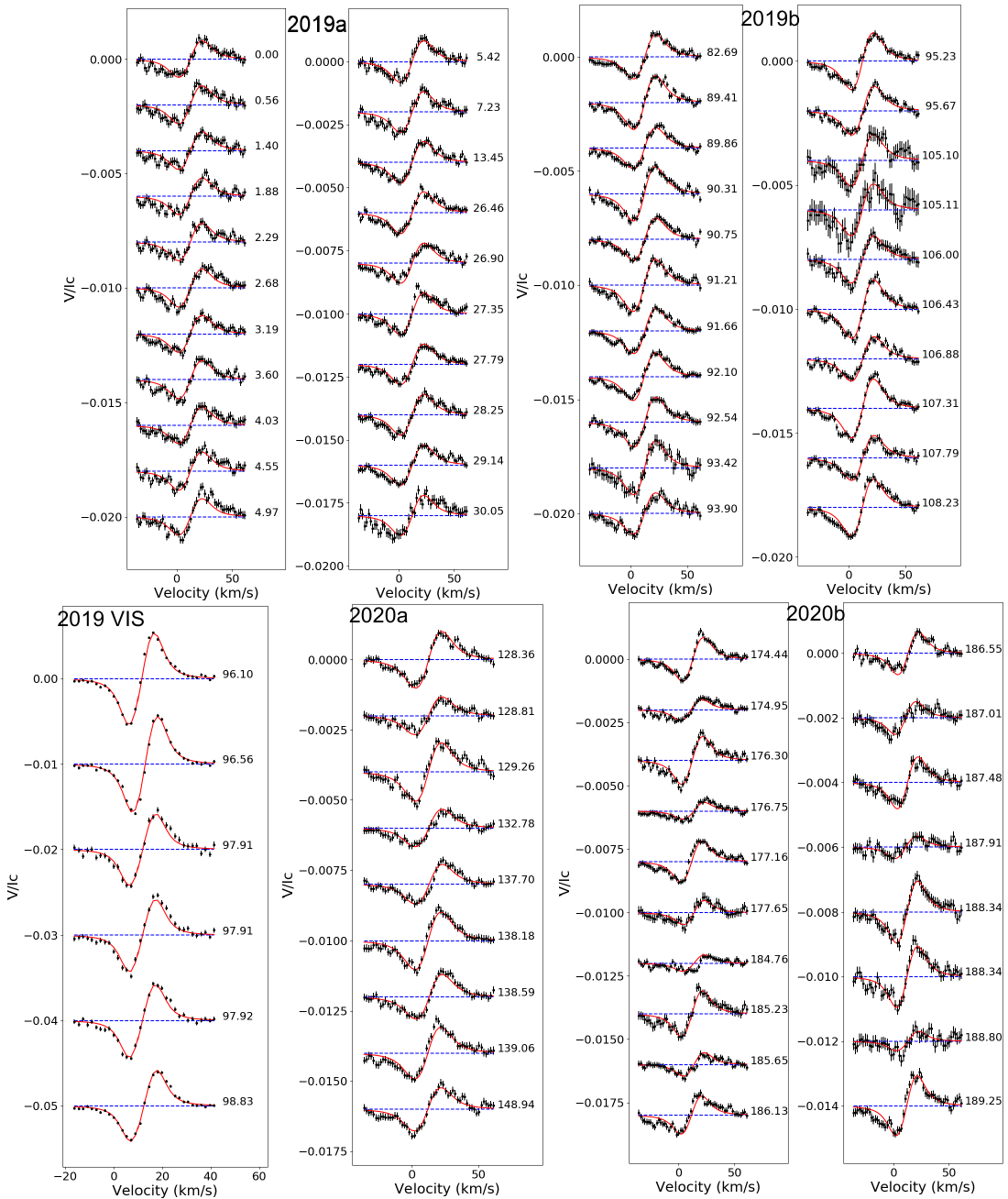}
    \caption{Full SPIRou and ESPaDOnS 2019 time series of Stokes $V$ profiles of AD~Leo normalised by the unpolarised continuum intensity. Two panels are present for 2019a (top left), 2019b (top right), one panel for 2019 in optical (bottom left) and 2020a (bottom middle) and two panels for 2020b (bottom right). In all panels, both the observed profiles (black) and Unno-Rachkovsky models (red) are shown. They are shifted vertically for visualisation purposes, and their associated rotational cycle is reported on the right (see Eq.~\ref{eq:ephemeris}). A remarkable feature is the increased intermittency of the profiles amplitude for the most recent observations, for example at cycle 174.44 and 174.95 of 2020b against cycle 1.40 and 1.88 in 2019a, suggesting an evolution of the magnetic field obliquity.}
    \label{fig:stokesV}%
\end{figure*}

We applied ZDI to the SPIRou and 2019 ESPaDOnS time series of Stokes~$V$ profiles to recover the large-scale magnetic field at the surface of AD~Leo. The magnetic geometry is modelled as the sum of a poloidal and a toroidal component, which are both expressed through spherical harmonics decomposition \citep{Donati2006, Lehmann2022}. The algorithm compares observed and synthetic Stokes~$V$ profiles iteratively, fitting the spherical harmonics coefficients $\alpha_{\ell,m}$, $\beta_{\ell,m}$, and $\gamma_{\ell,m}$ (with $\ell$ and $m$ the degree and order of the mode, respectively), until they match within a target reduced $\chi^2$. Because the inversion problem is ill-posed, a maximum-entropy regularisation scheme is applied to obtain the field map compatible with the data and with the lowest information content (for more details see \citealt{Skilling1984,DonatiBrown1997,Folsom2018}). 

In practice, we used the \texttt{zdipy} code described in \citet{Folsom2018}. In its initial version, the code performed tomographic inversion under weak-field approximation, for which Stokes~$V$ is proportional to the first derivative of Stokes~$I$ over velocity \citep{Landi2004}. For the present study, we have implemented the Unno-Rachkovsky's solutions to polarised radiative transfer equations in a Milne-Eddington atmosphere \citep{Unno1956,Rachkovsky1967,Landi2004} and incorporated the filling factor formalism outlined in \citet{Morin2008} and \citet{Donati2023}. The implementation of Unno-Rachkovsky's solutions was motivated by the need of a more general model for the observed Stokes~V profiles. Near-infrared observations of stars with intense magnetic fields are indeed more susceptible to distortions and broadening due to an enhanced Zeeman effect.

As input parameters for ZDI, we assumed $i=$ 20$^{\circ}$, $v_e\sin(i)=$ 3\,km\,s$^{-1}$, P$_\mathrm{rot}=$ 2.23\,days, and solid body rotation. We adopted a linear limb darkening coefficient in H band of 0.3 and V band of 0.7 \citep{Claret2011}. We set the maximum degree of the harmonic expansion $\ell_\mathrm{max}=$ 8 (considering the low $v_e\sin(i)$) and allowed an entropy weighting scheme proportional to $\ell$ during ZDI inversion, to favour simple geometries as in \citet{Lavail2018}. The SPIRou near-infrared time series was split similarly to Sec.~\ref{sec:B_lon}: 2019a (21 observations over 30 cycles), 2019b (21 observations over 26 cycles) 2020a (30 observations over 20 cycles), and 2020b (18 observations over 15 cycles). The Stokes~V time series of SPIRou and 2019 ESPaDOnS data are shown in Fig.~\ref{fig:stokesV}.

\begin{figure*}[!ht]
    \centering
    \includegraphics[width=\textwidth]{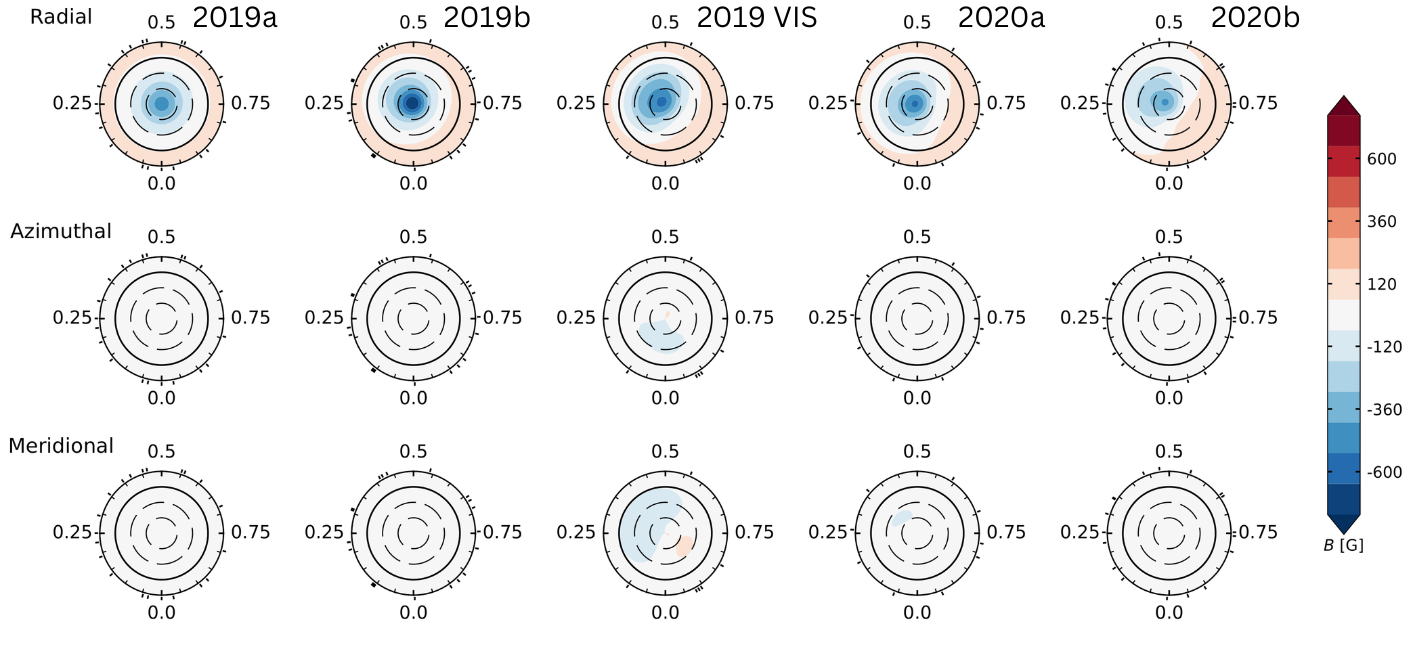}
    \caption{Reconstructed ZDI maps in flattened polar view of AD~Leo for the SPIRou and 2019 ESPaDOnS time series. The columns correspond to the epochs examined, from the left: 2019a, 2019b, 2019 optical, 2020a, and 2020b. In each column, the radial (top), azimuthal (middle) and meridional (bottom) components of the magnetic field vector are displayed. The radial ticks are located at the rotational phases when the observations were collected, while the concentric circles represent different stellar latitudes: -30\,$^{\circ}$, +30\,$^{\circ}$ and +60\,$^{\circ}$ (dashed lines) and equator (solid line). The colour bar range is set by the maximum (in absolute value) of the magnetic field and illustrates the positive (red) and negative (blue) polarity for each epoch. }
    \label{fig:zdi}%
\end{figure*}

\begin{figure}[!ht]
    \centering
    \includegraphics[width=\columnwidth]{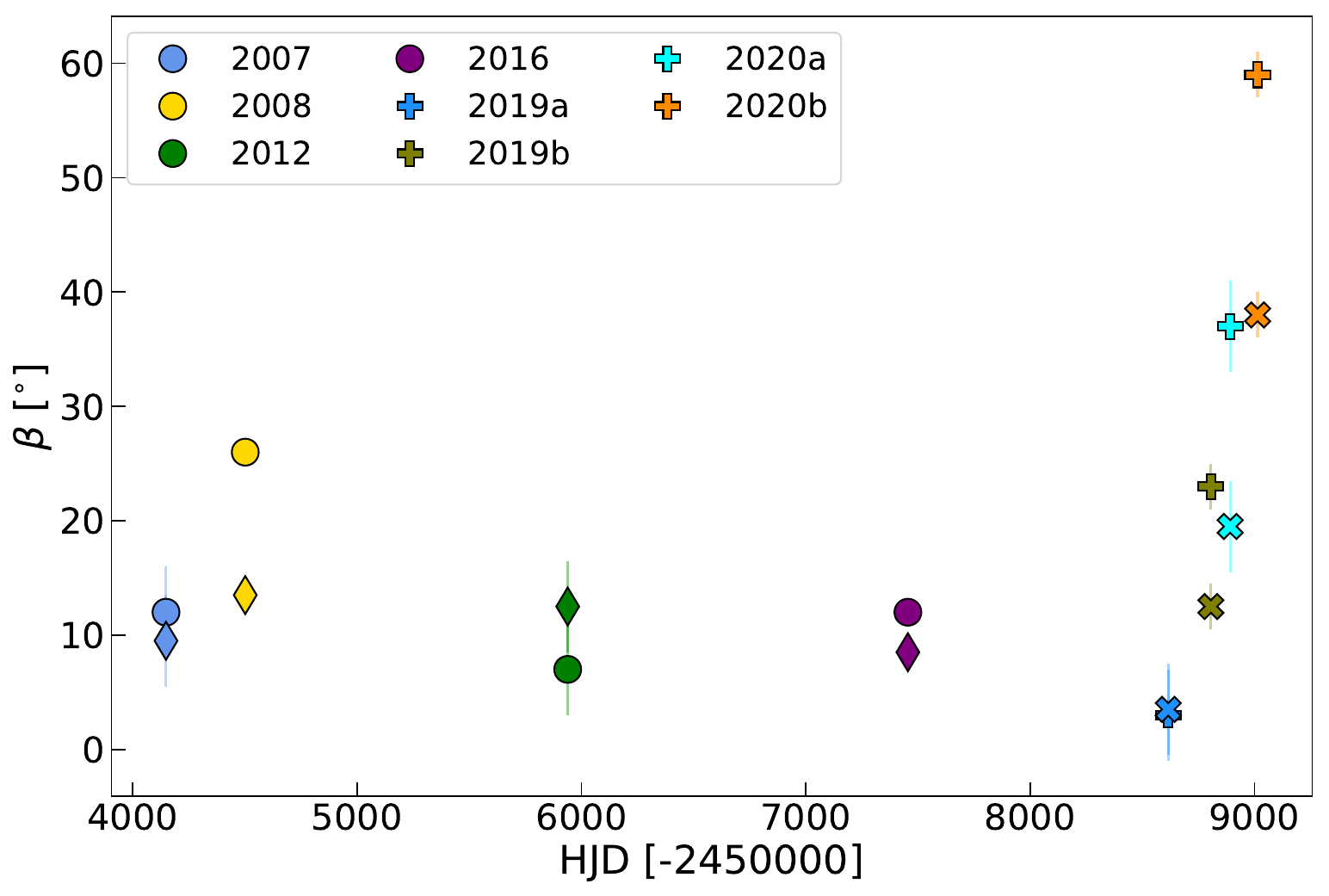}
    \caption{Evolution of the magnetic field obliquity with respect to the stellar rotation axis. Values estimated from Eq.~\ref{eq:stibbs} \citep{Stibbs1950} using optical and near-infrared data are illustrated as circles and pluses, colour-coded by epoch, whereas values from ZDI are shown as diamonds and crosses. The y axis is expressed in degrees of colatitude.}
    \label{fig:obliquity_evol}%
\end{figure}

For the 2019a, 2019b, 2020a, and 2020b epochs, we fitted the Stokes~$V$ profiles to a $\chi^2_r$ level of 1.2, 1.0, 1.1, and 1.1 from an initial value of 10.3, 15.8, 14.7, and 8.5, respectively. For the ESPaDOnS 2019 time series, we fitted down to $\chi^2_r=2.5$ from an initial value of 156.3. We attempted to merge the 2019b and 2020a epochs and reconstruct a single map, since they are separated by the shortest time gap. The quality of the final model is deteriorated ($\chi^2_r$=1.3) with respect to the two epochs separately, but the corresponding map and magnetic energies are consistently recovered. We therefore kept these two epochs separate.


From Fig.~\ref{fig:stokesV}, it is evident that the near-infrared Stokes~$V$ profiles manifest structures and stochastic variability in both lobes. This is not extended in the continuum, since the residuals with respect to the mean profile are compatible with the noise level, it is not rotationally-modulated, and it is not exhibited by Stokes~$N$. The presence of such variability was already suggested by the phase-folded variations in B$_l$ (Fig.~\ref{fig:Bl_evol}), as some data points featured a departure from a pure rotational modulation. Likewise, the residuals of the Stokes~$I$ profiles show clear variability, but the application of a 2D periodogram does not reveal any significant periodicity. While our ZDI model is capable of describing the general shape of Stokes~$V$ profiles, it is limited at reproducing the structures and at capturing all the information present. These considerations are also valid for optical observations in 2019, as the amplitude of Stokes~$V$ is not matched exactly by our ZDI model, and overall translate into an underestimate of the field strength. This motivates further the use of the PCA method described by \citet{Lehmann2022}, which is a data-driven approach offering a complementary view on the magnetic field evolution, as outlined in the next section.

We are able to constrain the filling factor $f_V$ following a $\chi^2$ minimisation prescription similar to \citep{Petit2002}. We found $f_V$ values oscillating between 9\% in 2019a, 16\% in 2019b and back to $\sim$11\% in the remaining epochs, compatible with \citet{Morin2008}, and larger by a factor of 1.7 than \citet{Lavail2018}. This would indicate a weakening of the local small-scale field since 2016, on top of a decrease in large-scale field intensity as seen in the reconstructions (Fig.~\ref{fig:zdi}). 

The filling factor $f_I$ was inspected by considering a grid of values between 0\% and 100\%; for each $f_I$ value, we synthesised a time series of model Stokes~$I$ profiles, computed the corresponding time series of $\chi^2_r$ with the observations, and phase-folded the $\chi^2_r$ curve at P$_\mathrm{rot}$. We then assessed at what value of $f_I$ the $\chi^2_r$ curve would start manifesting rotational modulation, because it would indicate that certain model profiles deviate from the observations. We noticed that values above 30\% deteriorate the fit of the profile core progressively, yielding variability and rotational modulation of the Stokes~$I$ profiles, which is not observed otherwise (see Sec.~\ref{sec:fwhm}). Values of $f_I=30\%$ are three times larger than $f_V$, in agreement with \citep{Morin2008}. Since the plausible $f_I$ values are consistent with 0\%, we adopted $f_I=0\%$ in the ZDI modelling.

The five maps of surface magnetic flux (one for each SPIRou epoch, and one for the ESPaDOnS 2019 epoch) are shown in Figure~\ref{fig:zdi} and their properties are reported in Table~\ref{tab:zdi_output}. In all cases, the configuration is predominantly poloidal, storing $>$95\% of the magnetic energy. The main modes are dipolar and quadrupolar, as they account for 70-90\% and 15-20\% of the magnetic energy. We report a weakening of the mean field strength ($\langle B \rangle$) of factor of 1.5 and 2.4 relative to the optical maps reconstructed by \citet{Morin2008} and \citet{Lavail2018}, respectively. The most remarkable feature is the reduction of magnetic energy contained in the axisymmetric mode, going from $>$ 99\% in 2019a to 60\% in 2020b, translating into an increase of the dipole obliquity relative to the rotation axis, from $3^\circ$ to $38^\circ$.

We note that the maximum field strength reconstructed with ZDI is between 1.2 and 2.4 times smaller than obtained via Eq.~\ref{eq:stibbs} \citep{Stibbs1950}. Likewise, the magnetic field obliquity is underestimated, as illustrated in Fig.~\ref{fig:obliquity_evol}. On one side, this difference stems from the limitation of the Stokes~$V$ ZDI model, since it does not encompass the full amplitude of the two lobes for some observations, and on the other side Eq.~\ref{eq:stibbs} assumes a purely dipolar field, contrarily to our reconstructions (the dipole accounts for 70-90\% of the energy). Nevertheless, both approaches allow us to observe an evident evolution of the obliquity, featuring a rapid increase in the most recent epochs.

Finally, we merged the 2019b, 2020a and 2020b data sets and attempted a joint rotation period and differential rotation search following \citet{Petit2002}. The results were inconclusive, likely due to the significant evolution of the surface magnetic field between each epoch.

\begin{figure*}[t]
    \centering
    \includegraphics[width=\textwidth]{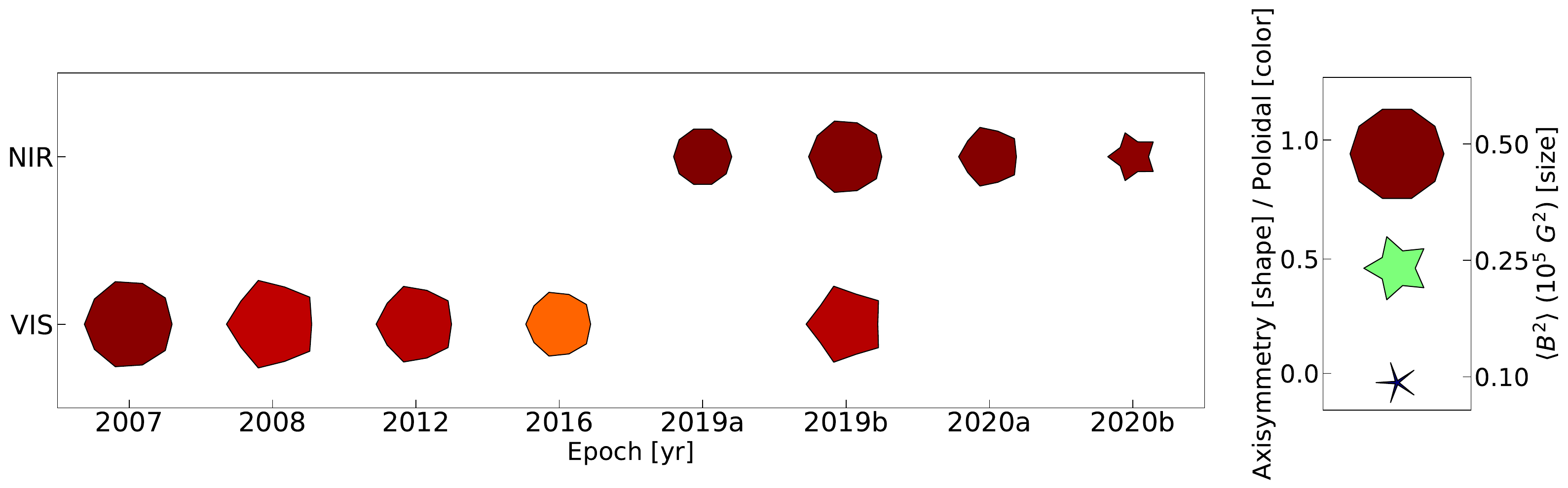}
    \caption{Evolution of the magnetic topology of AD~Leo over 14~yr. The time series includes optical ESPaDOnS and Narval data between 2007 and 2019, and near-infrared SPIRou data collected between 2019 and 2020. Field strength, level of axisymmetry and dominant field component (poloidal or toroidal) are encoded with the symbol size, shape, and colour, respectively, and were computed using the Unno-Rachkovsky implementation described in this work for consistency. The topology remained predominantly poloidal and dipolar, while the field strength has decreased over time. The most striking change is a decrease in axisymmetry in the most recent epoch.}
    \label{fig:zdievol}%
\end{figure*}

The summary of the magnetic field's evolution is illustrated in Fig.~\ref{fig:zdievol}. We performed ZDI reconstructions also for the archival ESPaDOnS and Narval data for consistency, finding reasonably compatible results with previous studies \citep{Morin2008,Lavail2018}. We observe a globally simple geometry (i.e. predominantly poloidal and dipolar) over 14~yr, with a decreasing strength. Our latest SPIRou observations revealed a clear evolution of the dipole obliquity in the form of a reduced axisymmetry, suggesting a potential dynamo magnetic cycle. These features are indeed compatible with the variations observed by \citet{Sanderson2003} and \citet{Lehmann2021} for the solar cycle.

\begin{table}[!t]
\caption{Properties of the magnetic maps for the 2019a, 2019b, 2020a, and 2020b SPIRou epochs and the 2019 ESPaDOnS data set. } 
\label{tab:zdi_output}     
\centering                       
\begin{tabular}{l r r r r r}      
\hline     
& 2019a & 2019b & 2019 & 2020a & 2020b\\ 
& NIR & NIR & VIS & NIR & NIR\\
\hline
f$_V$ & 9\% & 16\% & 12\% & 12\% & 11\%\\
$\langle B\rangle$ [G]   & 111.2 & 132.3 & 158.0 & 115.3 & 93.4 \\
B$_\mathrm{max}$ [G]    & 481.2 & 764.0 & 577.2 & 555.1 & 434.3\\
B$_\mathrm{pol}$ [\%]   & 100.0 & 99.9  & 95.0  & 99.3  & 98.7\\
B$_\mathrm{tor}$ [\%]   & 0.0   & 0.1   & 5.0   & 0.7   & 1.3\\
B$_\mathrm{dip}$ [\%]   & 81.3  & 71.1  & 81.7  & 75.7  & 70.1\\
B$_\mathrm{quad}$ [\%]  & 14.9  & 19.0  & 14.6  & 17.7  & 21.2\\
B$_\mathrm{oct}$ [\%]   & 2.8   & 6.2   & 2.7   & 4.4   & 5.9\\
B$_\mathrm{axisym}$ [\%]& 99.8  & 94.5  & 77.0  & 85.8  & 58.3\\
Obliquity [$^{\circ}$] & 2.5   & 12.5  & 21.5  & 19.5  & 38.0\\
\hline                                 
\end{tabular}
\tablefoot{The following quantities are listed: filling factor, mean magnetic strength, maximum magnetic strength, poloidal and toroidal magnetic energy as a fraction of the total energy, dipolar, quadrupolar and octupolar magnetic energy as a fraction of the poloidal energy, axisymmetric magnetic energy as a fraction of the total energy, and tilt of the magnetic axis relative to the rotation axis. The time span of the 2019 optical epoch is encompassed in the 2019b near-infrared epoch.}
\end{table}


%
%
%

\subsection{Diagnosing the large-scale field using PCA}\label{sec:PCA}


AD~Leo is an ideal target for analysing large-scale field evolution with the data-driven PCA method recently presented by \cite{Lehmann2022}, given its magnetic field strength and $v_e \sin(i)$. Principal component analysis allows us to uncover details about the stellar large-scale field directly from the LSD Stokes~$V$ profiles and to trace its magnetic field evolution across the observation run, without prior assumptions. Here, we analyse only the near-infrared time series, because the number of optical 2019 observations is not sufficient. 

\begin{figure}
	\centering
	\includegraphics[width=\columnwidth, trim={0 0 0 0}, clip]{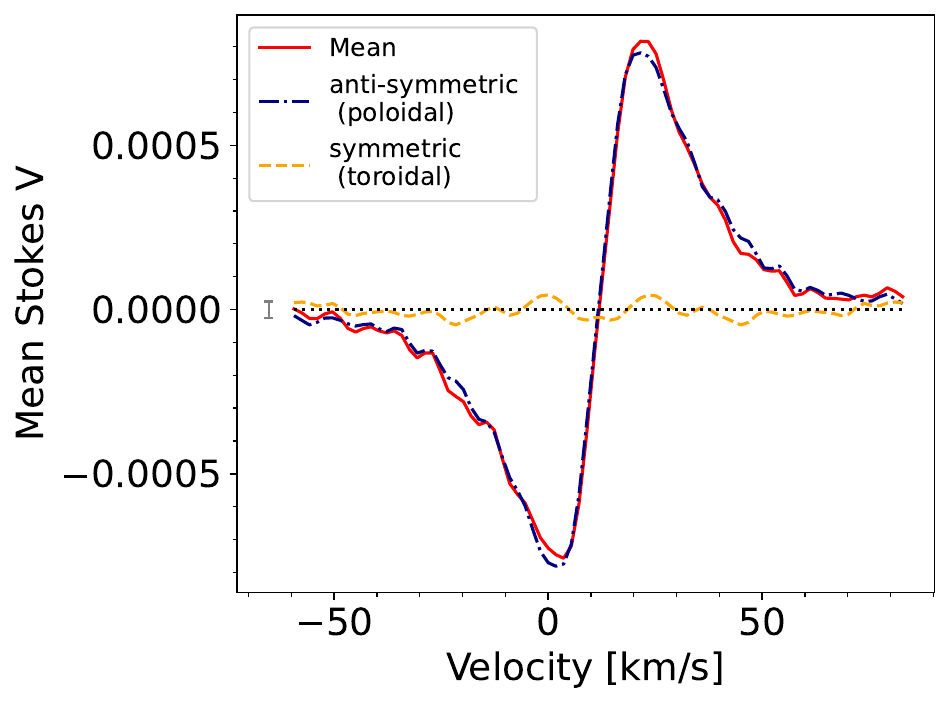}
    \caption{Decomposition of the Stokes~$V$ mean profile of AD~Leo SPIRou observations. The mean profile (solid red line) and its antisymmetric (dash dotted blue line) and symmetric (dashed yellow line) parts, related to the poloidal and toroidal components of the axisymmetric large-scale field, are shown.}
    \label{fig:MeanProfile}
\end{figure}

First, we can get insights about the star's axisymmetric large-scale field by analysing the mean Stokes~$V$ profile determined over all Stokes~$V$ LSD profiles (see \cite{Lehmann2022} for further details). Fig.~\ref{fig:MeanProfile} displays the mean profile and the decomposition into its antisymmetric and symmetric parts, denoting the poloidal and toroidal axisymmetric components, respectively. We clearly see that the mean profile is antisymmetric, which indicates a poloidal-dominated axisymmetric large-scale field. The amplitude of the symmetric part is comparable to the noise, and likely due to an artefact of uneven phase coverage rather than a true toroidal field signal \citep{Lehmann2022}. Compared to the mean-subtracted Stokes~$V$ profiles, the amplitude of the mean profile is generally strong, marking a dominant axisymmetric field. However, we observe an increase in the amplitude of the mean-subtracted Stokes~$V$ in the last two epochs 2020a and 2020b, which provides a first hint towards a less axisymmetric configuration.

\begin{figure}[t]
	\centering
	\includegraphics[width=0.5\textwidth, trim={1 400 2 1}, clip]{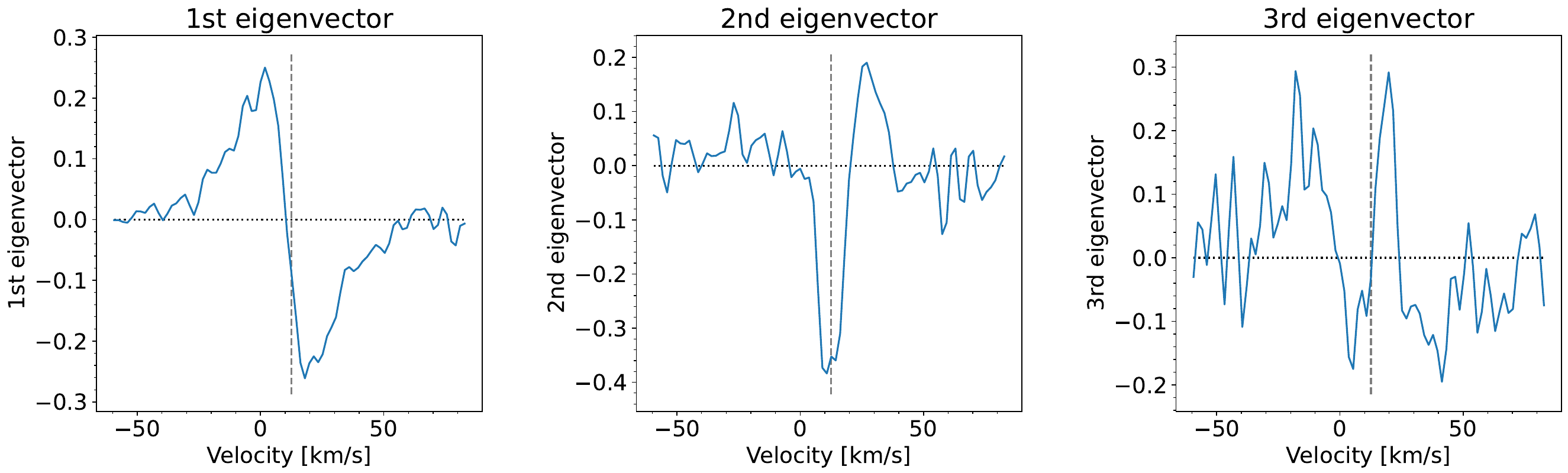}
	\includegraphics[width=0.5\textwidth, trim={0 400 0 0}, clip]{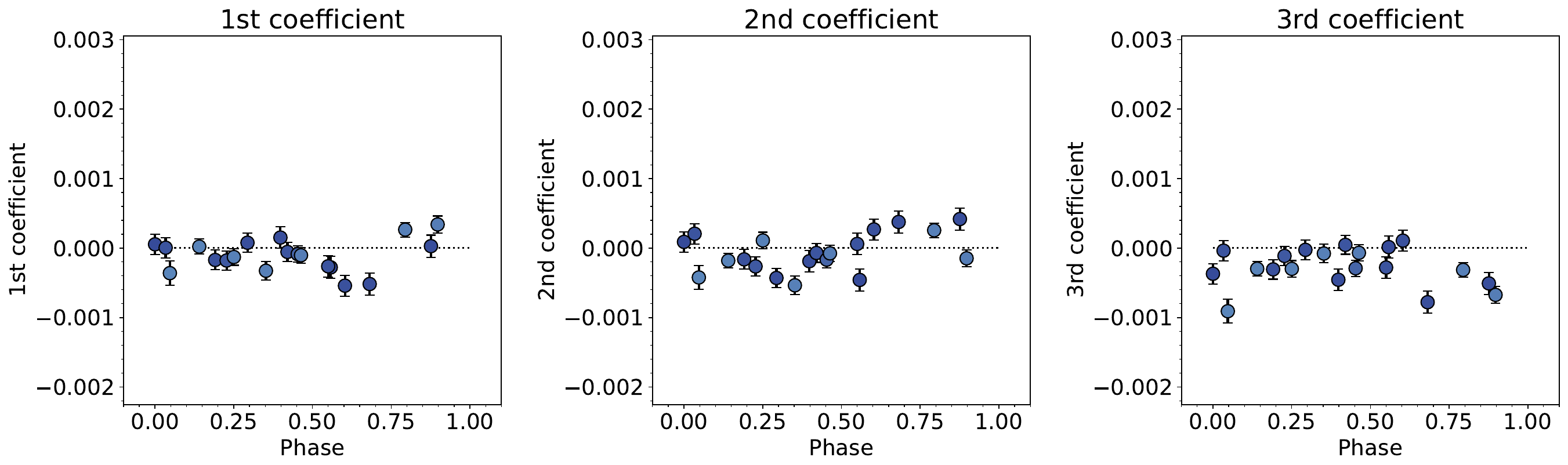}
	\includegraphics[width=0.5\textwidth, trim={0 400 0 0}, clip]{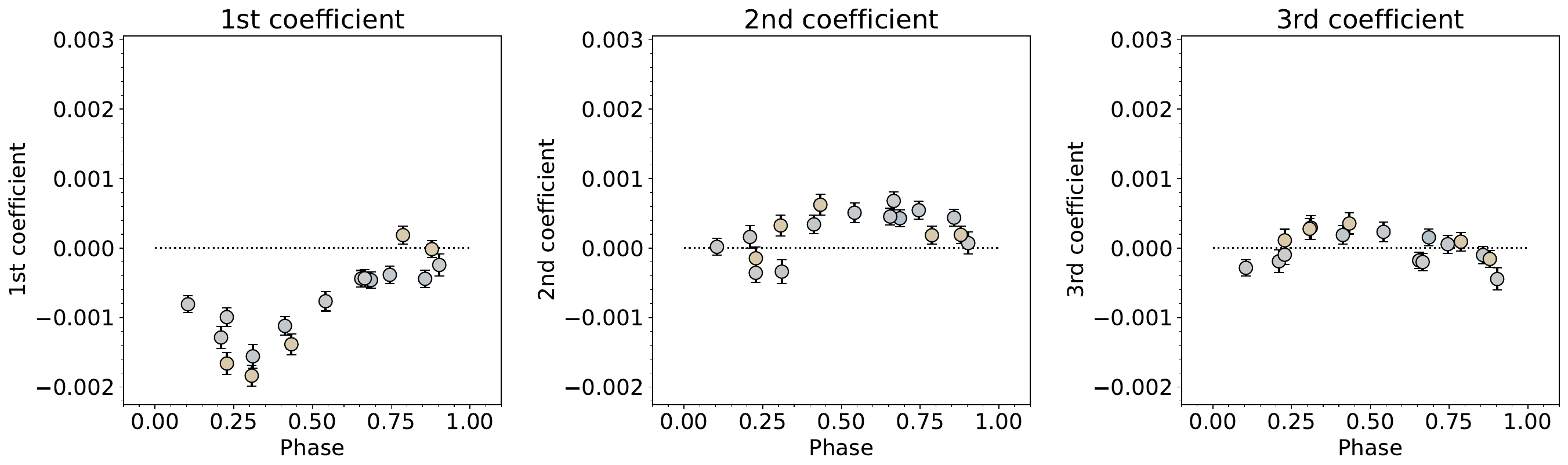}
    \includegraphics[width=0.5\textwidth, trim={0 400 0 0}, clip]{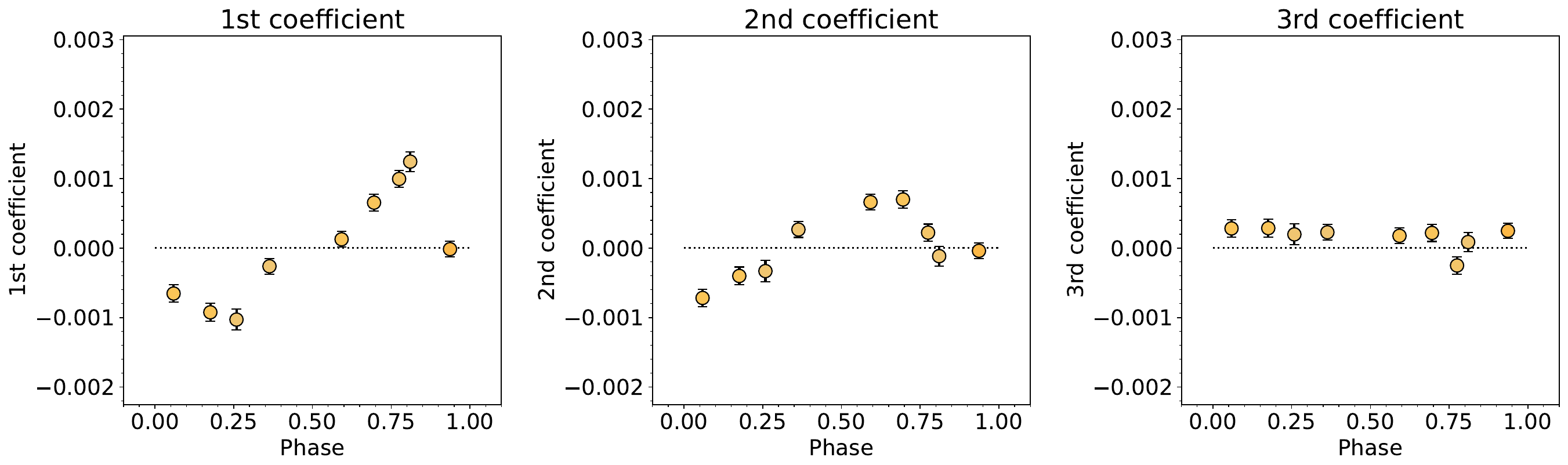}
	\includegraphics[width=0.5\textwidth, trim={0 400 0 0}, clip]{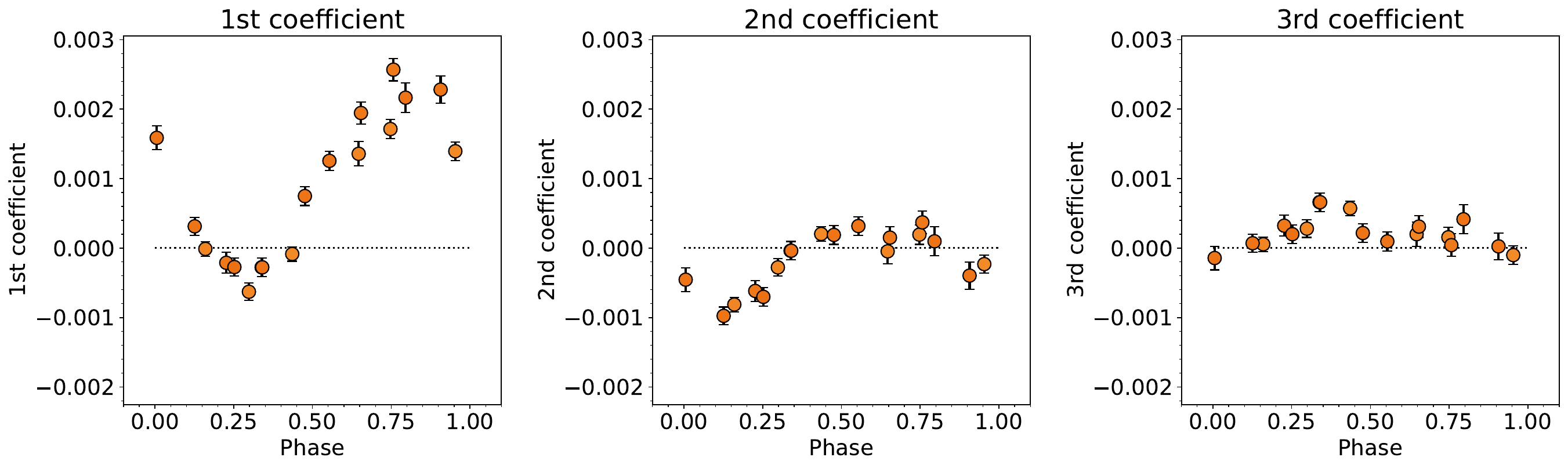}
    \caption{Principal component analysis applied to the mean-subtracted Stokes~$V$ profiles. The first three eigenvectors (top row) and their corresponding coefficients (2nd-4th row) are shown. The vertical line in the eigenvector panels indicates the radial velocity offset for AD~Leo, of approximately 12.4\,km\,s$^{-1}$, while the horizontal dotted line in all panels represents the null line. The coefficients are displayed by epoch: 2019a (2nd row), 2019b (3rd row), 2020a (4th row) and 2020b (5th row). They are phase folded using Eq.~\ref{eq:ephemeris} and colour-coded by rotation cycle.}
    \label{fig:EVCoeff}
\end{figure}

Second, the application of PCA to the mean-subtracted Stokes~$V$ profiles yields insights on the non-axisymmetric field, \citep{Lehmann2022}. For the mean-subtracted Stokes~$V$ profiles, we applied the mean profile computed across all epochs, which allows a direct reflection of the epoch-to-epoch variations in PCA coefficients (e.g. in amplitude and mean value). If the mean Stokes~$V$ profile were computed per epoch, we would miss such information, that is to say the mean value of the coefficients would be centred for each epoch, and the amplitudes could no longer be compared to each other. Fig.~\ref{fig:EVCoeff} presents the first three eigenvectors and their corresponding coefficients for the mean-subtracted Stokes~$V$ profiles separated by epoch and colour-coded by rotation cycle. The first eigenvector displays an antisymmetric shape proportional to the first derivative of the Stokes~$I$ profile and, together with the associated coefficient, scales mainly with the longitudinal magnetic field \citep{Landi2004}. The second eigenvector shows a more symmetric shape, more closely related to the second derivative of the Stokes~$I$ profile, and describes the temporal evolution of the Stokes~$V$ profiles between the maxima of the longitudinal field. According to \citet{Lehmann2022}, a strongly antisymmetric eigenvector traces the radial component and a symmetric eigenvector the azimuthal component for a dipole dominated field that is strongly poloidal, which is the case for AD~Leo. The third eigenvector features a signal as well (antisymmetric, and related to the third derivative of the Stokes~$I$ profile), which is detectable due to the high S/N of the data set, while the further eigenvectors are dominated by noise. Seeing three eigenvectors indicates that even if the axisymmetric field is likely to be dominant, we are able to detect and to analyse the non-axisymmetric field in great detail.

The coefficients of the eigenvectors suggest an evolving large-scale field as their trend changes for every epoch, see Fig.~\ref{fig:EVCoeff} 2nd-5th row.
In 2019a, the coefficients related to the first eigenvector show only a flat distribution around zero, implying a predominantly axisymmetric field. 
For the following epochs, 2019b, 2020a and 2020b, we see sine-like trends of the first two coefficients with rotational phase. The amplitude increases from epoch to epoch, indicating a growing obliquity of the dipole-dominated large-scale field. 
For the 2020b epoch, the obliquity becomes so large that the coefficients of the third eigenvector start to show a sine-like trend as well, which translates into a significant non-axisymmetric field.

Furthermore, the extremes of the coefficients associated to the antisymmetric and symmetric profile (first and second eigenvector) for the same epoch feature an apparent phase shift of $\approx 0.25$, which demonstrates that the dipolar component is poloidal dominated with little to no toroidal contribution \citep{Lehmann2022}. 
The extremes of the coefficients related to the antisymmetric eigenvector locate the pointing phase of the dipole \citep{Lehmann2022}. For the last three epochs, the maximum of this coefficient occurs at a pointing phase of $\approx 0.3$ for the northern pole of the dipole, and the sign of the eigenvector implies a negative polarity. The extremes of the coefficients occur at the same rotational phase throughout the whole observation run, designating a stable pointing phase of the dipole, in agreement with the $B_\ell$ measurements (see middle panel of Fig.~\ref{fig:Bl_evol}). 
 
By applying the PCA method on the time series of Stokes~$V$ \citep{Lehmann2022}, we confirm that AD~Leo features a dipolar large-scale field, whose obliquity increased during the latest epochs (2020a and 2020b). As the large-scale field became more non-axisymmetric, the pointing phase of the dipole remained stable.

\section{Achromaticity of the magnetic field}\label{sec:discussion}

The impact of stellar magnetic activity on radial velocity measurements features a chromatic dependence stemming from a combination of magnetic field and spot temperature contrast \citep{Reiners2013, Baroch2020}. Indeed, at near-infrared wavelengths the Zeeman broadening is expected to be stronger, while starspots contribute less owing to a lower contrast with the photosphere. The situation is reversed in the optical domain. For AD~Leo, recent work by \citet{Carmona2023} demonstrated the strong chromatic behaviour of radial velocity jitter, the latter being significantly weaker in the near-infrared domain than in optical. The combination of these effects becomes increasingly important with the activity level of the star, since the number of spots would be correspondingly larger \citep{Reinhold2019}, and it could possibly result in distinct contributions to the magnetic field strength, which can then be used to facilitate the modelling of stellar activity.

Fast-rotating stars are expected to feature high active latitudes and large polar spots \citep[e.g.,][]{Cang2020}, because the Coriolis force would overcome the buoyancy force, making the flux tubes ascend parallel to the stellar rotation axis \citep{Schuessler1992,Granzer2000}. There are some cases, however, in which fast rotation does not correlate with the presence of a polar spot \citep{Barnes2004,Morin2008a}. The fact that AD~Leo is a moderate rotator observed nearly pole-on makes it an interesting case to investigate whether longitudinal field measurements are chromatic, reflecting the behaviour of an underlying spot.

Previous studies dedicated to the Sun have shown that the magnetic field strength measured in individual lines varies significantly \citep{Demidov2008,Demidov2012}, and differences between optical and near-infrared domains have unveiled a dependence of the field strength on atmospheric height: the field increases while going towards deeper internal layers \citep{Zayer1989,Solanki1993}. For other stars, \citet{Valenti1995} reported a chromatic difference in magnetic field strength for the moderatively active K dwarf $\varepsilon$ Eri, but attributed its origin to incomplete modelling of the spectral lines used for the Zeeman broadening analysis. No wavelength dependence of the field strength was reported more recently, neither for $\varepsilon$ Eri \citep{Petit2021} nor for T Tauri stars \citep{Finociety2021}. The same conclusion was reached by \citet{Bellotti2021} when computing longitudinal field values for the active M~dwarf EV~Lac using blue ($<550$ nm) and red ($>550$ nm) lines of an optical line list. 

To investigate the longitudinal field chromaticity, we analyse the contemporaneous observations taken with SPIRou and ESPaDOnS in November 2019. We limit the LSD computation within successive wavelength bins of the line mask, and evaluate the longitudinal field for each case. Including both optical and near-infrared domains, we considered 11 subsets of lines in the following ranges: [350,390], [390,430], [430,480], [480,550], [550,650], [650,1100], [950,1100], [1100,1400], [1400,1600], [1600,1800], [1800,2500] nm. The [650,1100] and [950, 1100] nm ranges represent the red end of ESPaDOnS spectra and the blue end of SPIRou spectra, respectively. We adopt more wavelength regions than those presented in \citet{Bellotti2021}, allowing a finer search of chromatic trends. The number of lines used varies between 100 and 1000 in the optical, and between 120 and 300 in the near-infrared (see Fig.~\ref{fig:Stokes_chromatic}). In addition, we compute LSD using a 50-lines mask in the overlapping wavelength region of ESPaDOnS and SPIRou spectra ([950,1050] nm).

Stokes $I$ and $V$ profiles were computed for the simultaneous SPIRou and ESPaDOnS epochs, namely 2019b and 2019, respectively. To increase the S/N and allow a more precise estimate of B$_l$, the profiles obtained with a specific line list subset and belonging to the same epoch were co-added. This is reasonable considering the marginal amplitude variation over the epochs examined and the unchanged polarity of Stokes $V$. The longitudinal field was then computed with Eq.~\ref{eq:Bl} using the specific normalisation wavelength and Land\'e factor of each line subset, and adapting the velocity integration range according to the width of the co-added Stokes $V$ profile.

\begin{figure}[t]
    \centering
    \includegraphics[width=\columnwidth]{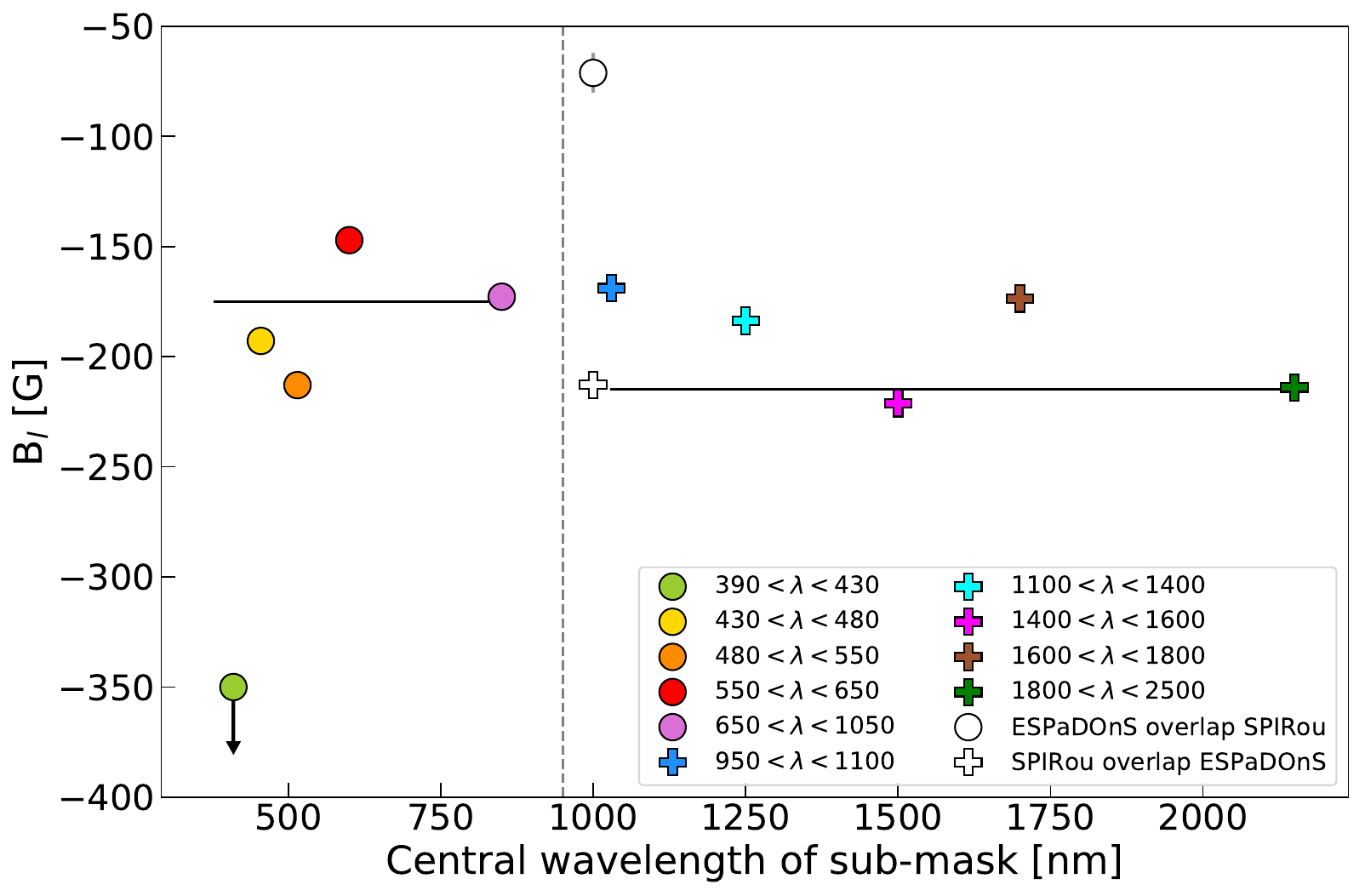}
    \caption{Test on B$_l$ measurements chromaticity. Shown are B$_l$ measurements obtained with line mask subsets based on wavelength of the optical (circles) and near-infrared (pluses) domain. The horizontal black lines indicate the values of B$_l$ computed with the full masks and the vertical dashed line separates arbitrarily optical from near-infrared measurements. The error bars are smaller than the symbol size. For visualisation purposes, the [350, 390] nm wavelength bin is not shown since it leads to an outlier data point at around $-$1~kG, and the [390, 430] nm bin yields a value at $-$750\,G so it is indicated with a downward arrow. Overall, no chromatic trend emerges from the data.}
    \label{fig:thermal_test}%
\end{figure}

From Fig.~\ref{fig:thermal_test}, we observe no clear chromaticity of B$_l$. The distribution of field strength is flat around $-$200\,G with a total scatter of 20\,G. Such dispersion is mainly due to LSD computations with a low number of lines, implying Stokes shapes more sensitive to variations in individual lines, blends and residuals of telluric correction. For the same reason, some profiles appear deformed and lead to evident outliers (see Fig.\ref{fig:Stokes_chromatic}). For instance, the B$_l$ value obtained from ESPaDOnS data in the spectral region overlapping with SPIRou is 100\,G weaker (in absolute value) than the B$_l$ value obtained from SPIRou data in the same wavelength region. This could be due to the low S/N at the very red edge of ESPaDOnS.

The case of [390,430] nm leads to a field value of $-$750\,G, despite the Stokes profiles do not show a particular deformation. We attribute this behaviour to an imprecise continuum normalisation of the spectra, likely due to a challenging identification of the continuum level in the blue part of the spectrum, where M~dwarfs feature forests of spectral lines. The effect is a smaller depth (and equivalent width) of the Stokes~$I$ profile relative to the other cases, which artificially increases the value of the field (in absolute value). Overall, although the [350,390] and [390,430] nm bins contain more than half of the lines in the optical mask, their weight in the LSD computation is small \citep{Kochukhov2010} making their effect in the computation of B$_l$ with the full mask negligible.

\begin{figure}[t]
    \centering
    \includegraphics[width=\columnwidth]{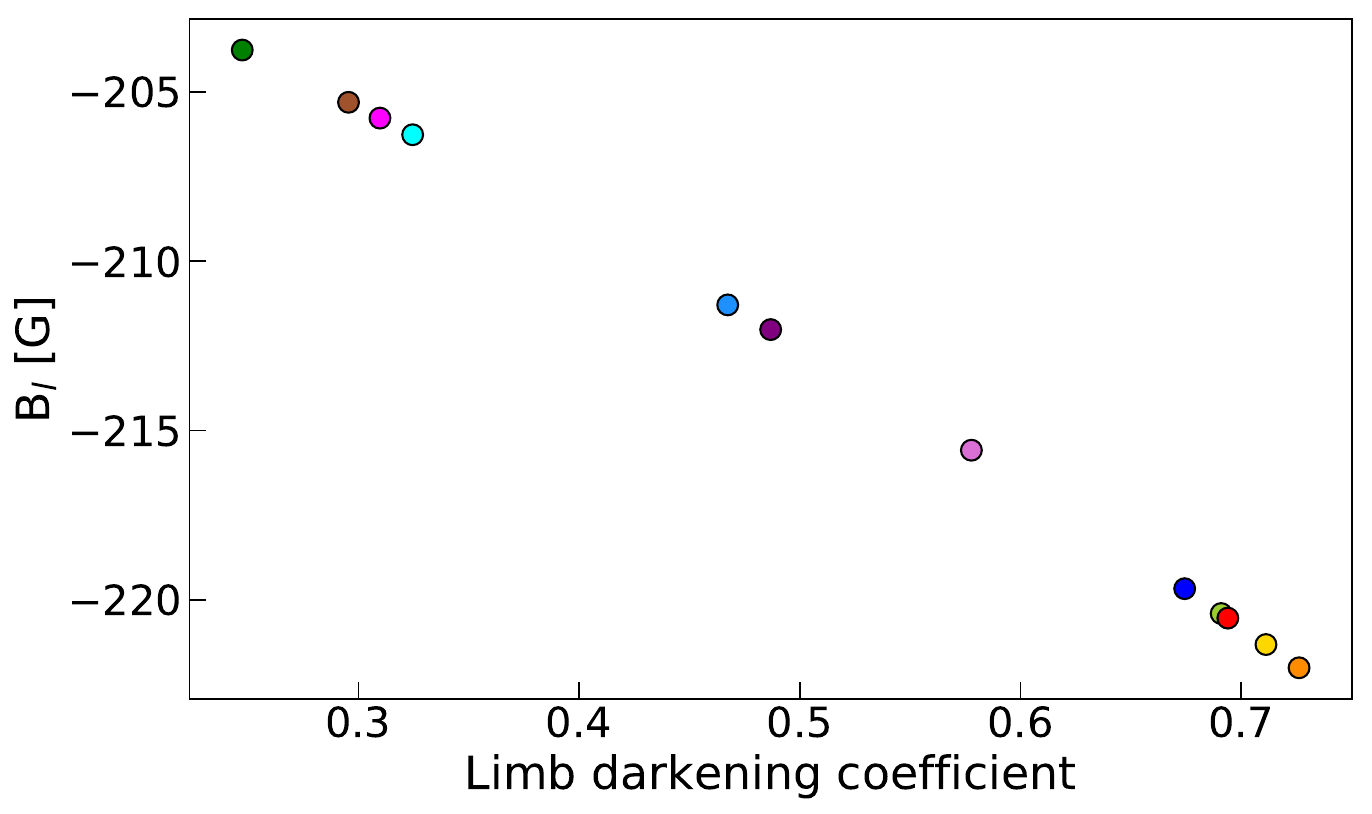}
    \caption{Dependence of the magnetic field strength on limb darkening. A monotonic decreasing (in absolute value) trend is observed from large (optical) to small (near-infrared) coefficient values. This effect is small, and negligible compared to noise in real observations.}
    \label{fig:limb_tests}%
\end{figure}

We repeated the same exercise for the other SPIRou epochs and found a similar behaviour, the only difference being 2020b data points shifting upwards because of the field global weakening. A possible implication of the lack of a chromatic trend may be the absence of a polar spot for AD~Leo. This would be justified considering that other faster-rotating M~dwarfs like V374~Peg \citep{Morin2008a} and HK~Aqr \citep{Barnes2004} do not show polar spots.

A potential source of chromaticity for B$_l$ values may come from limb darkening. This radial gradient in stellar brightness over the visible disk can be expressed as a linear function of the angle between the line of sight and the normal to a surface element ($\theta$)
\begin{equation}\label{eq:limb}
    \frac{I}{I_0} = 1 - \varepsilon(1 - \cos\theta),
\end{equation}
where $I_0$ is the brightness at disk centre ($\theta=0^{\circ}$) and $\varepsilon$ is the limb darkening coefficient. \citet{Claret2011} show that $\varepsilon$ decreases with wavelength, being 0.7 in V band and 0.3 in H band. The linear limb darkening law in Eq.\ref{eq:limb} is the one implemented in the ZDI reconstruction \citep{Folsom2018}. 

Owing to the stronger limb darkening in optical than in near-infrared, there is the possibility of additional polarity cancellation in the latter domain, which would lead to weaker field measurements. For the specific case of AD\,Leo, the low stellar inclination makes the equator appear at the limb and near-infrared observations would be more sensitive to this region. In particular, the sign of large-scale dipolar magnetic field lines exiting the pole would cancel out more with those at equator, compared to optical observations.

To verify this, we 1) linearly interpolated the limb darkening coefficients in \citet{Claret2011} at the wavelengths examined for the thermal contrast test (see Fig.~\ref{fig:thermal_test}), 2) synthesised Stokes profiles for the same coefficients assuming an axisymmetric dipole of 1\,kG seen pole-on (akin to AD\,Leo in 2019a) and infinite S/N, and 3) computed the associated field values with Eq.~\ref{eq:Bl}. The results are illustrated in Fig.~\ref{fig:limb_tests}. We observe a small (7\%) weakening of the field from optical to near-infrared, which is overwhelmed by noise in real observations.

\section{Discussion and conclusions}\label{sec:conclusion}

In this paper, we presented the results of an extended spectropolarimetric monitoring of the active M~dwarf AD~Leo, using near-infrared observations collected with SPIRou between 2019 and 2020 as part of the SLS survey. They add to the previous optical data obtained with ESPaDOnS and Narval between 2006 and 2019, making the entire time series encompass approximately 14~yr. To carry out our magnetic analysis, we computed the longitudinal magnetic field, tracked the variations of the Stokes $I$ FWHM, modelled Zeeman broadening on individual selected lines, reconstructed the large-scale field topology via ZDI, and assessed axisymmetry variations by means of a novel PCA method.

Initially, \citet{Morin2008} reported an axisymmetric, dipole-dominated structure that was stable over one year; later, \citet{Lavail2018} pointed out a large-scale weakening and small-scale enhancement of the field but no variation in the geometry. We found strong evidence of a large-scale field evolution, that is summarised as follows:

\begin{enumerate}
    \item The longitudinal magnetic field has weakened between 2006 and 2020, from $-300$ to $-50$\,G, with a rapid decrease of 100\,G in the 2020b. The dipolar longitudinal magnetic field evolved in the same time frame, starting from -850\,G in 2006, reaching -560\,G in 2016 and restoring back to -900\,G in 2020.
    \item The FWHM of Stokes~$I$ profiles does not show rotational modulation, but a dispersion that may partly be due to short-term variability. The epoch-averaged FWHM manifests a long-term variation both in optical and near-infrared, being wider in 2019b and 2020a, and narrower in 2019a and 2020b. The variations are enhanced when the Stokes profiles are computed with magnetically-sensitive lines, as opposed to the insensitive ones. The near-infrared data in particular feature a trend moderately correlated with B$_l$ (in absolute value). 
    \item The magnetic flux estimated from the modelling of Zeeman broadening exhibits a global increase over time, which is also correlated to the long-term trend of the longitudinal magnetic field (in absolute value). Moreover the epoch-averaged magnetic flux obtained for the near-infrared SPIRou time series oscillates in a similar manner to the FWHM of Stokes~$I$, demonstrating that the latter is capable of tracing secular evolution of the total, unsigned magnetic field.
    \item Zeeman-Doppler imaging reconstructions confirmed the same kind of topological evolution, with the axisymmetric level decreasing to $60\%$  and the obliquity between magnetic and rotation axis increasing to $38^\circ$. This already found support by the enhanced intermittency of the amplitude of Stokes $V$ profiles in late 2020. 
    \item The PCA method confirmed the predominantly poloidal and dipolar geometry of the large-scale field, as well as a lower axisymmetry in 2020a and 2020b. In addition, the pointing phase of the dipole remained stable during the evolution.
    \item Measurements of the magnetic field strength are overall achromatic, since they manifest only a marginal wavelength dependence due to limb darkening.
\end{enumerate}

Our results altogether suggest that AD~Leo may be entering a polarity reversal phase of a long-term magnetic cycle, analogous to the solar one. The combination of chromospheric activity studies and spectropolarimetric campaigns show that some Sun-like stars may manifest magnetic cycles and polarity reversals in phase with chromospheric cycles \citep{BoroSaikia2016,Jeffers2017}, while others have a more complex behaviour where very regular chromospheric oscillations have no straightforward polarimetric counterpart \citep{BoroSaikia2022}. 

Predicting when the polarity reversal may occur for AD~Leo is not a trivial task, as the B$_l$ data set does not feature a clear minimum or maximum. Recently, \citet{Fuhrmeister2023} did not report any evident trends from a long-term campaign of chromospheric indexes, whereas previous studies based on photometric observations reported either two co-existing timescales for cycles, namely 7\,yr and 2\,yr \citep{Buccino2014}, or an individual one of about 11\,yr \citep{Tuomi2018}. However, these time scales are not compatible with the variations in B$_l$ observed over 14~yr. The axisymmetric level of the large-scale topology is a more suitable proxy to track the cycle \citep{Lehmann2021}, but we recorded its change only in the most recent observations. 

A comparison between the magnetic field evolution described here and that of the radial velocity jitter obtained in \citet{Carmona2023} leads to a puzzling situation. \cite{Carmona2023} show that radial velocity variations in optical are essentially due to the presence of a spot and that this signal has changed only slightly (in phase and amplitude, the latter varies from 25.6$\pm$0.3\,m\,s$^{-1}$ to 23.6$\pm$0.5\,m\,s$^{-1}$) between 2005 and 2021. Such radial velocity signal is not detected in infrared with SPIRou, corroborating its strong chromaticity and therefore its origin due to stellar activity \citet{Carmona2023}. The fact that the dipolar field evolution is disjointed from a surface brightness evolution is not a surprise: \citet{Morin2008a} show that the mainly-dipolar topology of V374~Peg did not correlate with the complex brightness map reconstructed via Doppler imaging.

These considerations motivate long-term spectropolarimetric and velocimetric campaigns of active M~dwarfs. For AD~Leo in particular, additional monitoring is required to observe the polarity reversal and the cycle's extremes, to constrain a precise time scale. An extended temporal baseline could also give more insight on the link between topological variations and high-energy flaring events \citep[e.g.,][]{Stelzer2022}. At the same time, we could shed more light on the relation between the evolution of the large-scale magnetic field topology and the stability of the radial velocity jitter.

An additional detail we could infer about AD Leo's magnetic field is the helicity, which quantifies the linkage between poloidal and toroidal field lines and thus describes the complexity of the magnetic topology \citep{Lund2020,Lund2021}. For the Sun, \citet{Pipin2019} reported a temporal variation of the value correlated to the magnetic cycle. Indeed, helicity maxima and minima occur when the axis of symmetry of the poloidal and toroidal field components are aligned and orthogonal, respectively. 

For AD~Leo, the fraction of toroidal energy is only a negligible fraction of the total one, hence we should exert caution when deriving quantities from it. Over time, we observe that the poloidal axisymmetric ($m=0$) mode maintains $>$80\% of the magnetic energy and features a drop to 45\% in 2020b, while the energy in the toroidal axisymmetric mode decreases from 30\% to 6\%. As a result, the two components maintained an overall misaligned configuration, but in the most recent epoch, the poloidal component became more aligned with the toroidal one due to the axisymmetry decrease. Following the practical visualisation of \citet{Lund2021}, this evolution would correspond to an increase in field helicity.



The existence of a magnetic cycle for AD~Leo is in agreement with the observational evidence of such phenomena for M~dwarfs from radial velocity exoplanet searches \citep{GomesDaSilva2012,LopezSantiago2020}. In general, studies have shown that magnetic cycles introduce long-term signals in radial velocity data sets that can dominate over planetary signatures \citep{Meunier2010, Meunier2019}, as they modulate the appearance and number of heterogeneities on the stellar surface. It is therefore necessary to have an accurate constraint on the temporal variations of the cycle, in order to remove its contamination and allow a more reliable planetary detection and characterisation \citep{Lovis2011,Costes2021,Sairam2022}. 

Furthermore, activity cycles modulate the stellar radiation output and winds in which close-in planets are immersed \citep{Yeo2014,Hazra2020}. This leads to a temporal variation in the planetary atmospheric stripping with consequent alteration of the chemical properties and habitability \citep{Lanza2013,McCann2019,Louca2023,Konings2022}. Details on the occurrence of the cycle extremes can thus inform the most suitable interpretation framework and observing plans for missions dedicated to transmission spectroscopy like $Ariel$ \citep{Tinetti2021}. At the same time, periodic variations in the large-scale field geometry need to be considered for an accurate and updated modelling of the low-frequency radio emission discovered for M~dwarfs \citep{Callingham2021}, which has been recently proposed to potentially reveal the presence of close-in magnetic planets \citep{Vedantham2020,Kavanagh2021,Kavanagh2022}.

Finally, AD~Leo may not be an isolated case. To verify this, it is essential to explore the possibility for such cycles over a wider area of the stellar parameter space, namely mass and rotation period. 


%
%
%

\begin{acknowledgements}

We acknowledge funding from the French National Research Agency (ANR) under contract number ANR-18-CE31-0019 (SPlaSH). SB acknowledges funding from the European Space Agency (ESA), under the visiting researcher programme. LTL acknowledges funding from the European Research Council under the H2020 research \& innovation programme (grant \#740651 NewWorlds). XD and AC acknoweldge for funding  in the framework of the Investissements dAvenir programme (ANR-15-IDEX-02), through the funding of the `Origin of Life' project of the Univ. Grenoble-Alpes. OK acknowledges support by the Swedish Research Council (grant agreement no. 2019-03548), the Swedish National Space Agency, and the Royal Swedish Academy of Sciences. Based on observations obtained at the Canada-France-Hawaii Telescope (CFHT) which is operated by the National Research Council (NRC) of Canada, the Institut National des Sciences de l'Univers of the Centre National de la Recherche Scientifique (CNRS) of France, and the University of Hawaii. The observations at the CFHT were performed with care and respect from the summit of Maunakea which is a significant cultural and historic site. We gratefully acknowledge the CFHT QSO observers who made this project possible. This work has made use of the VALD database, operated at Uppsala University, the Institute of Astronomy RAS in Moscow, and the University of Vienna; Astropy, 12 a community-developed core Python package for Astronomy \citep{Astropy2013,Astropy2018}; NumPy \citep{VanderWalt2011}; Matplotlib: Visualization with Python \citep{Hunter2007}; SciPy \citep{Virtanen2020}. 

\end{acknowledgements}

%
%

\bibliographystyle{aa}
\bibliography{biblio}

\begin{thebibliography}{158}
\expandafter\ifx\csname natexlab\endcsname\relax\def\natexlab#1{#1}\fi

\bibitem[{{Artigau} {et~al.}(2014){Artigau}, {Astudillo-Defru}, {Delfosse},
  {Bouchy}, {Bonfils}, {Lovis}, {Pepe}, {Moutou}, {Donati}, {Doyon}, \&
  {Malo}}]{Artigau2014}
{Artigau}, {\'E}., {Astudillo-Defru}, N., {Delfosse}, X., {et~al.} 2014, in
  Society of Photo-Optical Instrumentation Engineers (SPIE) Conference Series,
  Vol. 9149, Observatory Operations: Strategies, Processes, and Systems V, ed.
  A.~B. {Peck}, C.~R. {Benn}, \& R.~L. {Seaman}, 914905

\bibitem[{{Astropy Collaboration} {et~al.}(2018){Astropy Collaboration},
  {Price-Whelan}, {Sip{\H{o}}cz}, {G{\"u}nther}, {Lim}, {Crawford}, {Conseil},
  {Shupe}, {Craig}, {Dencheva}, {Ginsburg}, {VanderPlas}, {Bradley},
  {P{\'e}rez-Su{\'a}rez}, {de Val-Borro}, {Aldcroft}, {Cruz}, {Robitaille},
  {Tollerud}, {Ardelean}, {Babej}, {Bach}, {Bachetti}, {Bakanov}, {Bamford},
  {Barentsen}, {Barmby}, {Baumbach}, {Berry}, {Biscani}, {Boquien}, {Bostroem},
  {Bouma}, {Brammer}, {Bray}, {Breytenbach}, {Buddelmeijer}, {Burke},
  {Calderone}, {Cano Rodr{\'\i}guez}, {Cara}, {Cardoso}, {Cheedella}, {Copin},
  {Corrales}, {Crichton}, {D'Avella}, {Deil}, {Depagne}, {Dietrich}, {Donath},
  {Droettboom}, {Earl}, {Erben}, {Fabbro}, {Ferreira}, {Finethy}, {Fox},
  {Garrison}, {Gibbons}, {Goldstein}, {Gommers}, {Greco}, {Greenfield},
  {Groener}, {Grollier}, {Hagen}, {Hirst}, {Homeier}, {Horton}, {Hosseinzadeh},
  {Hu}, {Hunkeler}, {Ivezi{\'c}}, {Jain}, {Jenness}, {Kanarek}, {Kendrew},
  {Kern}, {Kerzendorf}, {Khvalko}, {King}, {Kirkby}, {Kulkarni}, {Kumar},
  {Lee}, {Lenz}, {Littlefair}, {Ma}, {Macleod}, {Mastropietro}, {McCully},
  {Montagnac}, {Morris}, {Mueller}, {Mumford}, {Muna}, {Murphy}, {Nelson},
  {Nguyen}, {Ninan}, {N{\"o}the}, {Ogaz}, {Oh}, {Parejko}, {Parley}, {Pascual},
  {Patil}, {Patil}, {Plunkett}, {Prochaska}, {Rastogi}, {Reddy Janga},
  {Sabater}, {Sakurikar}, {Seifert}, {Sherbert}, {Sherwood-Taylor}, {Shih},
  {Sick}, {Silbiger}, {Singanamalla}, {Singer}, {Sladen}, {Sooley},
  {Sornarajah}, {Streicher}, {Teuben}, {Thomas}, {Tremblay}, {Turner},
  {Terr{\'o}n}, {van Kerkwijk}, {de la Vega}, {Watkins}, {Weaver}, {Whitmore},
  {Woillez}, {Zabalza}, \& {Astropy Contributors}}]{Astropy2018}
{Astropy Collaboration}, {Price-Whelan}, A.~M., {Sip{\H{o}}cz}, B.~M., {et~al.}
  2018, \aj, 156, 123

\bibitem[{{Astropy Collaboration} {et~al.}(2013){Astropy Collaboration},
  {Robitaille}, {Tollerud}, {Greenfield}, {Droettboom}, {Bray}, {Aldcroft},
  {Davis}, {Ginsburg}, {Price-Whelan}, {Kerzendorf}, {Conley}, {Crighton},
  {Barbary}, {Muna}, {Ferguson}, {Grollier}, {Parikh}, {Nair}, {Unther},
  {Deil}, {Woillez}, {Conseil}, {Kramer}, {Turner}, {Singer}, {Fox}, {Weaver},
  {Zabalza}, {Edwards}, {Azalee Bostroem}, {Burke}, {Casey}, {Crawford},
  {Dencheva}, {Ely}, {Jenness}, {Labrie}, {Lim}, {Pierfederici}, {Pontzen},
  {Ptak}, {Refsdal}, {Servillat}, \& {Streicher}}]{Astropy2013}
{Astropy Collaboration}, {Robitaille}, T.~P., {Tollerud}, E.~J., {et~al.} 2013,
  \aap, 558, A33

\bibitem[{{Babcock}(1961)}]{Babcock1961}
{Babcock}, H.~W. 1961, \apj, 133, 572

\bibitem[{{Bagnulo} {et~al.}(2009){Bagnulo}, {Landolfi}, {Landstreet}, {Landi
  Degl'Innocenti}, {Fossati}, \& {Sterzik}}]{Bagnulo2009}
{Bagnulo}, S., {Landolfi}, M., {Landstreet}, J.~D., {et~al.} 2009, \pasp, 121,
  993

\bibitem[{{Baliunas} {et~al.}(1995){Baliunas}, {Donahue}, {Soon}, {Horne},
  {Frazer}, {Woodard-Eklund}, {Bradford}, {Rao}, {Wilson}, {Zhang}, {Bennett},
  {Briggs}, {Carroll}, {Duncan}, {Figueroa}, {Lanning}, {Misch}, {Mueller},
  {Noyes}, {Poppe}, {Porter}, {Robinson}, {Russell}, {Shelton}, {Soyumer},
  {Vaughan}, \& {Whitney}}]{Baliunas1995}
{Baliunas}, S.~L., {Donahue}, R.~A., {Soon}, W.~H., {et~al.} 1995, \apj, 438,
  269

\bibitem[{{Barnes} {et~al.}(2004){Barnes}, {James}, \& {Collier
  Cameron}}]{Barnes2004}
{Barnes}, J.~R., {James}, D.~J., \& {Collier Cameron}, A. 2004, \mnras, 352,
  589

\bibitem[{{Baroch} {et~al.}(2020){Baroch}, {Morales}, {Ribas}, {Herrero},
  {Rosich}, {Perger}, {Anglada-Escud{\'e}}, {Reiners}, {Caballero},
  {Quirrenbach}, {Amado}, {Jeffers}, {Cifuentes}, {Passegger}, {Schweitzer},
  {Lafarga}, {Bauer}, {B{\'e}jar}, {Colom{\'e}}, {Cort{\'e}s-Contreras},
  {Dreizler}, {Galad{\'\i}-Enr{\'\i}quez}, {Hatzes}, {Henning}, {Kaminski},
  {K{\"u}rster}, {Montes}, {Rodr{\'\i}guez-L{\'o}pez}, \&
  {Zechmeister}}]{Baroch2020}
{Baroch}, D., {Morales}, J.~C., {Ribas}, I., {et~al.} 2020, \aap, 641, A69

\bibitem[{{Bellotti} {et~al.}(2022){Bellotti}, {Petit}, {Morin}, {Hussain},
  {Folsom}, {Carmona}, {Delfosse}, \& {Moutou}}]{Bellotti2021}
{Bellotti}, S., {Petit}, P., {Morin}, J., {et~al.} 2022, \aap, 657, A107

\bibitem[{{Bertaux} {et~al.}(2014){Bertaux}, {Lallement}, {Ferron}, {Boonne},
  \& {Bodichon}}]{Bertaux2014}
{Bertaux}, J.~L., {Lallement}, R., {Ferron}, S., {Boonne}, C., \& {Bodichon},
  R. 2014, \aap, 564, A46

\bibitem[{{Blackwell-Whitehead} {et~al.}(2006){Blackwell-Whitehead},
  {Lundberg}, {Nave}, {Pickering}, {Jones}, {Lyubchik}, {Pavlenko}, \&
  {Viti}}]{Blackwell-Whitehead2006-TiI-lines}
{Blackwell-Whitehead}, R.~J., {Lundberg}, H., {Nave}, G., {et~al.} 2006,
  \mnras, 373, 1603

\bibitem[{{Boro Saikia} {et~al.}(2016){Boro Saikia}, {Jeffers}, {Morin},
  {Petit}, {Folsom}, {Marsden}, {Donati}, {Cameron}, {Hall}, {Perdelwitz},
  {Reiners}, \& {Vidotto}}]{BoroSaikia2016}
{Boro Saikia}, S., {Jeffers}, S.~V., {Morin}, J., {et~al.} 2016, \aap, 594, A29

\bibitem[{{Boro Saikia} {et~al.}(2022){Boro Saikia}, {L{\"u}ftinger}, {Folsom},
  {Antonova}, {Alecian}, {Donati}, {Guedel}, {Hall}, {Jeffers}, {Kochukhov},
  {Marsden}, {Metodieva}, {Mittag}, {Morin}, {Perdelwitz}, {Petit}, {Schmid},
  \& {Vidotto}}]{BoroSaikia2022}
{Boro Saikia}, S., {L{\"u}ftinger}, T., {Folsom}, C.~P., {et~al.} 2022, \aap,
  658, A16

\bibitem[{{Boro Saikia} {et~al.}(2018){Boro Saikia}, {Marvin}, {Jeffers},
  {Reiners}, {Cameron}, {Marsden}, {Petit}, {Warnecke}, \&
  {Yadav}}]{BoroSaikia2018}
{Boro Saikia}, S., {Marvin}, C.~J., {Jeffers}, S.~V., {et~al.} 2018, \aap, 616,
  A108

\bibitem[{{Buccino} {et~al.}(2014){Buccino}, {Petrucci}, {Jofr{\'e}}, \&
  {Mauas}}]{Buccino2014}
{Buccino}, A.~P., {Petrucci}, R., {Jofr{\'e}}, E., \& {Mauas}, P. J.~D. 2014,
  \apjl, 781, L9

\bibitem[{{Callingham} {et~al.}(2021){Callingham}, {Vedantham}, {Shimwell},
  {Pope}, {Davis}, {Best}, {Hardcastle}, {R{\"o}ttgering}, {Sabater}, {Tasse},
  {van Weeren}, {Williams}, {Zarka}, {de Gasperin}, \&
  {Drabent}}]{Callingham2021}
{Callingham}, J.~R., {Vedantham}, H.~K., {Shimwell}, T.~W., {et~al.} 2021,
  Nature Astronomy, 5, 1233

\bibitem[{{Cang} {et~al.}(2020){Cang}, {Petit}, {Donati}, {Folsom}, {Jardine},
  {Villarreal D'Angelo}, {Vidotto}, {Marsden}, {Gallet}, \& {Zaire}}]{Cang2020}
{Cang}, T.~Q., {Petit}, P., {Donati}, J.~F., {et~al.} 2020, \aap, 643, A39

\bibitem[{{Carleo} {et~al.}(2020){Carleo}, {Malavolta}, {Lanza}, {Damasso},
  {Desidera}, {Borsa}, {Mallonn}, {Pinamonti}, {Gratton}, {Alei}, {Benatti},
  {Mancini}, {Maldonado}, {Biazzo}, {Esposito}, {Frustagli},
  {Gonz{\'a}lez-{\'A}lvarez}, {Micela}, {Scandariato}, {Sozzetti}, {Affer},
  {Bignamini}, {Bonomo}, {Claudi}, {Cosentino}, {Covino}, {Fiorenzano},
  {Giacobbe}, {Harutyunyan}, {Leto}, {Maggio}, {Molinari}, {Nascimbeni},
  {Pagano}, {Pedani}, {Piotto}, {Poretti}, {Rainer}, {Redfield}, {Baffa},
  {Baruffolo}, {Buchschacher}, {Billotti}, {Cecconi}, {Falcini}, {Fantinel},
  {Fini}, {Galli}, {Ghedina}, {Ghinassi}, {Giani}, {Gonzalez}, {Gonzalez},
  {Guerra}, {Hernandez Diaz}, {Hernandez}, {Iuzzolino}, {Lodi}, {Oliva},
  {Origlia}, {Perez Ventura}, {Puglisi}, {Riverol}, {Riverol}, {San Juan},
  {Sanna}, {Scuderi}, {Seemann}, {Sozzi}, \& {Tozzi}}]{Carleo2020}
{Carleo}, I., {Malavolta}, L., {Lanza}, A.~F., {et~al.} 2020, \aap, 638, A5

\bibitem[{{Carmona} {et~al.}(2023){Carmona}, {Delfosse}, {Bellotti},
  {Cort{\'e}s-Zuleta}, {Ould-Elhkim}, {Heidari}, {Mignon}, {Donati}, {Moutou},
  {Cook}, {Artigau}, {Fouqu{\'e}}, {Martioli}, {Cadieux}, {Morin}, {Forveille},
  {Boisse}, {H{\'e}brard}, {D{\'\i}az}, {Lafreni{\`e}re}, {Kiefer}, {Petit},
  {Doyon}, {Acu{\~n}a}, {Arnold}, {Bonfils}, {Bouchy}, {Bourrier}, {Dalal},
  {Deleuil}, {Demangeon}, {Dumusque}, {Hara}, {Hoyer}, {Mousis}, {Santerne},
  {S{\'e}grasan}, {Stalport}, \& {Udry}}]{Carmona2023}
{Carmona}, A., {Delfosse}, X., {Bellotti}, S., {et~al.} 2023, \aap, 674, A110

\bibitem[{{Chabrier} \& {Baraffe}(1997)}]{Chabrier1997}
{Chabrier}, G. \& {Baraffe}, I. 1997, \aap, 327, 1039

\bibitem[{{Charbonneau}(2010)}]{Charbonneau2010}
{Charbonneau}, P. 2010, Living Reviews in Solar Physics, 7, 3

\bibitem[{{Claret} \& {Bloemen}(2011)}]{Claret2011}
{Claret}, A. \& {Bloemen}, S. 2011, \aap, 529, A75

\bibitem[{{Cook} {et~al.}(2022){Cook}, {Artigau}, {Doyon}, {Hobson},
  {Martioli}, {Bouchy}, {Moutou}, {Carmona}, {Usher}, {Fouqu{\'e}}, {Arnold},
  {Delfosse}, {Boisse}, {Cadieux}, {Vandal}, {Donati}, \&
  {Desli{\`e}res}}]{Cook2022}
{Cook}, N.~J., {Artigau}, {\'E}., {Doyon}, R., {et~al.} 2022, \pasp, 134,
  114509

\bibitem[{{Costes} {et~al.}(2021){Costes}, {Watson}, {de Mooij}, {Saar},
  {Dumusque}, {Cameron}, {Phillips}, {G{\"u}nther}, {Jenkins}, {Mortier}, \&
  {Thompson}}]{Costes2021}
{Costes}, J.~C., {Watson}, C.~A., {de Mooij}, E., {et~al.} 2021, \mnras, 505,
  830

\bibitem[{{Cristofari} {et~al.}(2023){Cristofari}, {Donati}, {Folsom},
  {Masseron}, {Fouqu{\'e}}, {Moutou}, {Artigau}, {Carmona}, {Petit},
  {Delfosse}, {Martioli}, \& {the SLS consortium}}]{Cristofari2023}
{Cristofari}, P.~I., {Donati}, J.~F., {Folsom}, C.~P., {et~al.} 2023, \mnras,
  522, 1342

\bibitem[{{Demidov} \& {Balthasar}(2012)}]{Demidov2012}
{Demidov}, M.~L. \& {Balthasar}, H. 2012, \solphys, 276, 43

\bibitem[{{Demidov} {et~al.}(2008){Demidov}, {Golubeva}, {Balthasar}, {Staude},
  \& {Grigoryev}}]{Demidov2008}
{Demidov}, M.~L., {Golubeva}, E.~M., {Balthasar}, H., {Staude}, J., \&
  {Grigoryev}, V.~M. 2008, \solphys, 250, 279

\bibitem[{{Donati}(2003)}]{Donati2003}
{Donati}, J.~F. 2003, in Astronomical Society of the Pacific Conference Series,
  Vol. 307, Solar Polarization, ed. J.~{Trujillo-Bueno} \& J.~{Sanchez
  Almeida}, 41

\bibitem[{{Donati} \& {Brown}(1997)}]{DonatiBrown1997}
{Donati}, J.~F. \& {Brown}, S.~F. 1997, \aap, 326, 1135

\bibitem[{{Donati} {et~al.}(2023){Donati}, {Cristofari}, {Finociety}, {Klein},
  {Moutou}, {Gaidos}, {Cadieux}, {Artigau}, {Correia}, {Bou{\'e}}, {Cook},
  {Carmona}, {Lehmann}, {Bouvier}, {Martioli}, {Morin}, {Fouqu{\'e}},
  {Delfosse}, {Doyon}, {H{\'e}brard}, {Alencar}, {Laskar}, {Arnold}, {Petit},
  {K{\'o}sp{\'a}l}, {Vidotto}, {Folsom}, \& {SLS Collaboration}}]{Donati2023}
{Donati}, J.~F., {Cristofari}, P.~I., {Finociety}, B., {et~al.} 2023, \mnras
  [\eprint[arXiv]{2304.09642}]

\bibitem[{{Donati} {et~al.}(2006){Donati}, {Howarth}, {Jardine}, {Petit},
  {Catala}, {Landstreet}, {Bouret}, {Alecian}, {Barnes}, {Forveille},
  {Paletou}, \& {Manset}}]{Donati2006}
{Donati}, J.~F., {Howarth}, I.~D., {Jardine}, M.~M., {et~al.} 2006, \mnras,
  370, 629

\bibitem[{{Donati} {et~al.}(2020){Donati}, {Kouach}, {Moutou}, {Doyon},
  {Delfosse}, {Artigau}, {Baratchart}, {Lacombe}, {Barrick}, {H{\'e}brard},
  {Bouchy}, {Saddlemyer}, {Par{\`e}s}, {Rabou}, {Micheau}, {Dolon}, {Reshetov},
  {Challita}, {Carmona}, {Striebig}, {Thibault}, {Martioli}, {Cook},
  {Fouqu{\'e}}, {Vermeulen}, {Wang}, {Arnold}, {Pepe}, {Boisse}, {Figueira},
  {Bouvier}, {Ray}, {Feugeade}, {Morin}, {Alencar}, {Hobson}, {Castilho},
  {Udry}, {Santos}, {Hernandez}, {Benedict}, {Vall{\'e}e}, {Gallou}, {Dupieux},
  {Larrieu}, {Perruchot}, {Sottile}, {Moreau}, {Usher}, {Baril}, {Wildi},
  {Chazelas}, {Malo}, {Bonfils}, {Loop}, {Kerley}, {Wevers}, {Dunn}, {Pazder},
  {Macdonald}, {Dubois}, {Carri{\'e}}, {Valentin}, {Henault}, {Yan}, \&
  {Steinmetz}}]{Donati2020}
{Donati}, J.~F., {Kouach}, D., {Moutou}, C., {et~al.} 2020, \mnras, 498, 5684

\bibitem[{{Donati} \& {Landstreet}(2009)}]{Donati2009}
{Donati}, J.~F. \& {Landstreet}, J.~D. 2009, \araa, 47, 333

\bibitem[{{Donati} {et~al.}(2008{\natexlab{a}}){Donati}, {Morin}, {Petit},
  {Delfosse}, {Forveille}, {Auri{\`e}re}, {Cabanac}, {Dintrans}, {Fares},
  {Gastine}, {Jardine}, {Ligni{\`e}res}, {Paletou}, {Ramirez Velez}, \&
  {Th{\'e}ado}}]{Donati2008}
{Donati}, J.~F., {Morin}, J., {Petit}, P., {et~al.} 2008{\natexlab{a}}, \mnras,
  390, 545

\bibitem[{{Donati} {et~al.}(2008{\natexlab{b}}){Donati}, {Moutou}, {Far{\`e}s},
  {Bohlender}, {Catala}, {Deleuil}, {Shkolnik}, {Collier Cameron}, {Jardine},
  \& {Walker}}]{Donati2008b}
{Donati}, J.~F., {Moutou}, C., {Far{\`e}s}, R., {et~al.} 2008{\natexlab{b}},
  \mnras, 385, 1179

\bibitem[{{Donati} {et~al.}(1997){Donati}, {Semel}, {Carter}, {Rees}, \&
  {Collier Cameron}}]{Donati1997}
{Donati}, J.~F., {Semel}, M., {Carter}, B.~D., {Rees}, D.~E., \& {Collier
  Cameron}, A. 1997, \mnras, 291, 658

\bibitem[{{Fares} {et~al.}(2009){Fares}, {Donati}, {Moutou}, {Bohlender},
  {Catala}, {Deleuil}, {Shkolnik}, {Collier Cameron}, {Jardine}, \&
  {Walker}}]{Fares2009}
{Fares}, R., {Donati}, J.~F., {Moutou}, C., {et~al.} 2009, \mnras, 398, 1383

\bibitem[{{Feiden} {et~al.}(2021){Feiden}, {Skidmore}, \& {Jao}}]{Feiden2021}
{Feiden}, G.~A., {Skidmore}, K., \& {Jao}, W.-C. 2021, \apj, 907, 53

\bibitem[{{Finociety} {et~al.}(2021){Finociety}, {Donati}, {Klein}, {Zaire},
  {Lehmann}, {Moutou}, {Bouvier}, {Alencar}, {Yu}, {Grankin}, {Artigau},
  {Doyon}, {Delfosse}, {Fouqu{\'e}}, {H{\'e}brard}, {Jardine},
  {K{\'o}sp{\'a}l}, {M{\'e}nard}, {M{\'e}nard}, \& {SLS
  Consortium}}]{Finociety2021}
{Finociety}, B., {Donati}, J.~F., {Klein}, B., {et~al.} 2021, \mnras, 508, 3427

\bibitem[{{Folsom} {et~al.}(2018){Folsom}, {Bouvier}, {Petit}, {L{\`e}bre},
  {Amard}, {Palacios}, {Morin}, {Donati}, \& {Vidotto}}]{Folsom2018}
{Folsom}, C.~P., {Bouvier}, J., {Petit}, P., {et~al.} 2018, \mnras, 474, 4956

\bibitem[{{Folsom} {et~al.}(2016){Folsom}, {Petit}, {Bouvier}, {L{\`e}bre},
  {Amard}, {Palacios}, {Morin}, {Donati}, {Jeffers}, {Marsden}, \&
  {Vidotto}}]{Folsom2016}
{Folsom}, C.~P., {Petit}, P., {Bouvier}, J., {et~al.} 2016, \mnras, 457, 580

\bibitem[{{Foreman-Mackey} {et~al.}(2013){Foreman-Mackey}, {Hogg}, {Lang}, \&
  {Goodman}}]{ForemanMackey2013}
{Foreman-Mackey}, D., {Hogg}, D.~W., {Lang}, D., \& {Goodman}, J. 2013, \pasp,
  125, 306

\bibitem[{{Fouqu{\'e}} {et~al.}(2023){Fouqu{\'e}}, {Martioli}, {Donati},
  {Lehmann}, {Zaire}, {Bellotti}, {Gaidos}, {Morin}, {Moutou}, {Petit},
  {Alencar}, {Arnold}, {Artigau}, {Cang}, {Carmona}, {Cook},
  {Cort{\'e}s-Zuleta}, {Cristofari}, {Delfosse}, {Doyon}, {H{\'e}brard},
  {Malo}, {Reyl{\'e}}, \& {Usher}}]{Fouque2023}
{Fouqu{\'e}}, P., {Martioli}, E., {Donati}, J.~F., {et~al.} 2023, \aap, 672,
  A52

\bibitem[{{Fuhrmeister} {et~al.}(2023){Fuhrmeister}, {Czesla}, {Perdelwitz},
  {Nagel}, {Schmitt}, {Jeffers}, {Caballero}, {Zechmeister}, {Montes},
  {Reiners}, {L{\'o}pez-Gallifa}, {Ribas}, {Quirrenbach}, {Amado},
  {Galad{\'\i}-Enr{\'\i}quez}, {B{\'e}jar}, {Danielski}, {Hatzes}, {Kaminski},
  {K{\"u}rster}, {Morales}, \& {Zapatero Osorio}}]{Fuhrmeister2023}
{Fuhrmeister}, B., {Czesla}, S., {Perdelwitz}, V., {et~al.} 2023, \aap, 670,
  A71

\bibitem[{{Gaia Collaboration} {et~al.}(2021){Gaia Collaboration}, {Smart},
  {Sarro}, {Rybizki}, {Reyl{\'e}}, {Robin}, {Hambly}, {Abbas}, {Barstow}, {de
  Bruijne}, {Bucciarelli}, {Carrasco}, {Cooper}, {Hodgkin}, {Masana},
  {Michalik}, {Sahlmann}, {Sozzetti}, {Brown}, {Vallenari}, {Prusti},
  {Babusiaux}, {Biermann}, {Creevey}, {Evans}, {Eyer}, {Hutton}, {Jansen},
  {Jordi}, {Klioner}, {Lammers}, {Lindegren}, {Luri}, {Mignard}, {Panem},
  {Pourbaix}, {Randich}, {Sartoretti}, {Soubiran}, {Walton}, {Arenou},
  {Bailer-Jones}, {Bastian}, {Cropper}, {Drimmel}, {Katz}, {Lattanzi}, {van
  Leeuwen}, {Bakker}, {Casta{\~n}eda}, {De Angeli}, {Ducourant}, {Fabricius},
  {Fouesneau}, {Fr{\'e}mat}, {Guerra}, {Guerrier}, {Guiraud}, {Jean-Antoine
  Piccolo}, {Messineo}, {Mowlavi}, {Nicolas}, {Nienartowicz}, {Pailler},
  {Panuzzo}, {Riclet}, {Roux}, {Seabroke}, {Sordo}, {Tanga}, {Th{\'e}venin},
  {Gracia-Abril}, {Portell}, {Teyssier}, {Altmann}, {Andrae}, {Bellas-Velidis},
  {Benson}, {Berthier}, {Blomme}, {Brugaletta}, {Burgess}, {Busso}, {Carry},
  {Cellino}, {Cheek}, {Clementini}, {Damerdji}, {Davidson}, {Delchambre},
  {Dell'Oro}, {Fern{\'a}ndez-Hern{\'a}ndez}, {Galluccio}, {Garc{\'\i}a-Lario},
  {Garcia-Reinaldos}, {Gonz{\'a}lez-N{\'u}{\~n}ez}, {Gosset}, {Haigron},
  {Halbwachs}, {Harrison}, {Hatzidimitriou}, {Heiter}, {Hern{\'a}ndez},
  {Hestroffer}, {Holl}, {Jan{\ss}en}, {Jevardat de Fombelle}, {Jordan},
  {Krone-Martins}, {Lanzafame}, {L{\"o}ffler}, {Lorca}, {Manteiga}, {Marchal},
  {Marrese}, {Moitinho}, {Mora}, {Muinonen}, {Osborne}, {Pancino}, {Pauwels},
  {Recio-Blanco}, {Richards}, {Riello}, {Rimoldini}, {Roegiers}, {Siopis},
  {Smith}, {Ulla}, {Utrilla}, {van Leeuwen}, {van Reeven}, {Abreu Aramburu},
  {Accart}, {Aerts}, {Aguado}, {Ajaj}, {Altavilla}, {{\'A}lvarez}, {{\'A}lvarez
  Cid-Fuentes}, {Alves}, {Anderson}, {Anglada Varela}, {Antoja}, {Audard},
  {Baines}, {Baker}, {Balaguer-N{\'u}{\~n}ez}, {Balbinot}, {Balog}, {Barache},
  {Barbato}, {Barros}, {Bartolom{\'e}}, {Bassilana}, {Bauchet},
  {Baudesson-Stella}, {Becciani}, {Bellazzini}, {Bernet}, {Bertone}, {Bianchi},
  {Blanco-Cuaresma}, {Boch}, {Bombrun}, {Bossini}, {Bouquillon}, {Bragaglia},
  {Bramante}, {Breedt}, {Bressan}, {Brouillet}, {Burlacu}, {Busonero},
  {Butkevich}, {Buzzi}, {Caffau}, {Cancelliere}, {C{\'a}novas},
  {Cantat-Gaudin}, {Carballo}, {Carlucci}, {Carnerero}, {Casamiquela},
  {Castellani}, {Castro-Ginard}, {Castro Sampol}, {Chaoul}, {Charlot},
  {Chemin}, {Chiavassa}, {Cioni}, {Comoretto}, {Cornez}, {Cowell}, {Crifo},
  {Crosta}, {Crowley}, {Dafonte}, {Dapergolas}, {David}, {David}, {de Laverny},
  {De Luise}, {De March}, {De Ridder}, {de Souza}, {de Teodoro}, {de Torres},
  {del Peloso}, {del Pozo}, {Delgado}, {Delgado}, {Delisle}, {Di Matteo},
  {Diakite}, {Diener}, {Distefano}, {Dolding}, {Eappachen}, {Edvardsson},
  {Enke}, {Esquej}, {Fabre}, {Fabrizio}, {Faigler}, {Fedorets}, {Fernique},
  {Fienga}, {Figueras}, {Fouron}, {Fragkoudi}, {Fraile}, {Franke}, {Gai},
  {Garabato}, {Garcia-Gutierrez}, {Garc{\'\i}a-Torres}, {Garofalo}, {Gavras},
  {Gerlach}, {Geyer}, {Giacobbe}, {Gilmore}, {Girona}, {Giuffrida}, {Gomel},
  {Gomez}, {Gonzalez-Santamaria}, {Gonz{\'a}lez-Vidal}, {Granvik},
  {Guti{\'e}rrez-S{\'a}nchez}, {Guy}, {Hauser}, {Haywood}, {Helmi}, {Hidalgo},
  {Hilger}, {H{\l}adczuk}, {Hobbs}, {Holland}, {Huckle}, {Jasniewicz},
  {Jonker}, {Juaristi Campillo}, {Julbe}, {Karbevska}, {Kervella}, {Khanna},
  {Kochoska}, {Kontizas}, {Kordopatis}, {Korn}, {Kostrzewa-Rutkowska},
  {Kruszy{\'n}ska}, {Lambert}, {Lanza}, {Lasne}, {Le Campion}, {Le Fustec},
  {Lebreton}, {Lebzelter}, {Leccia}, {Leclerc}, {Lecoeur-Taibi}, {Liao},
  {Licata}, {Lindstr{\o}m}, {Lister}, {Livanou}, {Lobel}, {Madrero Pardo},
  {Managau}, {Mann}, {Marchant}, {Marconi}, {Marcos Santos}, {Marinoni},
  {Marocco}, {Marshall}, {Martin Polo}, {Mart{\'\i}n-Fleitas}, {Masip},
  {Massari}, {Mastrobuono-Battisti}, {Mazeh}, {McMillan}, {Messina}, {Millar},
  {Mints}, {Molina}, {Molinaro}, {Moln{\'a}r}, {Montegriffo}, {Mor},
  {Morbidelli}, {Morel}, {Morris}, {Mulone}, {Munoz}, {Muraveva}, {Murphy},
  {Musella}, {Noval}, {Ord{\'e}novic}, {Orr{\`u}}, {Osinde}, {Pagani},
  {Pagano}, {Palaversa}, {Palicio}, {Panahi}, {Pawlak}, {Pe{\~n}alosa
  Esteller}, {Penttil{\"a}}, {Piersimoni}, {Pineau}, {Plachy}, {Plum},
  {Poggio}, {Poretti}, {Poujoulet}, {Pr{\v{s}}a}, {Pulone}, {Racero},
  {Ragaini}, {Rainer}, {Raiteri}, {Rambaux}, {Ramos}, {Ramos-Lerate}, {Re
  Fiorentin}, {Regibo}, {Ripepi}, {Riva}, {Rixon}, {Robichon}, {Robin},
  {Roelens}, {Rohrbasser}, {Romero-G{\'o}mez}, {Rowell}, {Royer}, {Rybicki},
  {Sadowski}, {Sagrist{\`a} Sell{\'e}s}, {Salgado}, {Salguero}, {Samaras},
  {Sanchez Gimenez}, {Sanna}, {Santove{\~n}a}, {Sarasso}, {Schultheis},
  {Sciacca}, {Segol}, {Segovia}, {S{\'e}gransan}, {Semeux}, {Shahaf},
  {Siddiqui}, {Siebert}, {Siltala}, {Slezak}, {Solano}, {Solitro}, {Souami},
  {Souchay}, {Spagna}, {Spoto}, {Steele}, {Steidelm{\"u}ller}, {Stephenson},
  {S{\"u}veges}, {Szabados}, {Szegedi-Elek}, {Taris}, {Tauran}, {Taylor},
  {Teixeira}, {Thuillot}, {Tonello}, {Torra}, {Torra}, {Turon}, {Unger},
  {Vaillant}, {van Dillen}, {Vanel}, {Vecchiato}, {Viala}, {Vicente},
  {Voutsinas}, {Weiler}, {Wevers}, {Wyrzykowski}, {Yoldas}, {Yvard}, {Zhao},
  {Zorec}, {Zucker}, {Zurbach}, \& {Zwitter}}]{GaiaEDR3}
{Gaia Collaboration}, {Smart}, R.~L., {Sarro}, L.~M., {et~al.} 2021, \aap, 649,
  A6

\bibitem[{{Gastine} {et~al.}(2013){Gastine}, {Morin}, {Duarte}, {Reiners},
  {Christensen}, \& {Wicht}}]{Gastine2013}
{Gastine}, T., {Morin}, J., {Duarte}, L., {et~al.} 2013, \aap, 549, L5

\bibitem[{{Gomes da Silva} {et~al.}(2012){Gomes da Silva}, {Santos}, {Bonfils},
  {Delfosse}, {Forveille}, {Udry}, {Dumusque}, \& {Lovis}}]{GomesDaSilva2012}
{Gomes da Silva}, J., {Santos}, N.~C., {Bonfils}, X., {et~al.} 2012, \aap, 541,
  A9

\bibitem[{{Granzer} {et~al.}(2000){Granzer}, {Sch{\"u}ssler}, {Caligari}, \&
  {Strassmeier}}]{Granzer2000}
{Granzer}, T., {Sch{\"u}ssler}, M., {Caligari}, P., \& {Strassmeier}, K.~G.
  2000, \aap, 355, 1087

\bibitem[{{Gregory} {et~al.}(2012){Gregory}, {Donati}, {Morin}, {Hussain},
  {Mayne}, {Hillenbrand}, \& {Jardine}}]{Gregory2012}
{Gregory}, S.~G., {Donati}, J.~F., {Morin}, J., {et~al.} 2012, \apj, 755, 97

\bibitem[{{Gustafsson} {et~al.}(2008){Gustafsson}, {Edvardsson}, {Eriksson},
  {J{\o}rgensen}, {Nordlund}, \& {Plez}}]{Gustafsson2008}
{Gustafsson}, B., {Edvardsson}, B., {Eriksson}, K., {et~al.} 2008, \aap, 486,
  951

\bibitem[{{Hale} {et~al.}(1919){Hale}, {Ellerman}, {Nicholson}, \&
  {Joy}}]{Hale1919}
{Hale}, G.~E., {Ellerman}, F., {Nicholson}, S.~B., \& {Joy}, A.~H. 1919, \apj,
  49, 153

\bibitem[{{Hathaway}(2010)}]{Hathaway2010}
{Hathaway}, D.~H. 2010, Living Reviews in Solar Physics, 7, 1

\bibitem[{{Haywood} {et~al.}(2022){Haywood}, {Milbourne}, {Saar}, {Mortier},
  {Phillips}, {Charbonneau}, {Cameron}, {Cegla}, {Meunier}, \&
  {}}]{Haywood2022}
{Haywood}, R.~D., {Milbourne}, T.~W., {Saar}, S.~H., {et~al.} 2022, \apj, 935,
  6

\bibitem[{{Hazra} {et~al.}(2020){Hazra}, {Vidotto}, \& {D'Angelo}}]{Hazra2020}
{Hazra}, G., {Vidotto}, A.~A., \& {D'Angelo}, C.~V. 2020, \mnras, 496, 4017

\bibitem[{{H{\'e}brard} {et~al.}(2016){H{\'e}brard}, {Donati}, {Delfosse},
  {Morin}, {Moutou}, \& {Boisse}}]{Hebrard2016}
{H{\'e}brard}, {\'E}.~M., {Donati}, J.~F., {Delfosse}, X., {et~al.} 2016,
  \mnras, 461, 1465

\bibitem[{{Hill} {et~al.}(2019){Hill}, {Folsom}, {Donati}, {Herczeg},
  {Hussain}, {Alencar}, {Gregory}, \& {Matysse Collaboration}}]{Hill2019}
{Hill}, C.~A., {Folsom}, C.~P., {Donati}, J.~F., {et~al.} 2019, \mnras, 484,
  5810

\bibitem[{{Hunter}(2007)}]{Hunter2007}
{Hunter}, J.~D. 2007, Computing in Science and Engineering, 9, 90

\bibitem[{{Jeffers} {et~al.}(2017){Jeffers}, {Boro Saikia}, {Barnes}, {Petit},
  {Marsden}, {Jardine}, {Vidotto}, \& {BCool Collaboration}}]{Jeffers2017}
{Jeffers}, S.~V., {Boro Saikia}, S., {Barnes}, J.~R., {et~al.} 2017, \mnras,
  471, L96

\bibitem[{{Jeffers} {et~al.}(2022){Jeffers}, {Cameron}, {Marsden}, {Boro
  Saikia}, {Folsom}, {Jardine}, {Morin}, {Petit}, {See}, {Vidotto}, {Wolter},
  \& {Mittag}}]{Jeffers2022}
{Jeffers}, S.~V., {Cameron}, R.~H., {Marsden}, S.~C., {et~al.} 2022, \aap, 661,
  A152

\bibitem[{{Jeffers} {et~al.}(2018){Jeffers}, {Mengel}, {Moutou}, {Marsden},
  {Barnes}, {Jardine}, {Petit}, {Schmitt}, {See}, {Vidotto}, \& {BCool
  Collaboration}}]{Jeffers2018}
{Jeffers}, S.~V., {Mengel}, M., {Moutou}, C., {et~al.} 2018, \mnras, 479, 5266

\bibitem[{{Johns-Krull} {et~al.}(1999){Johns-Krull}, {Valenti}, \&
  {Koresko}}]{Johns-Krull1999}
{Johns-Krull}, C.~M., {Valenti}, J.~A., \& {Koresko}, C. 1999, \apj, 516, 900

\bibitem[{{Kausch} {et~al.}(2015){Kausch}, {Noll}, {Smette}, {Kimeswenger},
  {Barden}, {Szyszka}, {Jones}, {Sana}, {Horst}, \&
  {Kerber}}]{molecfit2015A&A...576A..78K}
{Kausch}, W., {Noll}, S., {Smette}, A., {et~al.} 2015, \aap, 576, A78

\bibitem[{{Kavanagh} {et~al.}(2021){Kavanagh}, {Vidotto}, {Klein}, {Jardine},
  {Donati}, \& {{\'O} Fionnag{\'a}in}}]{Kavanagh2021}
{Kavanagh}, R.~D., {Vidotto}, A.~A., {Klein}, B., {et~al.} 2021, \mnras, 504,
  1511

\bibitem[{{Kavanagh} {et~al.}(2022){Kavanagh}, {Vidotto}, {Vedantham},
  {Jardine}, {Callingham}, \& {Morin}}]{Kavanagh2022}
{Kavanagh}, R.~D., {Vidotto}, A.~A., {Vedantham}, H.~K., {et~al.} 2022, \mnras,
  514, 675

\bibitem[{{Kitchatinov} {et~al.}(2014){Kitchatinov}, {Moss}, \&
  {Sokoloff}}]{Kitchatinov2014}
{Kitchatinov}, L.~L., {Moss}, D., \& {Sokoloff}, D. 2014, \mnras, 442, L1

\bibitem[{{Klein} {et~al.}(2021){Klein}, {Donati}, {Moutou}, {Delfosse},
  {Bonfils}, {Martioli}, {Fouqu{\'e}}, {Cloutier}, {Artigau}, {Doyon},
  {H{\'e}brard}, {Morin}, {Rameau}, {Plavchan}, \& {Gaidos}}]{Klein2021}
{Klein}, B., {Donati}, J.-F., {Moutou}, C., {et~al.} 2021, \mnras, 502, 188

\bibitem[{{Kochukhov}(2021)}]{Kochukhov2021}
{Kochukhov}, O. 2021, \aapr, 29, 1

\bibitem[{{Kochukhov} {et~al.}(2009){Kochukhov}, {Heiter}, {Piskunov}, {Ryde},
  {Gustafsson}, {Bagnulo}, \& {Plez}}]{Kochukhov2009}
{Kochukhov}, O., {Heiter}, U., {Piskunov}, N., {et~al.} 2009, in American
  Institute of Physics Conference Series, Vol. 1094, 15th Cambridge Workshop on
  Cool Stars, Stellar Systems, and the Sun, ed. E.~{Stempels}, 124--129

\bibitem[{{Kochukhov} \& {Lavail}(2017)}]{Kochukhov-Lavail2017}
{Kochukhov}, O. \& {Lavail}, A. 2017, \apjl, 835, L4

\bibitem[{{Kochukhov} {et~al.}(2010){Kochukhov}, {Makaganiuk}, \&
  {Piskunov}}]{Kochukhov2010}
{Kochukhov}, O., {Makaganiuk}, V., \& {Piskunov}, N. 2010, \aap, 524, A5

\bibitem[{{Konings} {et~al.}(2022){Konings}, {Baeyens}, \&
  {Decin}}]{Konings2022}
{Konings}, T., {Baeyens}, R., \& {Decin}, L. 2022, \aap, 667, A15

\bibitem[{{Landi Degl'Innocenti} \& {Landolfi}(2004)}]{Landi2004}
{Landi Degl'Innocenti}, E. \& {Landolfi}, M. 2004, {Polarization in Spectral
  Lines}, Vol. 307

\bibitem[{{Landstreet}(1988)}]{Landstreet1988}
{Landstreet}, J.~D. 1988, \apj, 326, 967

\bibitem[{{Lanza}(2013)}]{Lanza2013}
{Lanza}, A.~F. 2013, \aap, 557, A31

\bibitem[{{Lavail} {et~al.}(2018){Lavail}, {Kochukhov}, \& {Wade}}]{Lavail2018}
{Lavail}, A., {Kochukhov}, O., \& {Wade}, G.~A. 2018, \mnras, 479, 4836

\bibitem[{{Lawler} {et~al.}(2013){Lawler}, {Guzman}, {Wood}, {Sneden}, \&
  {Cowan}}]{Lawler2013-TiI-lines}
{Lawler}, J.~E., {Guzman}, A., {Wood}, M.~P., {Sneden}, C., \& {Cowan}, J.~J.
  2013, \apjs, 205, 11

\bibitem[{{Lehmann} \& {Donati}(2022)}]{Lehmann2022}
{Lehmann}, L.~T. \& {Donati}, J.~F. 2022, \mnras, 514, 2333

\bibitem[{{Lehmann} {et~al.}(2021){Lehmann}, {Hussain}, {Vidotto}, {Jardine},
  \& {Mackay}}]{Lehmann2021}
{Lehmann}, L.~T., {Hussain}, G.~A.~J., {Vidotto}, A.~A., {Jardine}, M.~M., \&
  {Mackay}, D.~H. 2021, \mnras, 500, 1243

\bibitem[{{Leighton}(1969)}]{Leighton1969}
{Leighton}, R.~B. 1969, \apj, 156, 1

\bibitem[{{Lopez-Santiago} {et~al.}(2020){Lopez-Santiago}, {Martino},
  {M{\'\i}guez}, \& {V{\'a}zquez}}]{LopezSantiago2020}
{Lopez-Santiago}, J., {Martino}, L., {M{\'\i}guez}, J., \& {V{\'a}zquez}, M.~A.
  2020, \aj, 160, 273

\bibitem[{{Louca} {et~al.}(2023){Louca}, {Miguel}, {Tsai}, {Froning}, {Loyd},
  \& {France}}]{Louca2023}
{Louca}, A.~J., {Miguel}, Y., {Tsai}, S.-M., {et~al.} 2023, \mnras, 521, 3333

\bibitem[{{Lovis} {et~al.}(2011){Lovis}, {Dumusque}, {Santos}, {Bouchy},
  {Mayor}, {Pepe}, {Queloz}, {S{\'e}gransan}, \& {Udry}}]{Lovis2011}
{Lovis}, C., {Dumusque}, X., {Santos}, N.~C., {et~al.} 2011, arXiv e-prints,
  arXiv:1107.5325

\bibitem[{{Lund} {et~al.}(2020){Lund}, {Jardine}, {Lehmann}, {Mackay}, {See},
  {Vidotto}, {Donati}, {Fares}, {Folsom}, {Jeffers}, {Marsden}, {Morin}, \&
  {Petit}}]{Lund2020}
{Lund}, K., {Jardine}, M., {Lehmann}, L.~T., {et~al.} 2020, \mnras, 493, 1003

\bibitem[{{Lund} {et~al.}(2021){Lund}, {Jardine}, {Russell}, {Donati}, {Fares},
  {Folsom}, {Jeffers}, {Marsden}, {Morin}, {Petit}, \& {See}}]{Lund2021}
{Lund}, K., {Jardine}, M., {Russell}, A.~J.~B., {et~al.} 2021, \mnras, 502,
  4903

\bibitem[{{Maeder} \& {Meynet}(2000)}]{Maeder2000}
{Maeder}, A. \& {Meynet}, G. 2000, \araa, 38, 143

\bibitem[{{Mann} {et~al.}(2015){Mann}, {Feiden}, {Gaidos}, {Boyajian}, \& {von
  Braun}}]{Mann2015}
{Mann}, A.~W., {Feiden}, G.~A., {Gaidos}, E., {Boyajian}, T., \& {von Braun},
  K. 2015, \apj, 804, 64

\bibitem[{{Maunder}(1904)}]{Maunder1904}
{Maunder}, E.~W. 1904, \mnras, 64, 747

\bibitem[{{McCann} {et~al.}(2019){McCann}, {Murray-Clay}, {Kratter}, \&
  {Krumholz}}]{McCann2019}
{McCann}, J., {Murray-Clay}, R.~A., {Kratter}, K., \& {Krumholz}, M.~R. 2019,
  \apj, 873, 89

\bibitem[{{Meunier} {et~al.}(2010){Meunier}, {Desort}, \&
  {Lagrange}}]{Meunier2010}
{Meunier}, N., {Desort}, M., \& {Lagrange}, A.~M. 2010, \aap, 512, A39

\bibitem[{{Meunier} \& {Lagrange}(2019)}]{Meunier2019}
{Meunier}, N. \& {Lagrange}, A.~M. 2019, \aap, 628, A125

\bibitem[{{Mignon} {et~al.}(2023){Mignon}, {Meunier}, {Delfosse}, {Bonfils},
  {Santos}, {Forveille}, {Gaisn{\'e}}, {Astudillo-Defru}, {Lovis}, \&
  {Udry}}]{Mignon2023}
{Mignon}, L., {Meunier}, N., {Delfosse}, X., {et~al.} 2023, arXiv e-prints,
  arXiv:2303.03998

\bibitem[{{Mittag} {et~al.}(2017){Mittag}, {Robrade}, {Schmitt}, {Hempelmann},
  {Gonz{\'a}lez-P{\'e}rez}, \& {Schr{\"o}der}}]{Mittag2017}
{Mittag}, M., {Robrade}, J., {Schmitt}, J.~H.~M.~M., {et~al.} 2017, \aap, 600,
  A119

\bibitem[{{Morgenthaler} {et~al.}(2011){Morgenthaler}, {Petit}, {Morin},
  {Auri{\`e}re}, {Dintrans}, {Konstantinova-Antova}, \&
  {Marsden}}]{Morgenthaler2011}
{Morgenthaler}, A., {Petit}, P., {Morin}, J., {et~al.} 2011, Astronomische
  Nachrichten, 332, 866

\bibitem[{{Morin}(2012)}]{Morin2012}
{Morin}, J. 2012, in EAS Publications Series, Vol.~57, EAS Publications Series,
  ed. C.~{Reyl{\'e}}, C.~{Charbonnel}, \& M.~{Schultheis}, 165--191

\bibitem[{{Morin} {et~al.}(2011){Morin}, {Delfosse}, {Donati}, {Dormy},
  {Forveille}, {Jardine}, {Petit}, \& {Schrinner}}]{Morin2011}
{Morin}, J., {Delfosse}, X., {Donati}, J.~F., {et~al.} 2011, in SF2A-2011:
  Proceedings of the Annual meeting of the French Society of Astronomy and
  Astrophysics, ed. G.~{Alecian}, K.~{Belkacem}, R.~{Samadi}, \&
  D.~{Valls-Gabaud}, 503--508

\bibitem[{{Morin} {et~al.}(2008{\natexlab{a}}){Morin}, {Donati}, {Forveille},
  {Delfosse}, {Dobler}, {Petit}, {Jardine}, {Collier Cameron}, {Albert},
  {Manset}, {Dintrans}, {Chabrier}, \& {Valenti}}]{Morin2008a}
{Morin}, J., {Donati}, J.~F., {Forveille}, T., {et~al.} 2008{\natexlab{a}},
  \mnras, 384, 77

\bibitem[{{Morin} {et~al.}(2008{\natexlab{b}}){Morin}, {Donati}, {Petit},
  {Delfosse}, {Forveille}, {Albert}, {Auri{\`e}re}, {Cabanac}, {Dintrans},
  {Fares}, {Gastine}, {Jardine}, {Ligni{\`e}res}, {Paletou}, {Ramirez Velez},
  \& {Th{\'e}ado}}]{Morin2008}
{Morin}, J., {Donati}, J.~F., {Petit}, P., {et~al.} 2008{\natexlab{b}}, \mnras,
  390, 567

\bibitem[{{Morin} {et~al.}(2010){Morin}, {Donati}, {Petit}, {Delfosse},
  {Forveille}, \& {Jardine}}]{Morin2010}
{Morin}, J., {Donati}, J.~F., {Petit}, P., {et~al.} 2010, \mnras, 407, 2269

\bibitem[{{Morin} {et~al.}(2016){Morin}, {Hill}, \& {Watson}}]{Morin2016}
{Morin}, J., {Hill}, C.~A., \& {Watson}, C.~A. 2016, in Astrophysics and Space
  Science Library, Vol. 439, Astronomy at High Angular Resolution, ed. H.~M.~J.
  {Boffin}, G.~{Hussain}, J.-P. {Berger}, \& L.~{Schmidtobreick}, 223

\bibitem[{{Muheki} {et~al.}(2020){Muheki}, {Guenther}, {Mutabazi}, \&
  {Jurua}}]{Muheki2020}
{Muheki}, P., {Guenther}, E.~W., {Mutabazi}, T., \& {Jurua}, E. 2020, \aap,
  637, A13

\bibitem[{{Mullan} \& {MacDonald}(2001)}]{Mullan2001}
{Mullan}, D.~J. \& {MacDonald}, J. 2001, \apj, 559, 353

\bibitem[{{Namekata} {et~al.}(2020){Namekata}, {Maehara}, {Sasaki}, {Kawai},
  {Notsu}, {Kowalski}, {Allred}, {Iwakiri}, {Tsuboi}, {Murata}, {Niwano},
  {Shiraishi}, {Adachi}, {Iida}, {Oeda}, {Honda}, {Tozuka}, {Katoh}, {Onozato},
  {Okamoto}, {Isogai}, {Kimura}, {Kojiguchi}, {Wakamatsu}, {Tampo}, {Nogami},
  \& {Shibata}}]{Namekata2020}
{Namekata}, K., {Maehara}, H., {Sasaki}, R., {et~al.} 2020, \pasj, 72, 68

\bibitem[{{O'Brian} {et~al.}(1991){O'Brian}, {Wickliffe}, {Lawler}, {Whaling},
  \& {Brault}}]{OBrian1991-FeI-lines}
{O'Brian}, T.~R., {Wickliffe}, M.~E., {Lawler}, J.~E., {Whaling}, W., \&
  {Brault}, J.~W. 1991, Journal of the Optical Society of America B Optical
  Physics, 8, 1185

\bibitem[{{Parker}(1955)}]{Parker1955}
{Parker}, E.~N. 1955, \apj, 122, 293

\bibitem[{{Petit} {et~al.}(2009){Petit}, {Dintrans}, {Morgenthaler}, {Van
  Grootel}, {Morin}, {Lanoux}, {Auri{\`e}re}, \&
  {Konstantinova-Antova}}]{Petit2009}
{Petit}, P., {Dintrans}, B., {Morgenthaler}, A., {et~al.} 2009, \aap, 508, L9

\bibitem[{{Petit} {et~al.}(2008){Petit}, {Dintrans}, {Solanki}, {Donati},
  {Auri{\`e}re}, {Ligni{\`e}res}, {Morin}, {Paletou}, {Ramirez Velez},
  {Catala}, \& {Fares}}]{Petit2008}
{Petit}, P., {Dintrans}, B., {Solanki}, S.~K., {et~al.} 2008, \mnras, 388, 80

\bibitem[{{Petit} {et~al.}(2002){Petit}, {Donati}, \& {Collier
  Cameron}}]{Petit2002}
{Petit}, P., {Donati}, J.~F., \& {Collier Cameron}, A. 2002, \mnras, 334, 374

\bibitem[{{Petit} {et~al.}(2021){Petit}, {Folsom}, {Donati}, {Yu}, {do
  Nascimento}, {Jeffers}, {Marsden}, {Morin}, \& {Vidotto}}]{Petit2021}
{Petit}, P., {Folsom}, C.~P., {Donati}, J.~F., {et~al.} 2021, \aap, 648, A55

\bibitem[{{Petit} {et~al.}(2014){Petit}, {Louge}, {Th{\'e}ado}, {Paletou},
  {Manset}, {Morin}, {Marsden}, \& {Jeffers}}]{Petit2014}
{Petit}, P., {Louge}, T., {Th{\'e}ado}, S., {et~al.} 2014, \pasp, 126, 469

\bibitem[{{Petrovay}(2020)}]{Petrovay2020}
{Petrovay}, K. 2020, Living Reviews in Solar Physics, 17, 2

\bibitem[{{Pipin} {et~al.}(2019){Pipin}, {Pevtsov}, {Liu}, \&
  {Kosovichev}}]{Pipin2019}
{Pipin}, V.~V., {Pevtsov}, A.~A., {Liu}, Y., \& {Kosovichev}, A.~G. 2019,
  \apjl, 877, L36

\bibitem[{{Preston}(1967)}]{Preston1967}
{Preston}, G.~W. 1967, \apj, 150, 547

\bibitem[{{Rachkovsky}(1967)}]{Rachkovsky1967}
{Rachkovsky}, D.~N. 1967, Izvestiya Ordena Trudovogo Krasnogo Znameni Krymskoj
  Astrofizicheskoj Observatorii, 37, 56

\bibitem[{{Reiners}(2012)}]{Reiners2012}
{Reiners}, A. 2012, Living Reviews in Solar Physics, 9, 1

\bibitem[{{Reiners} \& {Basri}(2009)}]{ReinersBasri2009}
{Reiners}, A. \& {Basri}, G. 2009, \aap, 496, 787

\bibitem[{{Reiners} {et~al.}(2010){Reiners}, {Bean}, {Huber}, {Dreizler},
  {Seifahrt}, \& {Czesla}}]{Reiners2010}
{Reiners}, A., {Bean}, J.~L., {Huber}, K.~F., {et~al.} 2010, \apj, 710, 432

\bibitem[{{Reiners} {et~al.}(2013){Reiners}, {Shulyak}, {Anglada-Escud{\'e}},
  {Jeffers}, {Morin}, {Zechmeister}, {Kochukhov}, \& {Piskunov}}]{Reiners2013}
{Reiners}, A., {Shulyak}, D., {Anglada-Escud{\'e}}, G., {et~al.} 2013, \aap,
  552, A103

\bibitem[{{Reiners} {et~al.}(2022){Reiners}, {Shulyak}, {K{\"a}pyl{\"a}},
  {Ribas}, {Nagel}, {Zechmeister}, {Caballero}, {Shan}, {Fuhrmeister},
  {Quirrenbach}, {Amado}, {Montes}, {Jeffers}, {Azzaro}, {B{\'e}jar},
  {Chaturvedi}, {Henning}, {K{\"u}rster}, \& {Pall{\'e}}}]{Reiners2022}
{Reiners}, A., {Shulyak}, D., {K{\"a}pyl{\"a}}, P.~J., {et~al.} 2022, \aap,
  662, A41

\bibitem[{{Reinhold} {et~al.}(2019){Reinhold}, {Bell}, {Kuszlewicz}, {Hekker},
  \& {Shapiro}}]{Reinhold2019}
{Reinhold}, T., {Bell}, K.~J., {Kuszlewicz}, J., {Hekker}, S., \& {Shapiro},
  A.~I. 2019, \aap, 621, A21

\bibitem[{{Robertson} {et~al.}(2013){Robertson}, {Endl}, {Cochran}, \&
  {Dodson-Robinson}}]{Robertson2013}
{Robertson}, P., {Endl}, M., {Cochran}, W.~D., \& {Dodson-Robinson}, S.~E.
  2013, in American Astronomical Society Meeting Abstracts, Vol. 221, American
  Astronomical Society Meeting Abstracts \#221, 423.05

\bibitem[{{Ros{\'e}n} {et~al.}(2016){Ros{\'e}n}, {Kochukhov}, {Hackman}, \&
  {Lehtinen}}]{Rosen2016}
{Ros{\'e}n}, L., {Kochukhov}, O., {Hackman}, T., \& {Lehtinen}, J. 2016, \aap,
  593, A35

\bibitem[{{Route}(2016)}]{Route2016}
{Route}, M. 2016, \apjl, 830, L27

\bibitem[{{Ryabchikova} {et~al.}(2015){Ryabchikova}, {Piskunov}, {Kurucz},
  {Stempels}, {Heiter}, {Pakhomov}, \& {Barklem}}]{Ryabchikova2015}
{Ryabchikova}, T., {Piskunov}, N., {Kurucz}, R.~L., {et~al.} 2015, \physscr,
  90, 054005

\bibitem[{{Saar}(1994)}]{Saar1994}
{Saar}, S.~H. 1994, in Infrared Solar Physics, ed. D.~M. {Rabin}, J.~T.
  {Jefferies}, \& C.~{Lindsey}, Vol. 154, 493

\bibitem[{{Sairam} \& {Triaud}(2022)}]{Sairam2022}
{Sairam}, L. \& {Triaud}, A. H.~M.~J. 2022, \mnras, 514, 2259

\bibitem[{{Sanderson} {et~al.}(2003){Sanderson}, {Appourchaux}, {Hoeksema}, \&
  {Harvey}}]{Sanderson2003}
{Sanderson}, T.~R., {Appourchaux}, T., {Hoeksema}, J.~T., \& {Harvey}, K.~L.
  2003, Journal of Geophysical Research (Space Physics), 108, 1035

\bibitem[{{Schuessler} \& {Solanki}(1992)}]{Schuessler1992}
{Schuessler}, M. \& {Solanki}, S.~K. 1992, \aap, 264, L13

\bibitem[{{Schwabe}(1844)}]{Schwabe1844}
{Schwabe}, H. 1844, Astronomische Nachrichten, 21, 233

\bibitem[{{See} {et~al.}(2016){See}, {Jardine}, {Vidotto}, {Donati}, {Boro
  Saikia}, {Bouvier}, {Fares}, {Folsom}, {Gregory}, {Hussain}, {Jeffers},
  {Marsden}, {Morin}, {Moutou}, {do Nascimento}, {Petit}, \& {Waite}}]{See2016}
{See}, V., {Jardine}, M., {Vidotto}, A.~A., {et~al.} 2016, \mnras, 462, 4442

\bibitem[{{See} {et~al.}(2015){See}, {Jardine}, {Vidotto}, {Donati}, {Folsom},
  {Boro Saikia}, {Bouvier}, {Fares}, {Gregory}, {Hussain}, {Jeffers},
  {Marsden}, {Morin}, {Moutou}, {do Nascimento}, {Petit}, {Ros{\'e}n}, \&
  {Waite}}]{See2015}
{See}, V., {Jardine}, M., {Vidotto}, A.~A., {et~al.} 2015, \mnras, 453, 4301

\bibitem[{{Semel}(1989)}]{Semel1989}
{Semel}, M. 1989, \aap, 225, 456

\bibitem[{{Shkolnik} {et~al.}(2009){Shkolnik}, {Liu}, \& {Reid}}]{Shkolnik2009}
{Shkolnik}, E., {Liu}, M.~C., \& {Reid}, I.~N. 2009, \apj, 699, 649

\bibitem[{{Shulyak} {et~al.}(2017){Shulyak}, {Reiners}, {Engeln}, {Malo},
  {Yadav}, {Morin}, \& {Kochukhov}}]{Shulyak2017}
{Shulyak}, D., {Reiners}, A., {Engeln}, A., {et~al.} 2017, Nature Astronomy, 1,
  0184

\bibitem[{{Shulyak} {et~al.}(2019){Shulyak}, {Reiners}, {Nagel}, {Tal-Or},
  {Caballero}, {Zechmeister}, {B{\'e}jar}, {Cort{\'e}s-Contreras}, {Martin},
  {Kaminski}, {Ribas}, {Quirrenbach}, {Amado}, {Anglada-Escud{\'e}}, {Bauer},
  {Dreizler}, {Guenther}, {Henning}, {Jeffers}, {K{\"u}rster}, {Lafarga},
  {Montes}, {Morales}, \& {Pedraz}}]{Shulyak2019}
{Shulyak}, D., {Reiners}, A., {Nagel}, E., {et~al.} 2019, \aap, 626, A86

\bibitem[{{Shulyak} {et~al.}(2014){Shulyak}, {Reiners}, {Seemann}, {Kochukhov},
  \& {Piskunov}}]{Shulyak2014}
{Shulyak}, D., {Reiners}, A., {Seemann}, U., {Kochukhov}, O., \& {Piskunov}, N.
  2014, \aap, 563, A35

\bibitem[{{Skilling} \& {Bryan}(1984)}]{Skilling1984}
{Skilling}, J. \& {Bryan}, R.~K. 1984, \mnras, 211, 111

\bibitem[{{Smette} {et~al.}(2015){Smette}, {Sana}, {Noll}, {Horst}, {Kausch},
  {Kimeswenger}, {Barden}, {Szyszka}, {Jones}, {Gallenne}, {Vinther},
  {Ballester}, \& {Taylor}}]{molecfit2015A&A...576A..77S}
{Smette}, A., {Sana}, H., {Noll}, S., {et~al.} 2015, \aap, 576, A77

\bibitem[{{Solanki}(1993)}]{Solanki1993}
{Solanki}, S.~K. 1993, \ssr, 63, 1

\bibitem[{{Stelzer} {et~al.}(2022){Stelzer}, {Caramazza}, {Raetz}, {Argiroffi},
  \& {Coffaro}}]{Stelzer2022}
{Stelzer}, B., {Caramazza}, M., {Raetz}, S., {Argiroffi}, C., \& {Coffaro}, M.
  2022, \aap, 667, L9

\bibitem[{{Stibbs}(1950)}]{Stibbs1950}
{Stibbs}, D.~W.~N. 1950, \mnras, 110, 395

\bibitem[{{Tanner} {et~al.}(2013){Tanner}, {Basu}, \& {Demarque}}]{Tanner2013}
{Tanner}, J.~D., {Basu}, S., \& {Demarque}, P. 2013, \apj, 767, 78

\bibitem[{{Tinetti} {et~al.}(2021){Tinetti}, {Eccleston}, {Haswell}, {Lagage},
  {Leconte}, {L{\"u}ftinger}, {Micela}, {Min}, {Pilbratt}, {Puig}, {Swain},
  {Testi}, {Turrini}, {Vandenbussche}, {Rosa Zapatero Osorio}, {Aret},
  {Beaulieu}, {Buchhave}, {Ferus}, {Griffin}, {Guedel}, {Hartogh}, {Machado},
  {Malaguti}, {Pall{\'e}}, {Rataj}, {Ray}, {Ribas}, {Szab{\'o}}, {Tan},
  {Werner}, {Ratti}, {Scharmberg}, {Salvignol}, {Boudin}, {Halain}, {Haag},
  {Crouzet}, {Kohley}, {Symonds}, {Renk}, {Caldwell}, {Abreu}, {Alonso},
  {Amiaux}, {Berth{\'e}}, {Bishop}, {Bowles}, {Carmona}, {Coffey},
  {Colom{\'e}}, {Crook}, {D{\'e}sjonqueres}, {D{\'\i}az}, {Drummond},
  {Focardi}, {G{\'o}mez}, {Holmes}, {Krijger}, {Kovacs}, {Hunt}, {Machado},
  {Morgante}, {Ollivier}, {Ottensamer}, {Pace}, {Pagano}, {Pascale}, {Pearson},
  {M{\o}ller Pedersen}, {Pniel}, {Roose}, {Savini}, {Stamper}, {Szirovicza},
  {Szoke}, {Tosh}, {Vilardell}, {Barstow}, {Borsato}, {Casewell}, {Changeat},
  {Charnay}, {Civi{\v{s}}}, {Coud{\'e} du Foresto}, {Coustenis}, {Cowan},
  {Danielski}, {Demangeon}, {Drossart}, {Edwards}, {Gilli}, {Encrenaz}, {Kiss},
  {Kokori}, {Ikoma}, {Morales}, {Mendon{\c{c}}a}, {Moneti}, {Mugnai},
  {Garc{\'\i}a Mu{\~n}oz}, {Helled}, {Kama}, {Miguel}, {Nikolaou}, {Pagano},
  {Panic}, {Rengel}, {Rickman}, {Rocchetto}, {Sarkar}, {Selsis}, {Tennyson},
  {Tsiaras}, {Venot}, {Vida}, {Waldmann}, {Yurchenko}, {Szab{\'o}}, {Zellem},
  {Al-Refaie}, {Perez Alvarez}, {Anisman}, {Arhancet}, {Ateca}, {Baeyens},
  {Barnes}, {Bell}, {Benatti}, {Biazzo}, {B{\l}{\k{e}}cka}, {Bonomo}, {Bosch},
  {Bossini}, {Bourgalais}, {Brienza}, {Brucalassi}, {Bruno}, {Caines},
  {Calcutt}, {Campante}, {Canestrari}, {Cann}, {Casali}, {Casas}, {Cassone},
  {Cara}, {Carmona}, {Carone}, {Carrasco}, {Changeat}, {Chioetto},
  {Cortecchia}, {Czupalla}, {Chubb}, {Ciaravella}, {Claret}, {Claudi},
  {Codella}, {Garcia Comas}, {Cracchiolo}, {Cubillos}, {Da Peppo}, {Decin},
  {Dejabrun}, {Delgado-Mena}, {Di Giorgio}, {Diolaiti}, {Dorn}, {Doublier},
  {Doumayrou}, {Dransfield}, {Dumaye}, {Dunford}, {Jimenez Escobar}, {Van
  Eylen}, {Farina}, {Fedele}, {Fern{\'a}ndez}, {Fleury}, {Fonte}, {Fontignie},
  {Fossati}, {Funke}, {Galy}, {Garai}, {Garc{\'\i}a}, {Garc{\'\i}a-Rigo},
  {Garufi}, {Germano Sacco}, {Giacobbe}, {G{\'o}mez}, {Gonzalez},
  {Gonzalez-Galindo}, {Grassi}, {Griffith}, {Guarcello}, {Goujon}, {Gressier},
  {Grzegorczyk}, {Guillot}, {Guilluy}, {Hargrave}, {Hellin}, {Herrero},
  {Hills}, {Horeau}, {Ito}, {Jessen}, {Kabath}, {K{\'a}lm{\'a}n}, {Kawashima},
  {Kimura}, {Kn{\'\i}{\v{z}}ek}, {Kreidberg}, {Kruid}, {Kruijssen},
  {Kubel{\'\i}k}, {Lara}, {Lebonnois}, {Lee}, {Lefevre}, {Lichtenberg},
  {Locci}, {Lombini}, {Sanchez Lopez}, {Lorenzani}, {MacDonald}, {Magrini},
  {Maldonado}, {Marcq}, {Migliorini}, {Modirrousta-Galian}, {Molaverdikhani},
  {Molinari}, {Molli{\`e}re}, {Moreau}, {Morello}, {Morinaud}, {Morvan},
  {Moses}, {Mouzali}, {Nakhjiri}, {Naponiello}, {Narita}, {Nascimbeni},
  {Nikolaou}, {Noce}, {Oliva}, {Palladino}, {Papageorgiou}, {Parmentier},
  {Peres}, {P{\'e}rez}, {Perez-Hoyos}, {Perger}, {Cecchi Pestellini},
  {Petralia}, {Philippon}, {Piccialli}, {Pignatari}, {Piotto}, {Podio},
  {Polenta}, {Preti}, {Pribulla}, {Lopez Puertas}, {Rainer}, {Reess}, {Rimmer},
  {Robert}, {Rosich}, {Rossi}, {Rust}, {Saleh}, {Sanna}, {Schisano},
  {Schreiber}, {Schwartz}, {Scippa}, {Seli}, {Shibata}, {Simpson}, {Shorttle},
  {Skaf}, {Skup}, {Sobiecki}, {Sousa}, {Sozzetti}, {{\v{S}}poner}, {Steiger},
  {Tanga}, {Tackley}, {Taylor}, {Tecza}, {Terenzi}, {Tremblin}, {Tozzi},
  {Triaud}, {Trompet}, {Tsai}, {Tsantaki}, {Valencia}, {Carine Vandaele}, {Van
  der Swaelmen}, {Adibekyan}, {Vasisht}, {Vazan}, {Del Vecchio}, {Waltham},
  {Wawer}, {Widemann}, {Wolkenberg}, {Hou Yip}, {Yung}, {Zilinskas},
  {Zingales}, \& {Zuppella}}]{Tinetti2021}
{Tinetti}, G., {Eccleston}, P., {Haswell}, C., {et~al.} 2021, arXiv e-prints,
  arXiv:2104.04824

\bibitem[{{Tuomi} {et~al.}(2018){Tuomi}, {Jones}, {Barnes},
  {Anglada-Escud{\'e}}, {Butler}, {Kiraga}, \& {Vogt}}]{Tuomi2018}
{Tuomi}, M., {Jones}, H. R.~A., {Barnes}, J.~R., {et~al.} 2018, \aj, 155, 192

\bibitem[{{Unno}(1956)}]{Unno1956}
{Unno}, W. 1956, \pasj, 8, 108

\bibitem[{{Valenti} {et~al.}(1995){Valenti}, {Marcy}, \& {Basri}}]{Valenti1995}
{Valenti}, J.~A., {Marcy}, G.~W., \& {Basri}, G. 1995, \apj, 439, 939

\bibitem[{{van der Walt} {et~al.}(2011){van der Walt}, {Colbert}, \&
  {Varoquaux}}]{VanderWalt2011}
{van der Walt}, S., {Colbert}, S.~C., \& {Varoquaux}, G. 2011, Computing in
  Science and Engineering, 13, 22

\bibitem[{{van Saders} \& {Pinsonneault}(2012)}]{VanSaders2012}
{van Saders}, J.~L. \& {Pinsonneault}, M.~H. 2012, \apj, 746, 16

\bibitem[{{Vedantham} {et~al.}(2020){Vedantham}, {Callingham}, {Shimwell},
  {Tasse}, {Pope}, {Bedell}, {Snellen}, {Best}, {Hardcastle}, {Haverkorn},
  {Mechev}, {O'Sullivan}, {R{\"o}ttgering}, \& {White}}]{Vedantham2020}
{Vedantham}, H.~K., {Callingham}, J.~R., {Shimwell}, T.~W., {et~al.} 2020,
  Nature Astronomy, 4, 577

\bibitem[{{Vidotto} \& {Cleary}(2020)}]{Vidotto2020}
{Vidotto}, A.~A. \& {Cleary}, A. 2020, \mnras, 494, 2417

\bibitem[{{Virtanen} {et~al.}(2020){Virtanen}, {Gommers}, {Burovski},
  {Oliphant}, {Weckesser}, {Cournapeau}, {Alexbrc}, {Peterson}, {Reddy},
  {Wilson}, {Haberland}, {Mayorov}, {Endolith}, {Nelson}, {Der Van Walt},
  {Laxalde}, {Brett}, {Polat}, {Larson}, {Millman}, {Lars}, {Van Mulbregt},
  {Eric-Jones}, {Carey}, {Moore}, {Kern}, {Leslie}, {Perktold}, {Striega}, \&
  {Feng}}]{Virtanen2020}
{Virtanen}, P., {Gommers}, R., {Burovski}, E., {et~al.} 2020, {scipy/scipy:
  SciPy 1.5.3}

\bibitem[{{Wade} {et~al.}(2001){Wade}, {Bagnulo}, {Kochukhov}, {Landstreet},
  {Piskunov}, \& {Stift}}]{Wade2001}
{Wade}, G.~A., {Bagnulo}, S., {Kochukhov}, O., {et~al.} 2001, \aap, 374, 265

\bibitem[{{Wilson}(1968)}]{Wilson1968}
{Wilson}, O.~C. 1968, \apj, 153, 221

\bibitem[{{Wright} {et~al.}(2011){Wright}, {Drake}, {Mamajek}, \&
  {Henry}}]{Wright2011}
{Wright}, N.~J., {Drake}, J.~J., {Mamajek}, E.~E., \& {Henry}, G.~W. 2011,
  \apj, 743, 48

\bibitem[{{Yeo} {et~al.}(2014){Yeo}, {Krivova}, \& {Solanki}}]{Yeo2014}
{Yeo}, K.~L., {Krivova}, N.~A., \& {Solanki}, S.~K. 2014, \ssr, 186, 137

\bibitem[{{Zacharias} {et~al.}(2013){Zacharias}, {Finch}, {Girard}, {Henden},
  {Bartlett}, {Monet}, \& {Zacharias}}]{Zacharias2013}
{Zacharias}, N., {Finch}, C.~T., {Girard}, T.~M., {et~al.} 2013, \aj, 145, 44

\bibitem[{{Zaire} {et~al.}(2022){Zaire}, {Jouve}, {Gastine}, {Donati}, {Morin},
  {Landin}, \& {Folsom}}]{Zaire2022}
{Zaire}, B., {Jouve}, L., {Gastine}, T., {et~al.} 2022, \mnras, 517, 3392

\bibitem[{{Zayer} {et~al.}(1989){Zayer}, {Solanki}, \& {Stenflo}}]{Zayer1989}
{Zayer}, I., {Solanki}, S.~K., \& {Stenflo}, J.~O. 1989, \aap, 211, 463

\bibitem[{{Zeeman}(1897)}]{Zeeman1897}
{Zeeman}, P. 1897, \nat, 55, 347

\end{thebibliography}

\appendix

\section{FWHM of Stokes I}\label{sec:FWHM_StokesI}

In this appendix, the Stokes~$I$ profiles computed using different line selections are shown. We compare the width of the profiles obtained with the full atomic mask, high-Land\'e factor (g$_\mathrm{eff}>1.2$) lines and low-Land\'e factor (g$_\mathrm{eff}<1.2$) lines.

\begin{figure}[b]
	\centering
	\includegraphics[width=\columnwidth]{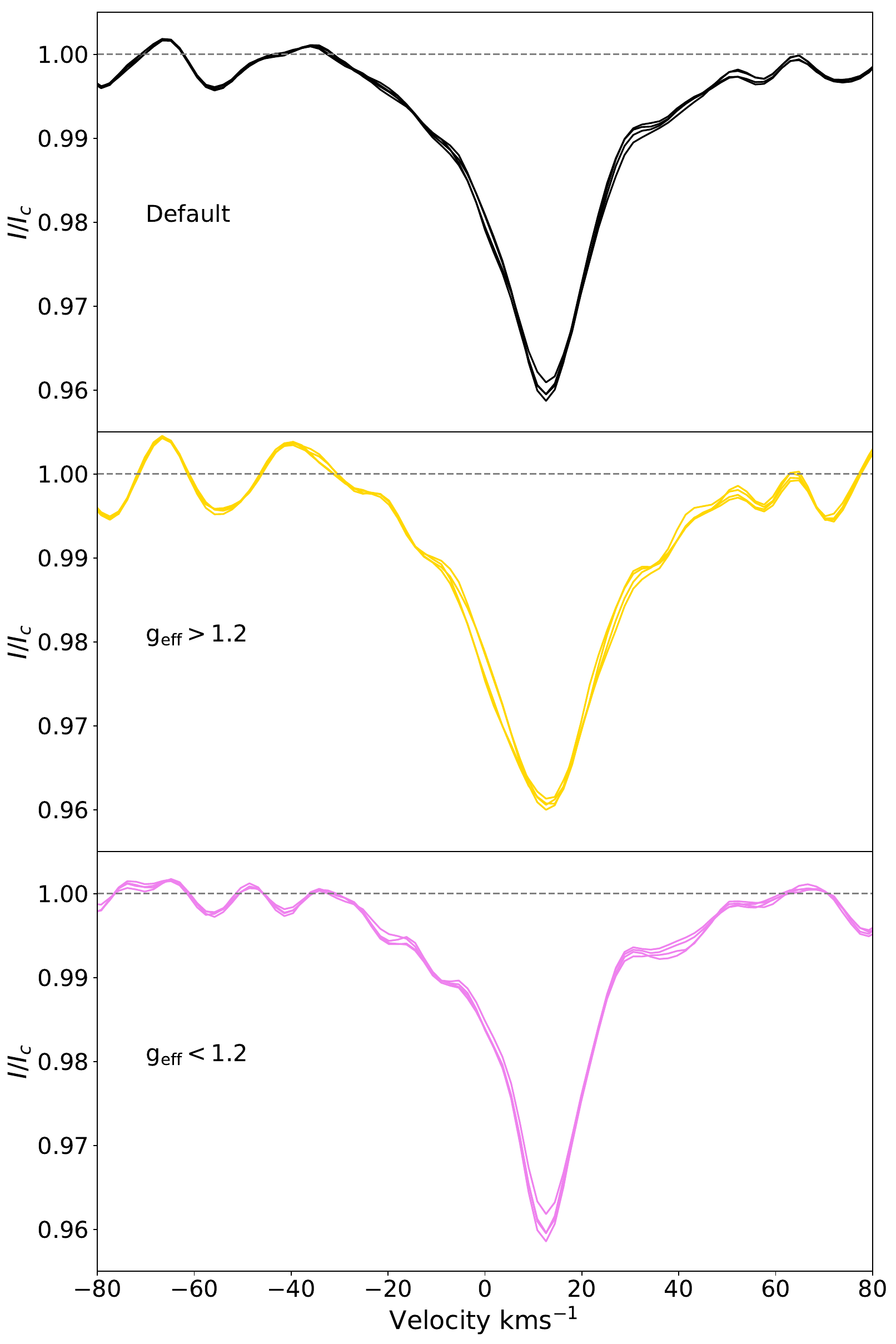}
    \caption{Median Stokes~$I$ profiles computed for the four near-infrared epochs: 2019a, 2019b, 2020a, and 2020b. The profiles are obtained using the full mask (top) magnetically sensitive (middle) and insensitive (bottom) line lists in LSD. Following the linear dependence of the Zeeman effect on the Land\'e factor, high-g$_\mathrm{eff}$ lines are broader than low-g$_\mathrm{eff}$ lines on average. The width of the profiles changes over time in the high-g$_\mathrm{eff}$ case, while it remains reasonably constant in the low-g$_\mathrm{eff}$ case.}
    \label{fig:StokesI_fwhm}
\end{figure}

\section{Chromatic Stokes profiles}\label{sec:chromatic_Stokes}

The various Stokes profiles computed with different wavelength-based line lists for LSD are reported. The wavelength intervals of the line subsets are: [350,390], [390,430], [430,480], [480,550], [550,650], [650,1100], [950,1100], [1100,1400], [1400,1600], [1600,1800], [1800,2500]\,nm.

\begin{figure*}[t]
	\centering
	\includegraphics[width=\textwidth]{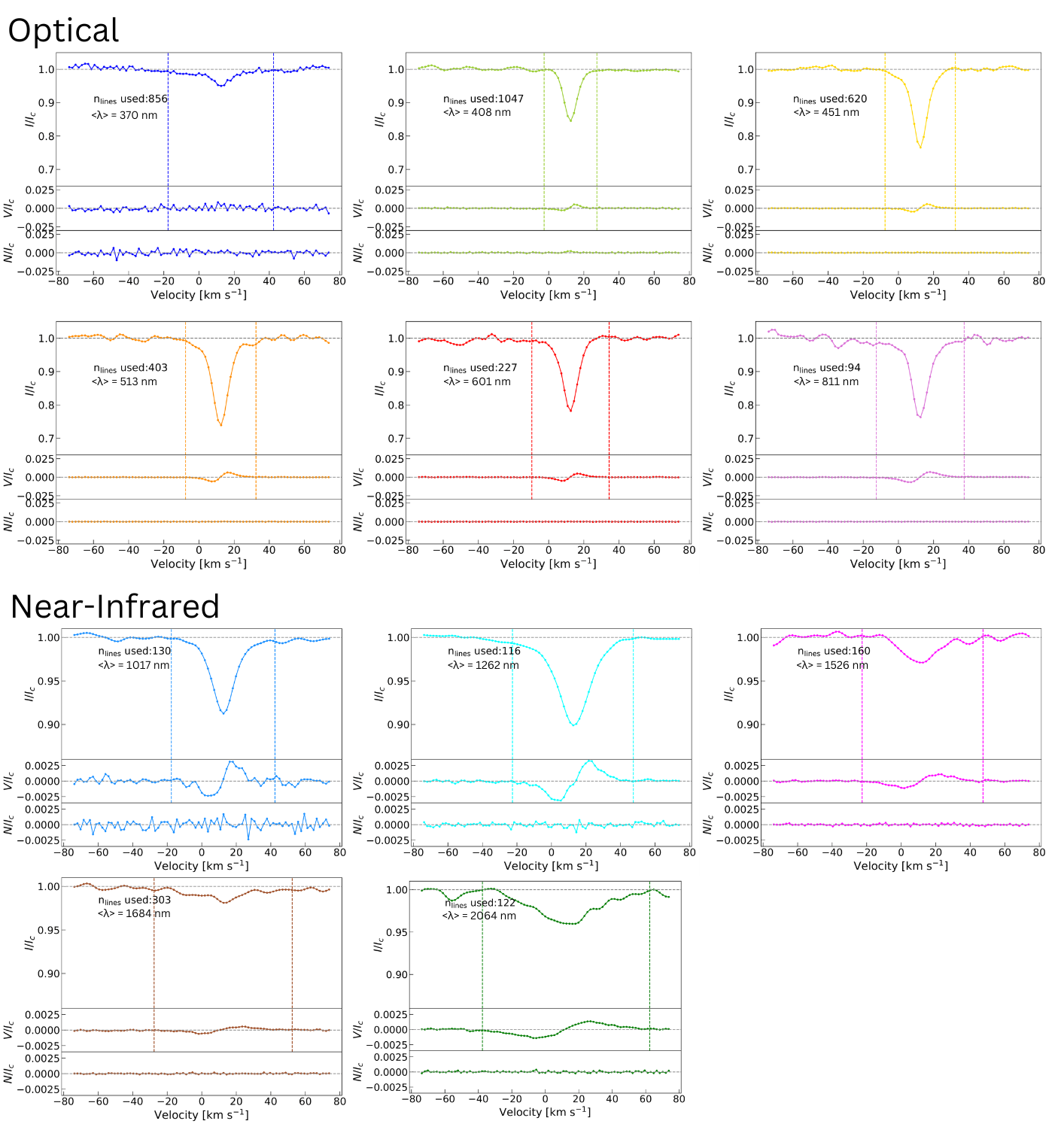}
    \caption{Series of Stokes profiles computed with subsets of the line list used for LSD, as described in Sec.\ref{sec:discussion}. Every panel displays Stokes~$I$ (top), $V$ (middle) and $N$ (bottom) for each line subset and vertical dashed lines indicate the adjusted velocity range over which B$_l$ was estimated. The optical profiles are obtained by stacking the ESPaDOnS observations from 2019, whereas the near-infrared ones from SPIRou 2019b time series, for each panel. The number of lines used in the LSD computation as well as the mean wavelength are displayed.}
    \label{fig:Stokes_chromatic}
\end{figure*}

\section{Zeeman broadening examples}\label{sec:zbroad_appendix}

\begin{figure*}[htb]
    \centering
    \includegraphics[width=\textwidth]{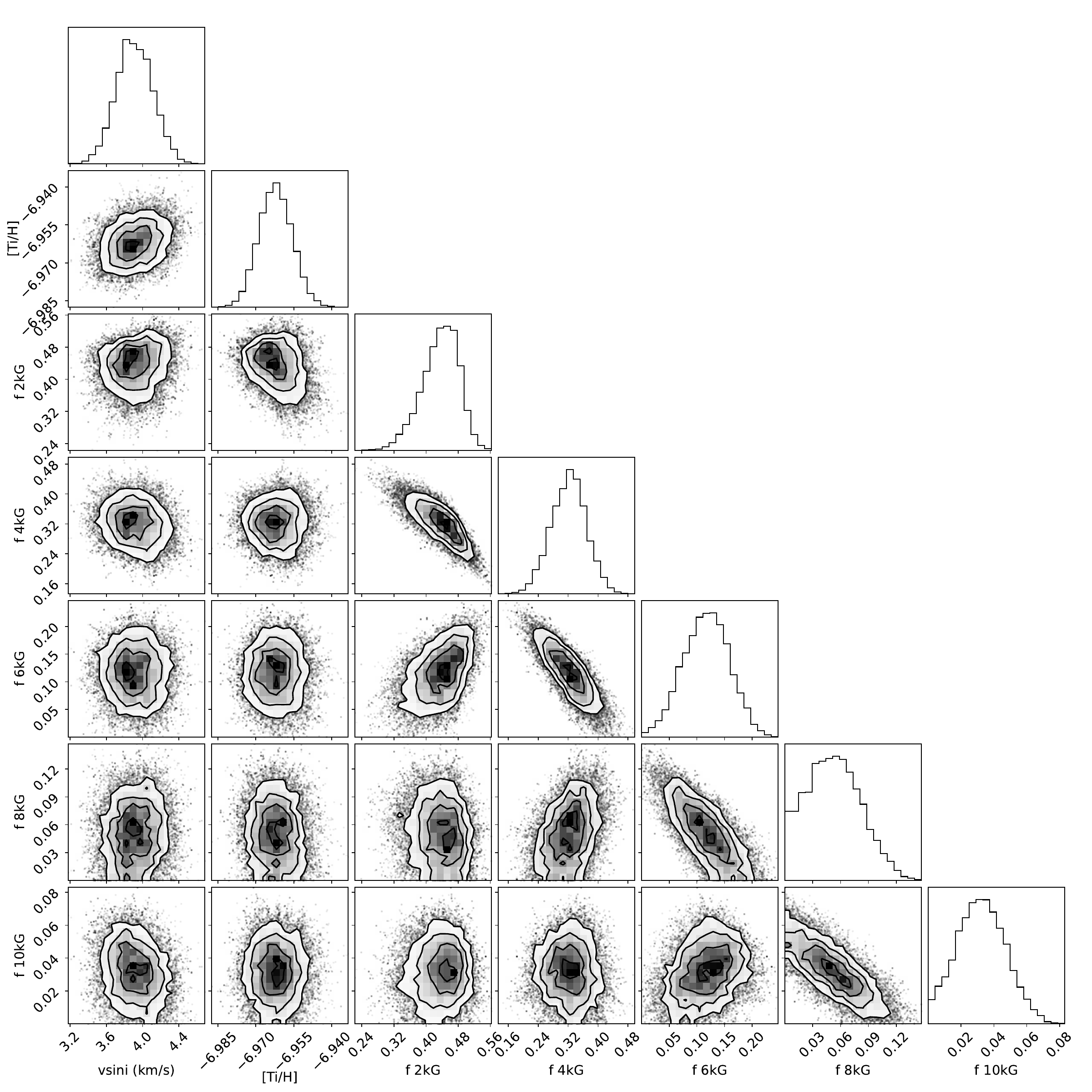}
    \caption{Posterior distribution for parameters from the Zeeman broadening analysis of the ESPaDOnS observation on 24 Febuary 2016.  The filling factor for a given magnetic field strength is $f$, and abundances are in $\log n_X / n_{\rm H}$ units.}
    \label{fig:zbroad-cornerVis}%
\end{figure*}

\begin{figure*}[htb]
    \centering
    \includegraphics[width=0.85\textwidth]{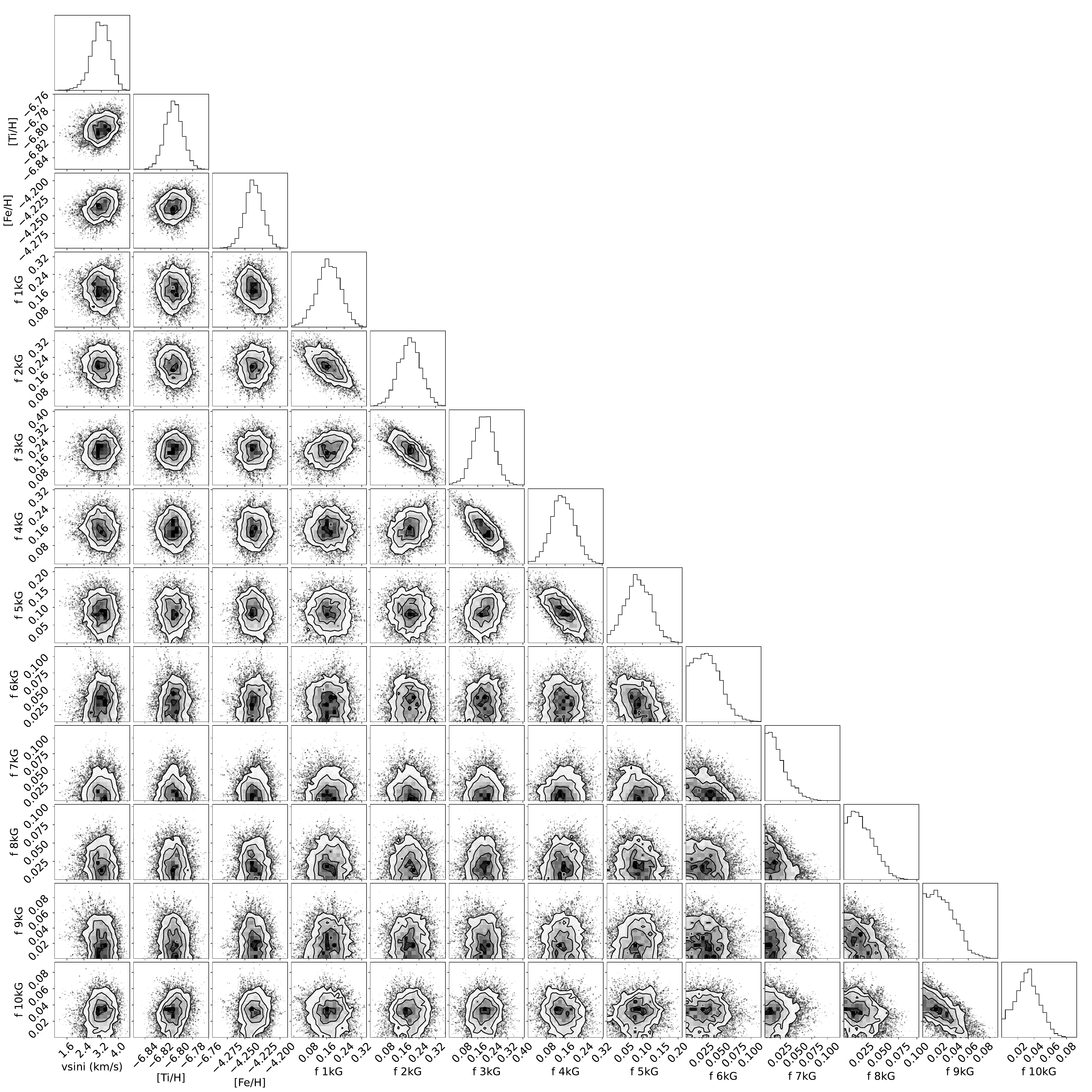}
    \caption{Posterior distribution for parameters from the Zeeman broadening analysis of the SPIRou observation on 3 November 2020.  }
    \label{fig:zbroad-cornerIR}%
\end{figure*}

Example plots of the posterior distributions from the Zeeman broadening MCMC analysis are shown in Fig.~\ref{fig:zbroad-cornerVis} for a ESPaDOnS observation and in Fig.~\ref{fig:zbroad-cornerIR} for a SPIRou observation.  Summaries of the results of the MCMC analysis for each epoch are provided in Tables \ref{tab:zbroad-vis} and \ref{tab:zbroad-ir}.

\begin{table*}
\caption{Parameters from the Zeeman broadening analysis, averaged for each epoch, for ESPaDOnS and Narval optical observations. Uncertainties are the standard deviations for each epoch.}
\centering
\begin{tabular}{cccccccc}
\hline
epoch         & 2006 & 2007 & 2008 & 2011 & 2012 & 
2016 & 2019\\
\hline
$v_e\sin(i)$ [km\,s$^{-1}$]     & 4.27$\pm$0.38 & 4.25$\pm$0.17 & 4.16$\pm$0.29 & 3.46$\pm$0.50 & 2.65$\pm$0.48  & 3.57$\pm$0.21  & 3.39$\pm$0.45 \\
$[$Ti/H$]$    & $-$6.97$\pm$0.01 & $-$6.90$\pm$0.01 & $-$6.93$\pm$0.02 & $-$6.99$\pm$0.01 & $-$7.00$\pm$0.01 & $-$6.96$\pm$0.01 & $-$6.95$\pm$0.01\\
$f_{0\rm kG}$ & 0.06$\pm$0.02 & 0.24$\pm$0.06 & 0.18$\pm$0.06 & 0.05$\pm$0.06 & 0.02$\pm$0.01 & 0.05$\pm$0.02 & 0.06$\pm$0.01\\
$f_{2 \rm kG}$& 0.46$\pm$0.01 & 0.32$\pm$0.07 & 0.39$\pm$0.07 & 0.60$\pm$0.06 & 0.52$\pm$0.02 & 0.44$\pm$0.01 & 0.41$\pm$0.01\\
$f_{4 \rm kG}$& 0.33$\pm$0.02 & 0.31$\pm$0.02 & 0.28$\pm$0.04 & 0.19$\pm$0.07 & 0.29$\pm$0.01 & 0.33$\pm$0.01 & 0.36$\pm$0.01\\
$f_{6 \rm kG}$& 0.09$\pm$0.01 & 0.06$\pm$0.01 & 0.08$\pm$0.01 & 0.11$\pm$0.05 & 0.09$\pm$0.01 & 0.11$\pm$0.01 & 0.09$\pm$0.01\\
$f_{8 \rm kG}$& 0.05$\pm$0.01 & 0.03$\pm$0.01 & 0.04$\pm$0.01 & 0.04$\pm$0.03 & 0.05$\pm$0.01 & 0.05$\pm$0.01 & 0.05$\pm$0.01\\
$f_{10\rm kG}$& 0.04$\pm$0.01 & 0.02$\pm$0.01 & 0.02$\pm$0.01 & 0.03$\pm$0.03 & 0.02$\pm$0.01 & 0.04$\pm$0.01 & 0.04$\pm$0.01\\
$\sum Bf$ [kG] & 3.59$\pm$0.04 & 2.82$\pm$0.13 & 2.97$\pm$0.13 & 3.22$\pm$0.15 & 3.37$\pm$0.06 & 3.65$\pm$0.04 & 3.62$\pm$0.03\\
\hline
\end{tabular}
\label{tab:zbroad-vis}
\end{table*}

\begin{table*}
\caption{Parameters from the Zeeman broadening analysis, averaged for each epoch, for SPIRou observations. Uncertainties are the standard deviations for each epoch.}
\centering
\begin{tabular}{ccccc}
\hline
epoch         & 2019a & 2019b & 2020a & 2020b\\
\hline
$v_\mathrm{eq}\sin(i)$ [km\,s$^{-1}$]     & 3.42$\pm$0.23 & 3.34$\pm$0.18 & 3.48$\pm$0.17 & 3.48$\pm$0.21\\
$[$Ti/H$]$    & $-$6.78$\pm$0.02 & $-$6.78$\pm$0.02 & $-$6.79$\pm$0.02 & $-$6.78$\pm$0.01\\
$[$Fe/H$]$    & $-$4.20$\pm$0.01 & $-$4.22$\pm$0.01 & $-$4.20$\pm$0.03 & $-$4.21$\pm$0.02\\
$f_{0\rm kG}$ & 0.06$\pm$0.02 & 0.05$\pm$0.02 & 0.05$\pm$0.02 & 0.06$\pm$0.01\\
$f_{1 \rm kG}$& 0.15$\pm$0.02 & 0.12$\pm$0.03 & 0.15$\pm$0.02 & 0.17$\pm$0.02\\
$f_{2 \rm kG}$& 0.20$\pm$0.04 & 0.23$\pm$0.05 & 0.19$\pm$0.04 & 0.19$\pm$0.04\\
$f_{3 \rm kG}$& 0.24$\pm$0.04 & 0.19$\pm$0.05 & 0.23$\pm$0.05 & 0.24$\pm$0.04\\
$f_{4 \rm kG}$& 0.12$\pm$0.04 & 0.16$\pm$0.04 & 0.14$\pm$0.03 & 0.13$\pm$0.03\\
$f_{5 \rm kG}$& 0.06$\pm$0.02 & 0.08$\pm$0.02 & 0.06$\pm$0.02 & 0.06$\pm$0.01\\
$f_{6 \rm kG}$& 0.05$\pm$0.01 & 0.05$\pm$0.02 & 0.05$\pm$0.01 & 0.04$\pm$0.01\\
$f_{7 \rm kG}$& 0.03$\pm$0.01 & 0.02$\pm$0.01 & 0.03$\pm$0.01 & 0.03$\pm$0.01\\
$f_{8 \rm kG}$& 0.02$\pm$0.01 & 0.02$\pm$0.01 & 0.02$\pm$0.01 & 0.02$\pm$0.01\\
$f_{9 \rm kG}$& 0.03$\pm$0.01 & 0.03$\pm$0.01 & 0.03$\pm$0.01 & 0.03$\pm$0.01\\
$f_{10\rm kG}$& 0.03$\pm$0.01 & 0.03$\pm$0.01 & 0.04$\pm$0.01 & 0.03$\pm$0.01\\
$\sum Bf$ [kG] & 3.40$\pm$0.09 & 3.49$\pm$0.09 & 3.44$\pm$0.04 & 3.29$\pm$0.07\\
\hline
\end{tabular}
\label{tab:zbroad-ir}
\end{table*}

\section{Observing log}

This appendix contains the journal of observations of AD Leo, for both optical and near-infrared observations. It also includes all measurements of longitduinal magnetic field and $\Sigma Bf$.

\onecolumn
\begin{longtable}{lcrccc}
\caption{\label{tab:log} List of AD~Leo observations collected with SPIRou. The columns are: (1 and 2) date and universal time of the observations, (3) rotational cycle of the observations found using Eq.~\ref{eq:ephemeris}, (4) exposure time of a polarimetric sequence, (5) signal-to-noise ratio at 1650 nm per spectral element, (6) RMS noise level of Stokes $V$ signal in units of unpolarised continuum.} \\
\hline\hline
Date & UT & $n_\mathrm{cyc}$ & $t_{exp}$ & S/N & $\sigma_\mathrm{LSD}$\\
 & [hh:mm:ss] & & [s] & & [$10^{-4}I_c$]\\
\hline
\endfirsthead
\caption{continued.}\\
\hline\hline
Date & UT & $n_\mathrm{cyc}$ & $t_{exp}$ & S/N & $\sigma_\mathrm{LSD}$\\
 & [hh:mm:ss] & & [s] & & [$10^{-4}I_c$]\\
\hline
\endhead
\hline
\endfoot
\multicolumn{6}{c}{2019}\\
\hline
April 15       &    06:11:02.35    & 0.00 &     4x61     & 151 & 1.9\\
April 16       &    12:03:07.61    & 0.56 &     4x61     & 130 & 1.8\\
April 18       &    09:00:59.77    & 1.40 &     4x61     & 138 & 1.9\\
April 19       &    10:36:14.55    & 1.88 &     4x61     & 143 & 1.8\\
April 20       &    08:56:19.44    & 2.29 &     4x61     & 143 & 1.8\\
April 21       &    05:42:43.43    & 2.68 &     4x61     & 165 & 1.5\\
April 22       &    08:58:10.67    & 3.19 &     4x61     & 147 & 1.9\\
April 23       &    07:01:28.73    & 3.60 &     4x61     & 154 & 1.9\\
April 24       &    06:04:06.74    & 4.03 &     4x61     & 147 & 1.9\\
April 25       &    09:41:45.90    & 4.55 &     4x61     & 152 & 1.9\\
April 26       &    08:13:21.75    & 4.97 &     4x61     & 162 & 1.9\\
April 27       &    08:17:48.22    & 5.42 &     4x61     & 139 & 2.1\\
May 01         &    08:59:11.25    & 7.23 &     4x61     & 153 & 1.8\\
May 15         &    06:11:38.01    & 13.45 &     4x61     & 165 & 1.6\\
June 13        &    06:30.59.49    & 26.46 &     4x61     & 186 & 2.1\\
June 14        &    05:44:29.79    & 26.89 &     4x61     & 193 & 2.0\\
June 15        &    06:04:31.16    & 27.35 &     4x61     & 192 & 2.1\\
June 16        &    05:44:52.21    & 27.79 &     4x61     & 173 & 1.8\\
June 17        &    06:08:46.48    & 28.25 &     4x61     & 150 & 1.8\\
June 19        &    05:47:00.42    & 29.14 &     4x61     & 175 & 1.5\\
June 21        &    06:18:29.14    & 30.05 &     4x61     & 169 & 1.7\\
October 16     &    15:31:29.14    & 82.69 &     4x61     & 180 & 1.4\\
October 31     &    15:32:56.51    & 89.41 &     4x61     & 169 & 1.6\\
November 01    &    15:22:10.08    & 89.86 &     4x61     & 159 & 1.5\\
November 02    &    15:36:52.90    & 90.31 &     4x61     & 170 & 1.4\\
November 03    &    14:53:22.07    & 90.75 &     4x61     & 151 & 1.5\\
November 04    &    15:43:18.06    & 91.21 &     4x61     & 137 & 1.9\\
November 05    &    15:33:34.93    & 91.66 &     4x61     & 164 & 1.3\\
November 06    &    15:37:27.47    & 92.10 &     4x61     & 190 & 1.5\\
November 07    &    15:01:11.73    & 92.54 &     4x61     & 165 & 1.5\\
November 09    &    14:03:02.38    & 93.42 &     4x61     & 116 & 1.6\\
November 10    &    15:51:32.68    & 93.90 &     4x61     & 151 & 1.5\\
November 13    &    14:47:09.89    & 95.22 &     4x61     & 201 & 1.8\\
November 14    &    14:13:58.20    & 95.66 &     4x61     & 197 & 1.4\\
December 05    &    15:12:55.68    & 105.10 &     4x61     & 116 & 1.4\\
December 05    &    15:35:03.10    & 105.10 &     4x61     & 68 & 1.5\\
December 07    &    15:08:39.83    & 105.99 &     4x61     & 133 & 2.2\\
December 08    &    14:29:19.93    & 106.43 &     4x61     & 181 & 1.8\\
December 09    &    14:21:48.74    & 106.88 &     4x61     & 193 & 1.5\\
December 10    &    13:15:39.48    & 107.31 &     4x61     & 194 & 1.5\\
December 11    &    14:58:07.44    & 107.79 &     4x61     & 189 & 3.0\\
December 12    &    14:33:27.45    & 108.23 &     4x61     & 190 & 3.6\\
\hline
\multicolumn{6}{c}{2020}\\
\hline
January 26     &    12:11:48.90    & 128.36 &     4x61     & 218 & 2.0\\
January 27     &    12:07:16.45    & 128.81 &     4x61     & 166 & 1.5\\
January 28     &    12:06:25.46    & 129.26 &     4x61     & 193 & 1.5\\
February 05    &    08:18:51.61    & 132.78 &     4x61     & 174 & 1.4\\
February 16    &    07:39:02.87    & 137.70 &     4x61     & 193 & 1.8\\
February 17    &    09:20:45.46    & 138.18 &     4x61     & 171 & 1.9\\
February 18    &    07:39:13.56    & 138.59 &     4x61     & 191 & 1.2\\
February 19    &    08:37:09.84    & 139.06 &     4x61     & 210 & 1.5\\
March 12       &    09:16:56.32    & 148.94 &     4x61     & 217 & 2.0\\
May 08         &    05:59:01.58    & 174.43 &     4x61     & 194 & 1.4\\
May 09         &    09:42:30.22    & 174.95 &     4x61     & 206 & 1.5\\
May 12         &    09:39:45.19    & 176.30 &     4x61     & 177 & 1.4\\
May 13         &    09:43:17.10    & 176.75 &     4x61     & 204 & 1.4\\
May 14         &    07:46:35.66    & 177.16 &     4x61     & 208 & 1.5\\
May 15         &    09:50:31.27    & 177.65 &     4x61     & 164 & 1.6\\
May 31         &    06:22:44.29    & 184.76 &     4x61     & 215 & 1.2\\
June 01        &    07:29:57.32    & 185.23 &     4x61     & 189 & 1.3\\
June 02        &    06:23:31.70    & 185.65 &     4x61     & 202 & 1.4\\
June 03        &    07:38:43.10    & 186.13 &     4x61     & 196 & 1.3\\
June 04        &    06:32:43.13    & 186.55 &     4x61     & 197 & 1.3\\
June 05        &    06:42:42.88    & 187.01 &     4x61     & 139 & 1.8\\
June 06        &    07:53:54.53    & 187.48 &     4x61     & 167 & 1.2\\
June 07        &    06:58:07.88    & 187.91 &     4x61     & 154 & 1.6\\
June 08        &    06:03:47.85    & 188.34 &     4x61     & 135 & 1.3\\
June 08        &    06:10:09.48    & 188.34 &     4x61     & 143 & 1.4\\
June 09        &    06:31:50.17    & 188.79 &     4x61     & 120 & 1.3\\
June 10        &    06:57:13.44    & 189.25 &     4x61     & 200 & 1.8\\
October 31         &    15:06:44.49    & 253.75 &     4x61     & 205 & 1.5\\
November 03         &    15:29:52.78    & 255.11 &     4x61     & 220 & 1.5\\
\hline
\end{longtable}

\begin{longtable}{lcrccc}
\caption{\label{tab:log_opt} List of AD~Leo observations collected with ESPaDOnS in 2019. The columns are: (1 and 2) date and universal time of the observations, (3) rotational cycle of the observations found using Eq.~\ref{eq:ephemeris}, (4) exposure time of a polarimetric sequence, (5) signal-to-noise ratio at 650 nm per spectral element, (6) RMS noise level of Stokes $V$ signal in units of unpolarised continuum.} \\
\hline\hline
Date & UT & $n_\mathrm{cyc}$ & $t_{exp}$ & S/N & $\sigma_\mathrm{LSD}$\\
 & [hh:mm:ss] & & [s] & & [$10^{-4}I_c$]\\
\hline
November 15   &   13:24:01.40   & 96.32 &    4x300  & 234 & 1.7\\
November 16   &   14:20:01.30   & 96.79 &    4x300  & 230 & 1.9\\
November 19   &   14:08:04.00   & 98.13 &    4x300  & 151 & 3.6\\
November 19   &   14:33:03.00   & 98.14 &    4x300  & 159 & 3.2\\
November 19   &   14:56:04.00   & 98.15 &    4x300  & 180 & 2.8\\
November 21   &   15:43:01.00   & 99.06 &    4x300  & 265 & 1.8\\

\hline
\end{longtable}

\renewcommand{\arraystretch}{1.5}
\begin{longtable}{cccc}
\caption{\label{tab:Bl_log} List of optical and near-infrared measurements of longitudinal magnetic field and magnetic flux. The columns are: (1) Heliocentric Julian date of the observation, (2) B$_l$ with formal error bar (see Eq.~\ref{eq:Bl}), and (3) magnetic flux from Zeeman broadening modelling, when a reliable measurement was possible, and (4) the instrument used.} \\
\hline\hline
HJD & B$_l$ & $Bf$ & Instrument\\
 $[-2450000]$ & [G]  & [kG] &\\
\hline
\endfirsthead
\caption{continued.}\\
\hline\hline
HJD & B$_l$ & $Bf$ & Instrument\\
 $[-2450000]$ & [G] & [kG]&\\
\hline
\endhead
\hline
\endfoot
3747.0876   & $-$269.4 $\pm$ 26.4 & $3.65^{+0.07}_{-0.08}$ & ESPaDOnS \\
3748.8868   & $-$272.4 $\pm$ 15.1 & $3.52^{+0.09}_{-0.10}$ & ESPaDOnS \\
3780.0705   & $-$280.7 $\pm$ 13.1 & $3.55^{+0.10}_{-0.10}$ & ESPaDOnS \\
3895.8047   & $-$291.2 $\pm$ 10.7 & $3.62^{+0.08}_{-0.09}$ & ESPaDOnS \\
3896.8124   & $-$260.1 $\pm$ 7.6  & $3.61^{+0.08}_{-0.09}$ & ESPaDOnS \\
3897.8005   & $-$266.7 $\pm$ 7.4  & $3.63^{+0.08}_{-0.09}$ & ESPaDOnS \\
3898.7785   & $-$294.2 $\pm$ 8.1  & $3.58^{+0.08}_{-0.10}$ & ESPaDOnS \\
4127.5975   & $-$294.6 $\pm$ 16.5 & $\ldots$ & Narval \\
4128.6088   & $-$248.6 $\pm$ 10.8 & $2.95^{+0.16}_{-0.16}$ & Narval \\
4129.5717   & $-$296.2 $\pm$ 11.8 & $2.85^{+0.21}_{-0.22}$ & Narval \\
4130.6084   & $-$253.8 $\pm$ 9.4  & $2.65^{+0.21}_{-0.20}$ & Narval \\
4133.6312   & $-$274.8 $\pm$ 10.4 & $\ldots$ & Narval \\
4134.6112   & $-$271.7 $\pm$ 11.4 & $\ldots $ & Narval \\
4135.6217   & $-$231.4 $\pm$ 10.4 & $\ldots $ & Narval \\
4136.5925   & $-$294.9 $\pm$ 13.0 & $\ldots $ & Narval \\
4276.7715   & $-$261.0 $\pm$ 8.2  & $\ldots $ & ESPaDOns \\
4485.5177   & $-$290.1 $\pm$ 13.2 & $3.18^{+0.17}_{-0.20}$ & Narval \\
4489.5683   & $-$248.6 $\pm$ 10.3 & $3.10^{+0.17}_{-0.18}$ & Narval \\
4492.5379   & $-$285.3 $\pm$ 10.7 & $\ldots $ & Narval \\
4493.5486   & $-$204.6 $\pm$ 10.6 & $3.03^{+0.18}_{-0.19}$ & Narval \\
4495.5611   & $-$227.2 $\pm$ 12.6 & $2.81^{+0.19}_{-0.19}$ & Narval \\
4499.5675   & $-$256.8 $\pm$ 11.1 & $\ldots $ & Narval \\
4501.5473   & $-$288.6 $\pm$ 12.1 & $\ldots $ & Narval \\
4502.5475   & $-$202.8 $\pm$ 9.7  & $\ldots $ & Narval \\
4506.5576   & $-$218.9 $\pm$ 10.0 & $2.84^{+0.19}_{-0.20}$ & Narval \\
4508.5516   & $-$265.5 $\pm$ 11.0 & $\ldots $ & Narval \\
4509.5564   & $-$219.6 $\pm$ 12.2 & $2.90^{+0.18}_{-0.20}$ & Narval \\
4510.5523   & $-$286.6 $\pm$ 15.3 & $\ldots $ & Narval \\
4511.5694   & $-$200.2 $\pm$ 10.5 & $2.86^{+0.17}_{-0.17}$ & Narval \\
4512.5537   & $-$296.7 $\pm$ 10.8 & $2.99^{+0.16}_{-0.16}$ & Narval \\
5896.7560   & $-$249.3 $\pm$ 11.5 & $3.22^{+0.15}_{-0.16}$ & Narval \\
5934.6407   & $-$247.0 $\pm$ 9.5  & $3.39^{+0.11}_{-0.12}$ & Narval \\
5935.6765   & $-$225.9 $\pm$ 9.3  & $3.50^{+0.09}_{-0.11}$ & Narval \\
5936.6050   & $-$241.7 $\pm$ 10.9 & $3.38^{+0.09}_{-0.10}$ & Narval \\
5937.7575   & $-$254.6 $\pm$ 11.2 & $3.44^{+0.08}_{-0.10}$ & Narval \\
5938.6659   & $-$243.9 $\pm$ 8.8  & $3.31^{+0.08}_{-0.10}$ & Narval \\
5939.6031   & $-$234.2 $\pm$ 10.6 & $3.34^{+0.08}_{-0.10}$ & Narval \\
5940.6416   & $-$203.8 $\pm$ 9.1  & $3.31^{+0.09}_{-0.09}$ & Narval \\
5941.6411   & $-$232.7 $\pm$ 9.7  & $3.36^{+0.11}_{-0.13}$ & Narval \\
5942.6256   & $-$220.5 $\pm$ 10.2 & $3.30^{+0.09}_{-0.10}$ & Narval \\
7435.7957   & $-$177.0 $\pm$ 5.5  & $3.68^{+0.07}_{-0.08}$ & ESPaDOnS \\
7436.8831   & $-$165.8 $\pm$ 5.3  & $3.67^{+0.08}_{-0.09}$ & ESPaDOnS \\
7441.8954   & $-$181.4 $\pm$ 5.3  & $3.57^{+0.06}_{-0.07}$ & ESPaDOnS \\
7443.0074   & $-$166.1 $\pm$ 5.3  & $3.60^{+0.09}_{-0.10}$ & ESPaDOnS \\
7447.9103   & $-$160.7 $\pm$ 6.4  & $3.65^{+0.14}_{-0.16}$ & ESPaDOnS \\
7449.0011   & $-$182.9 $\pm$ 7.7  & $3.66^{+0.11}_{-0.13}$ & ESPaDOnS \\
7449.9154   & $-$154.6 $\pm$ 6.5  & $3.66^{+0.12}_{-0.15}$ & ESPaDOnS \\
7450.8328   & $-$190.5 $\pm$ 6.9  & $3.70^{+0.09}_{-0.11}$ & ESPaDOnS \\
7495.8253   & $-$192.9 $\pm$ 8.8  & $3.66^{+0.12}_{-0.15}$ & ESPaDOnS \\
7498.7442   & $-$170.6 $\pm$ 5.8  & $3.68^{+0.09}_{-0.09}$ & ESPaDOnS \\
8588.7573  &  $-$194.7$\pm$17.2  & $3.40^{+0.08}_{-0.09}$ & SPIRou \\
8590.0016 & $-$212.3$\pm$19.9 & $\ldots$ & SPIRou \\
8592.9410  &  $-$206.5$\pm$17.4  & $3.33^{+0.09}_{-0.11}$ &  SPIRou \\
8593.8715  &  $-$202.5$\pm$16.5  & $3.37^{+0.11}_{-0.11}$ &  SPIRou \\
8594.7370  &  $-$234.3$\pm$16.8  & $3.42^{+0.10}_{-0.11}$ &  SPIRou \\
8595.8726  &  $-$211.7$\pm$16.4  & $3.47^{+0.10}_{-0.13}$ &  SPIRou \\
8596.7915  &  $-$217.9$\pm$16.0  & $3.37^{+0.09}_{-0.09}$ &  SPIRou \\
8597.7516  &  $-$219.1$\pm$18.6  & $3.33^{+0.10}_{-0.09}$ &  SPIRou \\
8598.9026  &  $-$240.0$\pm$16.9  & $3.49^{+0.09}_{-0.09}$ &  SPIRou \\
8599.8411  &  $-$263.6$\pm$17.6  & $3.36^{+0.09}_{-0.10}$ &  SPIRou \\
8600.8441  &  $-$182.9$\pm$16.7  & $3.36^{+0.10}_{-0.11}$ &  SPIRou \\
8604.8725  &  $-$190.5$\pm$16.3  & $3.27^{+0.11}_{-0.12}$ &  SPIRou \\
8618.7549  &  $-$216.5$\pm$15.1  & $3.27^{+0.08}_{-0.09}$ &  SPIRou \\
8647.7665  &  $-$217.0$\pm$12.9  & $3.26^{+0.09}_{-0.09}$ &  SPIRou \\
8648.7341  &  $-$221.3$\pm$13.1  & $3.38^{+0.09}_{-0.09}$ &  SPIRou \\
8649.7480  &  $-$230.0$\pm$15.3  & $3.23^{+0.10}_{-0.10}$ &  SPIRou \\
8650.7343  &  $-$194.7$\pm$13.6  & $3.49^{+0.09}_{-0.10}$ &  SPIRou \\
8651.7509  &  $-$209.4$\pm$16.0  & $3.49^{+0.08}_{-0.10}$ &  SPIRou \\
8653.7357  &  $-$227.2$\pm$13.6  & $3.38^{+0.09}_{-0.09}$ &  SPIRou \\
8655.7575  &  $-$250.3$\pm$20.3  & $3.42^{+0.09}_{-0.10}$ &  SPIRou \\
8773.1472  &  $-$223.6$\pm$14.4  & $3.39^{+0.10}_{-0.10}$ &  SPIRou \\
8788.1497  &  $-$223.7$\pm$13.4  & $3.32^{+0.09}_{-0.10}$ &  SPIRou \\
8789.1423  &  $-$184.8$\pm$13.7  & $3.43^{+0.10}_{-0.09}$ &  SPIRou \\
8790.1526  &  $-$245.2$\pm$15.4  & $3.50^{+0.09}_{-0.10}$ &  SPIRou \\
8791.1225  &  $-$192.7$\pm$13.8  & $3.50^{+0.12}_{-0.12}$ &  SPIRou \\
8792.1572  &  $-$256.3$\pm$16.3  & $3.46^{+0.09}_{-0.10}$ &  SPIRou \\
8793.1506  &  $-$204.8$\pm$13.6  & $3.50^{+0.09}_{-0.10}$ &  SPIRou \\
8794.1534  &  $-$246.5$\pm$13.5  & $3.53^{+0.09}_{-0.09}$ &  SPIRou \\
8795.1283  &  $-$209.1$\pm$15.4  & $3.48^{+0.09}_{-0.09}$ &  SPIRou \\
8797.0881  &  $-$189.3$\pm$25.7  & $3.51^{+0.10}_{-0.09}$ &  SPIRou \\
8798.1635  &  $-$202.8$\pm$17.8  & $3.63^{+0.10}_{-0.10}$ &  SPIRou \\
8801.1190  &  $-$254.3$\pm$14.6  & $3.56^{+0.11}_{-0.12}$ &  SPIRou \\
8802.0961  &  $-$186.1$\pm$14.1  & $3.58^{+0.08}_{-0.09}$ &  SPIRou \\
8803.0577  &  $-$208.3$\pm$7.0   & $3.67^{+0.13}_{-0.14}$ & ESPaDOnS\\
8804.0967  &  $-$189.1$\pm$7.2   & $3.58^{+0.10}_{-0.11}$ & ESPaDOnS\\
8807.0890  &  $-$178.9$\pm$12.9  & $3.65^{+0.07}_{-0.09}$ & ESPaDOnS\\
8807.1062  &  $-$198.8$\pm$11.5  & $3.61^{+0.08}_{-0.09}$ & ESPaDOnS\\
8807.1223  &  $-$169.3$\pm$10.1  & $3.63^{+0.07}_{-0.07}$ & ESPaDOnS\\
8809.1547  &  $-$161.1$\pm$6.4   & $3.57^{+0.08}_{-0.09}$ & ESPaDOnS\\
8823.1385  &  $-$254.6$\pm$30.7  & $3.48^{+0.08}_{-0.09}$ &  SPIRou \\
8823.1539  &  $-$242.9$\pm$42.2  & $3.69^{+0.10}_{-0.09}$ &  SPIRou \\
8825.1357  &  $-$193.2$\pm$22.1  & $3.60^{+0.08}_{-0.08}$ &  SPIRou \\
8826.1084  &  $-$236.2$\pm$14.4  & $3.41^{+0.09}_{-0.09}$ &  SPIRou \\
8827.1033  &  $-$178.9$\pm$13.7  & $3.42^{+0.09}_{-0.09}$ &  SPIRou \\
8828.0574  &  $-$249.2$\pm$13.9  & $3.40^{+0.11}_{-0.11}$ &  SPIRou \\
8829.1286  &  $-$152.6$\pm$14.4  & $3.44^{+0.10}_{-0.10}$ &  SPIRou \\
8830.1115  &  $-$248.1$\pm$14.8  & $3.35^{+0.10}_{-0.09}$ &  SPIRou \\
8875.0136  &  $-$176.0$\pm$12.2  & $3.43^{+0.09}_{-0.10}$ &  SPIRou \\
8876.0104  &  $-$118.6$\pm$13.3  & $3.54^{+0.08}_{-0.09}$ &  SPIRou \\
8877.0098  &  $-$238.7$\pm$14.5  & $3.40^{+0.10}_{-0.11}$ &  SPIRou \\
8884.8515  &  $-$122.9$\pm$12.6  & $3.42^{+0.10}_{-0.11}$ &  SPIRou \\
8895.8233  &  $-$132.6$\pm$11.9  & $3.40^{+0.10}_{-0.10}$ &  SPIRou \\
8896.8939  &  $-$219.3$\pm$13.7  & $3.44^{+0.11}_{-0.10}$ &  SPIRou \\
8897.8233  &  $-$157.3$\pm$12.7  & $3.43^{+0.09}_{-0.11}$ &  SPIRou \\
8898.8635  &  $-$216.3$\pm$13.3  & $3.44^{+0.10}_{-0.11}$ &  SPIRou \\
8920.8895  &  $-$176.4$\pm$12.6  & $3.42^{+0.09}_{-0.10}$ &  SPIRou \\
8977.7467  &  $-$177.7$\pm$11.8  & $3.27^{+0.09}_{-0.10}$ &  SPIRou \\
8978.9018  &  $-$114.9$\pm$11.5  & $3.33^{+0.09}_{-0.08}$ &  SPIRou \\
8981.8996  &  $-$214.6$\pm$14.0  & $3.28^{+0.10}_{-0.11}$ &  SPIRou \\
8982.9020  &  $-$81.2$\pm$11.5   & $3.43^{+0.11}_{-0.10}$ &  SPIRou \\
8983.8209  &  $-$190.0$\pm$11.2  & $3.27^{+0.08}_{-0.09}$ &  SPIRou \\
8984.9069  &  $-$100.5$\pm$16.8  & $3.44^{+0.11}_{-0.11}$ &  SPIRou \\
9000.7614  &  $-$50.4$\pm$10.5   & $3.26^{+0.12}_{-0.11}$ &  SPIRou \\
9001.8080  &  $-$217.7$\pm$17.7  & $3.32^{+0.09}_{-0.10}$ &  SPIRou \\
9002.7618  &  $-$61.7$\pm$11.9   & $3.20^{+0.09}_{-0.11}$ &  SPIRou \\
9003.8140  &  $-$199.2$\pm$12.7  & $3.18^{+0.10}_{-0.11}$ &  SPIRou \\
9004.7681  &  $-$110.4$\pm$11.8  & $3.26^{+0.09}_{-0.09}$ &  SPIRou \\
9005.7750  &  $-$96.4$\pm$15.5   & $3.30^{+0.09}_{-0.09}$ &  SPIRou \\
9006.8244  &  $-$119.8$\pm$14.9  & $3.30^{+0.09}_{-0.10}$ &  SPIRou \\
9007.7856  &  $-$70.0$\pm$15.1   & $3.30^{+0.10}_{-0.10}$ &  SPIRou \\
9008.7478  &  $-$178.4$\pm$16.2  & $3.28^{+0.10}_{-0.10}$ &  SPIRou \\
9008.7522  &  $-$185.2$\pm$16.6  & $3.30^{+0.09}_{-0.10}$ &  SPIRou \\
9009.7672  &  $-$64.5$\pm$18.5   & $3.35^{+0.10}_{-0.11}$ &  SPIRou \\
9010.7848  &  $-$194.7$\pm$13.2  & $3.17^{+0.10}_{-0.09}$ &  SPIRou \\
9154.1315  &  $-$62.2$\pm$10.1   & $3.31^{+0.07}_{-0.07}$ &  SPIRou \\
9157.1479  &  $-$46.1$\pm$9.9    & $3.28^{+0.08}_{-0.09}$ &  SPIRou \\
\hline
\end{longtable}

\end{document}